\documentclass{ws-ijgmmp}
\usepackage{graphicx,amssymb,amsmath,amsfonts}
\usepackage{subfigure}
\usepackage[linktocpage,pagebackref]{hyperref}
\usepackage{color}
\begin{document}

\markboth{Parthapratim Pradhan}{(Circular Geodesics in Tidal Charged Black Hole)}

\title{Circular Geodesics in Tidal Charged Black Hole}
\author{Parthapratim Pradhan\footnote{pppradhan77@gmail.com.}}

\address{Department of Physics, Hiralal Mazumdar Memorial College For Women,
Dakshineswar, Kolkata-700035,  India.}

\maketitle

\begin{history}
\received{Day Month Year}
\revised{Day Month Year}
\comby{Managing Editor}
\end{history}

\begin{abstract}
We study the existence and stability criteria  for circular geodesics of spherically symmetric tidal 
charged  black hole~(BH). We investigate in details the equatorial causal geodesics of the tidal charged 
BH in comparison with spherically symmetric Reissner-Nordstr\"{o}m BH  spacetime.  
We particularly  focused on both the null circular geodesics and time-like circular geodesics. Using the 
effective potential diagram, we have compared the geodesic structure between two spacetimes. We have derived the 
ISCO~(innermost stable circular orbit), MBCO~(marginally bound circular orbit) and CPO~(circular photon orbit) for both 
the space-times.  Moreover, we have derived the \emph{quasi-normal modes~(QNM) frequency} in the \emph{eikonal 
limit} for both the spacetimes via Lyapunov exponent. In the Appendix section, we have shown that a 
spherically symmetric tidal-charged BH  can act as  particle accelerators with ultra-high 
center-of-mass~(CM) energy in the limiting case of \emph{maximal BH tidal charge~($q$)} and it is possible when 
two neutral particles are colliding near the horizon. 
\end{abstract}
\maketitle
\tableofcontents

\section{Introduction}
Black Holes are  the most curious as well as  {the} most fascinating objects in 
the universe. They have played an interesting key role for prediction of different kind of observables  
near the BHs in the universe but till to date it has not been observationally  verified. 
For instance, the gravitational bending of light, the gravitational redshift, gravitational time-delay or 
Shapiro time-delay, Perihelion precession of light, gravitational lensing and Lense-Thirring effect  etc. all 
are the physical  phenomenon that are directly linked to study of {the} geodesic structure of BH.
It should be mentioned that the four classical tests of Einstein's general relativity were 
observationally verified particularly for the solar system. The most important effect that has been predicted 
and observed very recently for BHs are gravitational waves. 
The investigation of geodesic motion also  {determines} important feature of the space-time. 
Therefore, it is necessary to study the proper geodesic structure of the BH spacetime. 

One aspect is that the geodesic properties of a given spacetime can be determined by studying the motion of 
time-like and light-like test particles. The motion of the time-like test particles are governed by the  geodesic 
equation whose form is given by 
\begin{eqnarray}
 \frac{d^2x^{a}}{d\tau^2}=-\Gamma^{a}_{bc} \frac{dx^{b}}{d\tau}\frac{dx^{c}}{d\tau}.
\end{eqnarray}
where $\Gamma^{a}_{bc}$  {are} Christoffel symbols.

For light-like test particle, the geodesic equation is given by 
\begin{eqnarray}
 \frac{d^2x^{a}}{d\lambda^2}=-\Gamma^{a}_{bc} \frac{dx^{b}}{d\lambda}\frac{dx^{c}}{d\lambda}.
\end{eqnarray}
where $\lambda$ is affine parameter.

It should be noted that ISCO plays an important role in the context of accretion 
disk theory~\cite{hart}. For example to compute the binding energy and to estimate the temperature of accretion 
disk by using the  {Eddington} luminosity, it is  {very} important to know the ISCO radius~\cite{st}.  {On the other hand},
CPO is  {played an important role} to determine the QNM frequency in the the eikonal approximation 
which is related to the Lyapunov exponent and which determines the instability time scale of the null circular 
{geodesics}.

BH solutions  {arose} not only in general relativity and string theory, but also
in brane-world gravity models. There has been many brane-world models with extra
dimensions in which the standard model fields live on a three-brane and only gravity
propagate in the bulk~\cite{arkani},\cite{dvali},\cite{rs},\cite{roy},\cite{land},\cite{gingrich}.  
Such large extra spatial dimensions and TeV scale quantum gravity may 
 {opened} up the possibility that microscopic BHs could be produced and detected at the CERN 
Large Hadron Collider (LHC)~\cite{casadio},\cite{gingrich}.
Thus it is  {quite natural} to investigate all of the implications of the
Randall-Sundrum model~\cite{rs} for particle accelerators  {which are related to the 
study of the geodesic motion of the test particle}. A class of Randall-Sundrum~(RS)  
solutions have been found in which the solution is a Reissner Nordstr{\o}m~(RN) metric with the electric charge 
replaced by a \emph{tidal charge}~\footnote{ A tidal charge is a dimensionless quantity which 
generates via gravitional effects from fifth dimension which is localized on a 3-brane. This quantity may have
positive or negative values.}  Now we shall called it Dadhich-Maartens-Papadopoulos-Rezania~(DMPR) BH~\cite{roy} 
or RS tidal-charged BH~\cite{narit} or simply tidal charged BH. 

This class of BH solution has the same form as RN BH solution but without 
``electric charge'', instead the RN types correction of Schwarzschild potential which has the form 
$\Phi_{Sch}=-\frac{M}{M_{p}^2r}$ modified to the form of the potential 
$\Phi_{tidal}=-\frac{M}{M_{p}^2r}+\frac{q}{2r^2}$, where $q$
is called ``tidal charge'' and $M_{p}$ is the effective Plank mass on the brane. This is arising from the 
``projection to the brane of free gravitional field effects in the bulk''. Again this effects are propagated 
via the bulk Weyl tensor ~\cite{roy}. It can also be produced from purely scalar effects~(Coulomb like) of the free 
gravitional field in the bulk.

In Einstein's gravity, circular orbits of arbitrary radii are not possible because there exists a minimum
radii below which no circular geodesics are possible. The stability criteria for the existence of ISCO, MBCO  
and CPO have been described for the Schwarzschild BH~\cite{sch}, RN BH~\cite{sch,pp1,pqr}, 
spherically symmetric string BH~\cite{bb,fer}. Previous studies have not been 
considered the stability conditions for the DMPR spherically symmetric BHs characterized by their mass~($M$) and 
tidal charge~($q$) localized on a three-brane in 5D gravity in the RS scenario. In this work, we are interested to study the 
complete geodesic structure of the  said BH. Next, we derived the innermost stable circular orbit (ISCO) or 
last stable circular orbit (LSCO), marginally bound circular orbit (MBCO) and circular photon orbit (CPO) of 
the said BH.  Moreover, we compute the QNM frequency of the null circular geodesics in the 
eikonal approximation via Lyapunov exponent and which determines the instability time scale 
of the null circular orbit. 

Finally in appendix section, we shortly investigate the Ba\~{n}ados, Silk and West (BSW) mechanism for DMPR BH. 
Next we compute the CM energy for neutral test particles. When the value of tidal charge $q>0$, they are formally 
identical to the RN BH of general theory of relativity. Thus the CM energy is diverging in the \emph{extremal} limit. 
In this case the BH  can act as particle accelerators with arbitrarily high CM energy.
When the value of tidal charge $q<0$, there exists only one horizon then the CM energy is finite 
 and the BH can not act as particle accelerators. Lastly for $M=0$ and $q<0$, we get 
the CM energy is also finite and the BH also can not  act as particle accelerators.

The paper is organized as follows. In Sec. 2, we shall study the basic features of  {a} tidal charged BH. 
In Sec. 3, we shall study in details the complete geodesic structure of the said BH. In Sec. 4,  we compute 
the QNM frequency for DMPR BH in  the eikonal approximation.  Finally, we conclude our discussions 
in section 5. In the Appendix section, we compute the CM energy of the tidal-charged BH.

\section{DMPR BH metric and its Properties}

\begin{figure}
\begin{center}
{\includegraphics[width=0.45\textwidth]{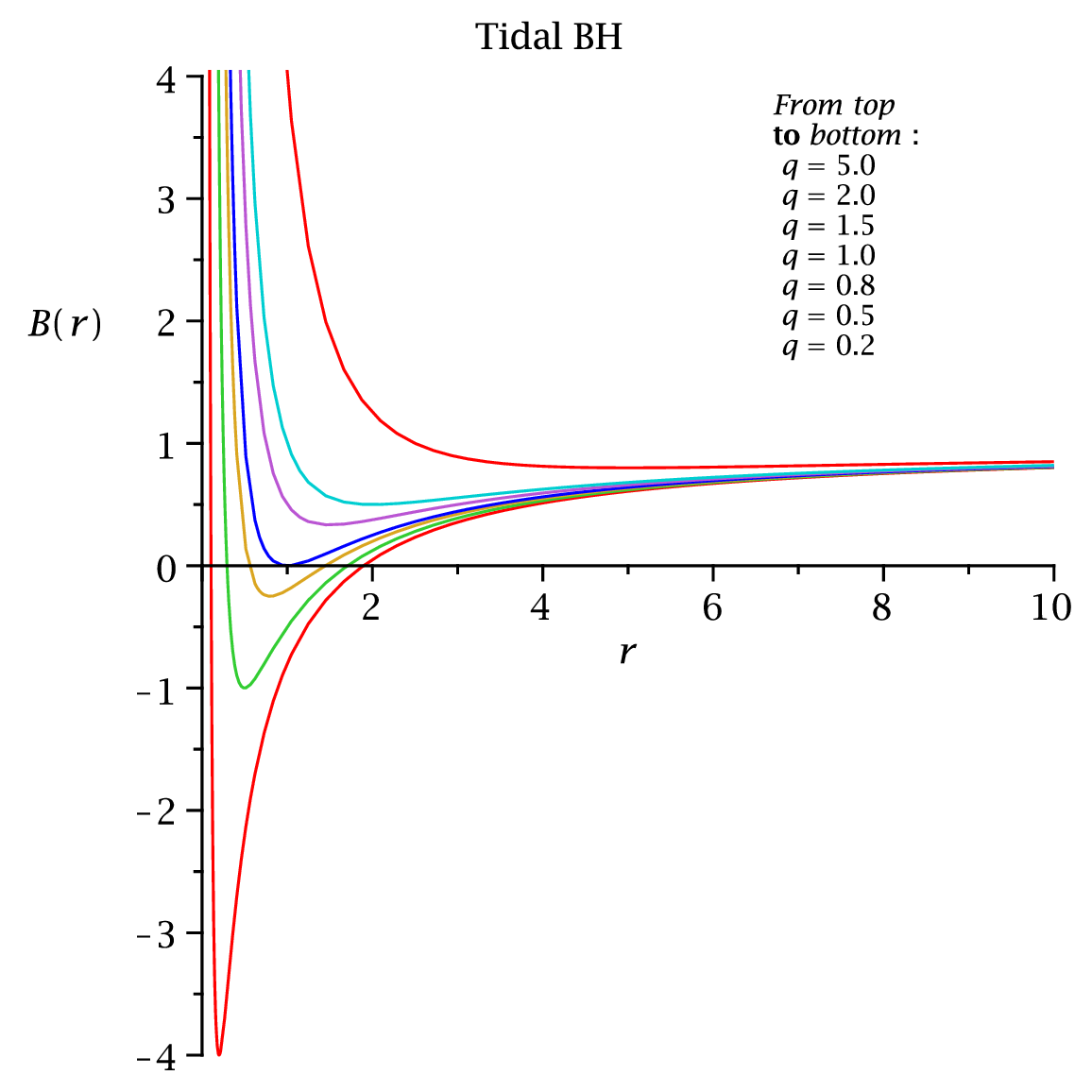}}
{\includegraphics[width=0.45\textwidth]{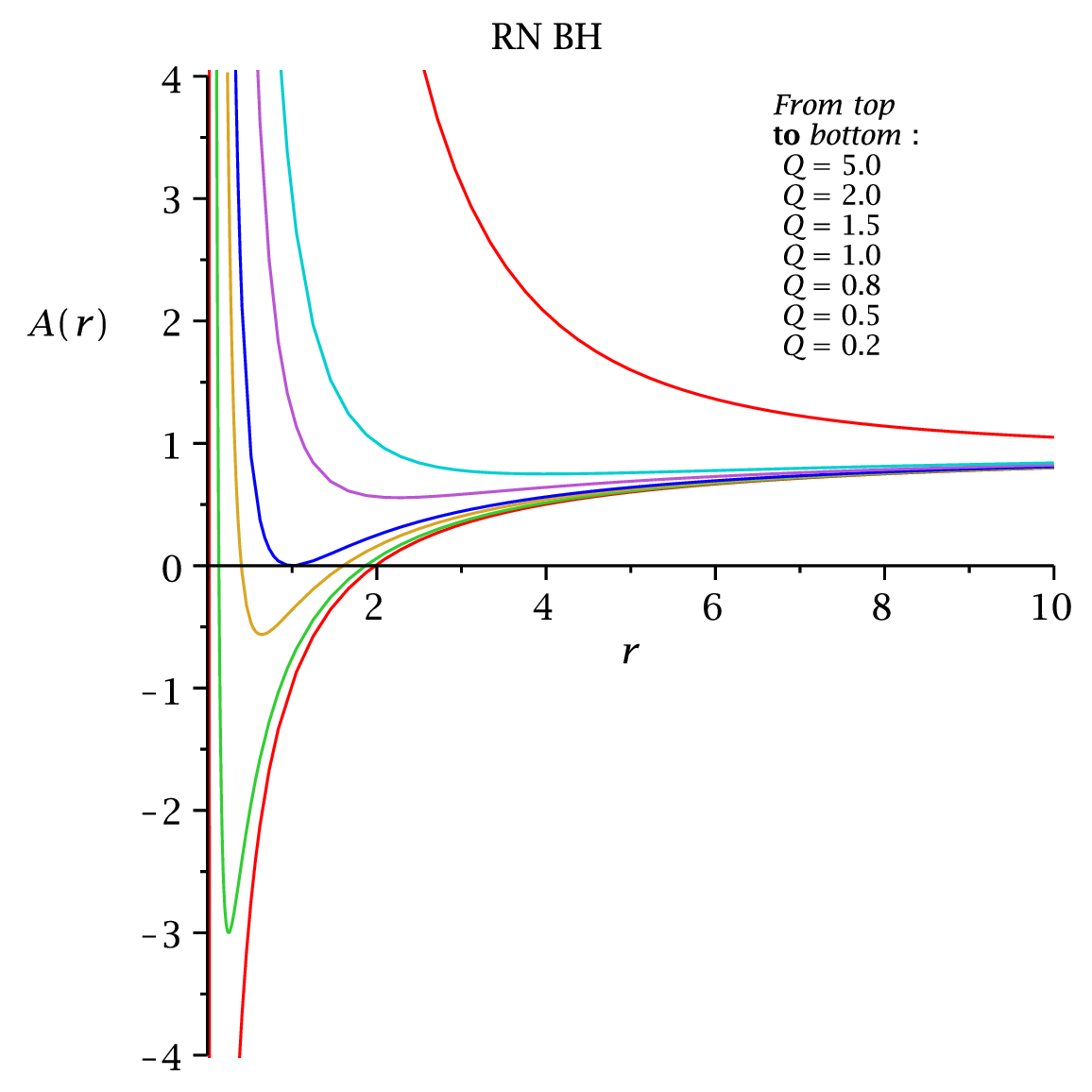}}
\end{center}
\caption{The figure shows the variation of ${\cal B}(r)$ and ${\cal A}(r)$ with $r$ for 
Brane-world BH and RN BH with $M=M_{p}=M_{5}=1$. For RN BH the metric coefficient is 
${\cal A}(r) = \left(1-\frac{r_{+}}{r}\right)\left(1-\frac{r_{-}}{r}\right)$ and the BH 
has event horizon is  at $r_{+}=M+\sqrt{M^{2}-Q^{2}}$ and the Cauchy horizon is at 
$r_{-}=M-\sqrt{M^{2}-Q^{2}}$. \label{mf}}
\end{figure}

The metric of a static, spherically symmetric DMPR BH \cite{roy} is given by
\begin{eqnarray}
ds^2=-{\cal B}(r)dt^{2}+\frac{dr^{2}}{{\cal B}(r)}+ r^{2}\left(d\theta^{2}+
\sin^{2}\theta d\phi^{2}\right) ~\label{sph}
\end{eqnarray}
where the function ${\cal B}(r)$ is defined by
\begin{eqnarray}
{\cal B}(r) &=& \left(1-\frac{2M}{M_{p}^2r}+\frac{q}{M_{5}^2r^2}\right)
\end{eqnarray}
where $M$ is defined as the BH mass, $q$ is the dimensionless tidal charge, $M_{p}(=1.2 \times 10^{16} Tev)$ is 
the effective Plank mass on the brane and $M_{5}$ is the fundamental Planck scale in the 5D bulk.
It should be noted that the above metric is nearly identical to the RN metric, except that
the RN term $\frac{Q^2}{r^2}$, which is necessarily non-negative, is replaced by $\frac{q}{M_{5}^2r^2}$
that can have any sign. In Fig. \ref{mf}, we show how the metric coefficient changes for DMPR BH in
contrast with RN BH. From the figure, it is clear that the change of metric coefficients look similar both for
DMPR BH and RN BH in the extremal limit i.e. $q=Q=1$.

Eq. (\ref{sph}) is an exact solution of effective Einstein equation on the brane i.e.
\begin{eqnarray}
G_{ab} &=& -\Lambda g_{ab}+\frac{8\pi}{M_{p}^2}T_{ab}+\left(\frac{8\pi}{M_{5}^3}\right)^2 S_{ab}-{\cal E}_{ab}~.
\label{eee}
\end{eqnarray}
where,
\begin{eqnarray}
M_{p} &=& \sqrt{\frac{3}{4\pi}}\left(\frac{M_{5}^2}{\sqrt{\lambda}} \right)M_{5},\\
\Lambda &=& \frac{4\pi}{M_{5}^3}\left[\Lambda_{5}+\left( \frac{4\pi}{3M_{5}^3}\lambda^2\right) \right],\\
\Lambda_{5} &=& - \frac{4\pi \lambda^2}{3M_{5}^3}. \label{emp}
\end{eqnarray}
Generally, $M_{5}<<M_{p}$ i.e. the fundamental Planck scale is much lower than the
effective scale in the brane-world. For the vacuum case, it should be noted that
$T_{ab}=S_{ab}=0=\Lambda$. Then Eq.(\ref{eee}) is reduced to
\begin{eqnarray}
R_{ab}&=& -{\cal E}_{ab}, R_{a} ^{a}=0={\cal E}_{a} ^{a}. \label{veq}
\end{eqnarray}
Now the four dimensional horizon structure of the RS tidal-charged BH strictly depends
on the sign of $q$.

{\bf Case I:}
When $q\geq 0$, then the component $g_{tt}={\cal B}(r)=0$ gives two horizons
\begin{eqnarray}
r_{\pm} &=& \frac{M}{M_{p}^2}\left[1\pm \sqrt{1-\frac{qM_{p}^4}{M^2M_{5}^2}}\right]
~\label{hor}
\end{eqnarray}
which is analogous to the spherically symmetric RN BH. Here $r_{+}$ and  $r_{-}$ are called
event horizon (${\cal H}^+$)  or outer horizon and Cauchy horizon (${\cal H}^-$)  or
inner horizon respectively.  The $r_{+}=r_{-}$ or $M^{2}=q \left(\frac{M_{p}^4}{M_{5}^2}\right)$
corresponds to the extremal case and in this case the extremal horizon is situated at
\begin{eqnarray}
r_{+}=r_{-}=\frac{M}{M_{p}^2}~\label{exh}
\end{eqnarray}
Since in general theory of relativity(GTR), both horizons lie inside the Schwarzschild horizon
i.e. $0\leq r_{-}\leq r_{+}\leq r_{s}=\frac{2M}{M_{p}^2}$, and there is an upper bound on $q$
i.e. $0\leq q \leq q_{max}=\left(\frac{M_{5}}{M_{p}}\right) \left(\frac{M}{M_{p}}\right)^2$.

\begin{figure}
\begin{center}
{\includegraphics[width=0.45\textwidth]{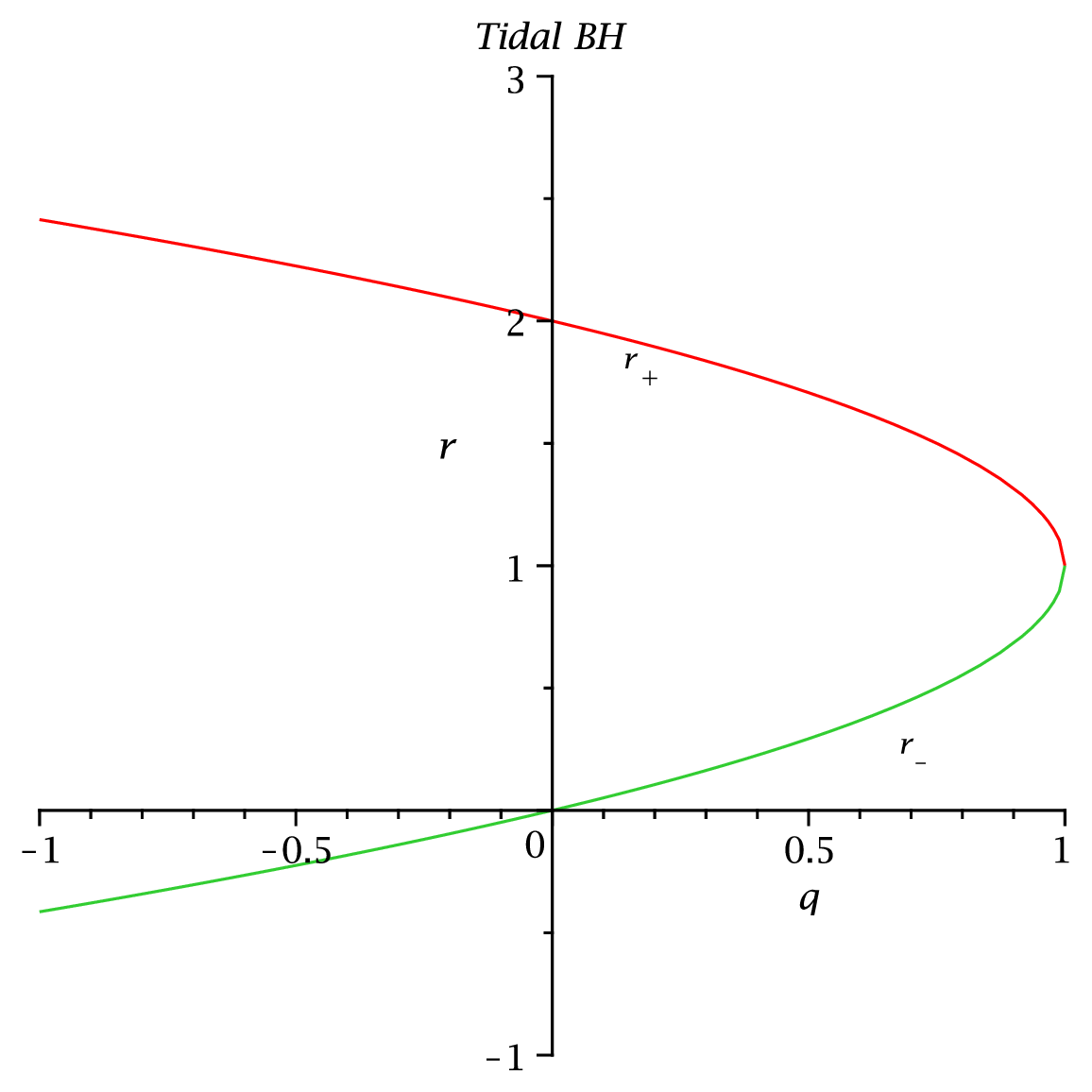}}
{\includegraphics[width=0.45\textwidth]{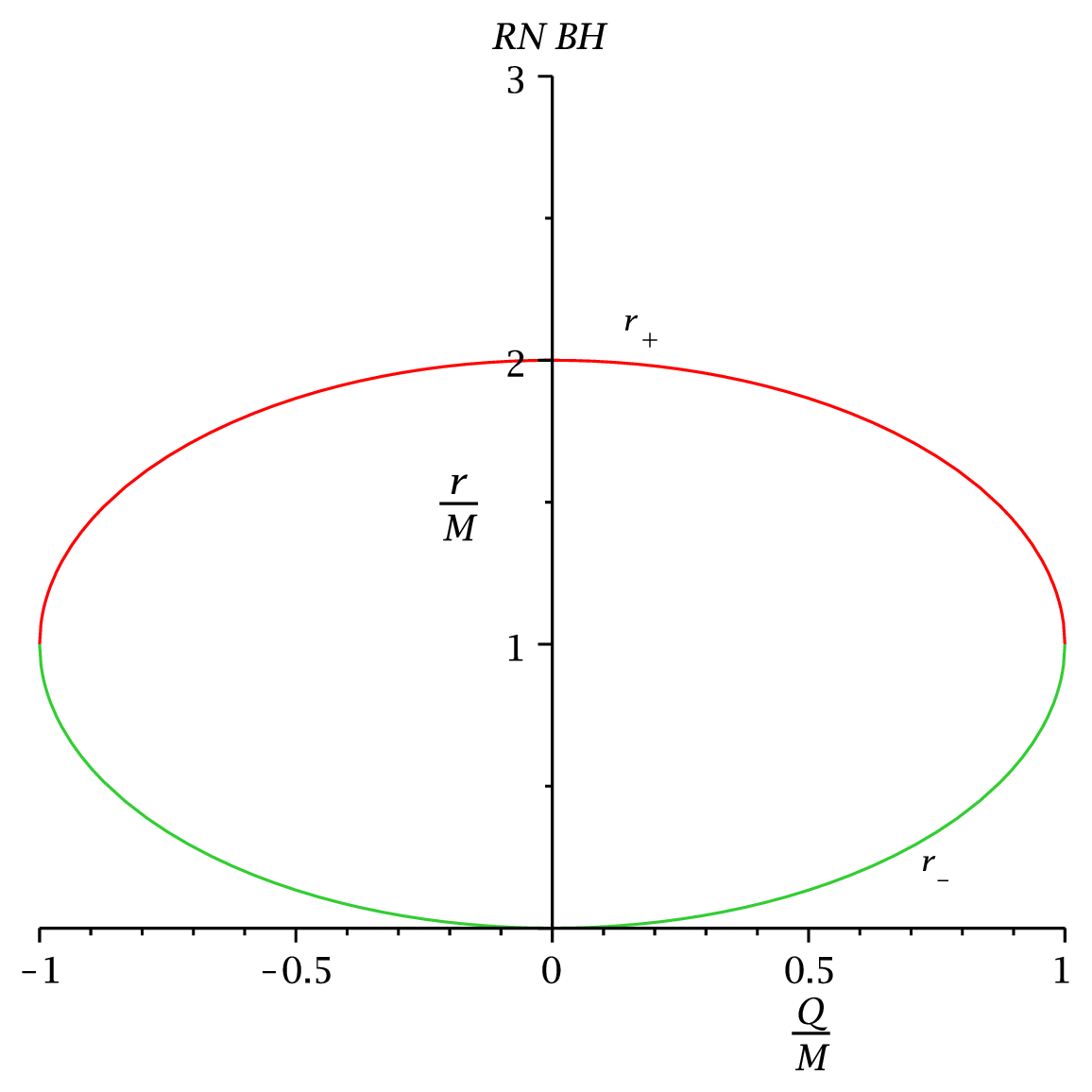}}
\end{center}
\caption{The figure shows the qualitative behavior of horizons for RS tidal-charged BH and
RN BH with $M=M_{p}=M_{5}=1$.
\label{hori}}
\end{figure}

We  {can observed} that the event horizon and Cauchy horizons are qualitatively different 
for DMPR BH and RN BH, which can be seen from Fig. \ref{hori}.

{\bf Case II:}
When $q<0$, which is impossible in GTR RN case leads to single horizon, lying exterior to the
Schwarzschild horizon:
\begin{eqnarray}
r_{+} &=& \frac{M}{M_{p}^2}\left[1 + \sqrt{1-\frac{qM_{p}^4}{M^2M_{5}^2}}\right]>r_{s}
~\label{hor1}
\end{eqnarray}

{\bf Case III:}
When $M=0$ and $q<0$, in this case the metric becomes
\begin{eqnarray}
ds^2=-\left[1+\frac{q}{M_{5}^2r^2}\right]dt^{2}+\frac{dr^{2}}{\left[1+\frac{q}{M_{5}^2r^2}\right]}+
r^{2}\left(d\theta^{2}+\sin^{2}\theta d\phi^{2}\right) ~\label{sph1}
\end{eqnarray}
The horizon still exists without the gravitational sources and on the brane it is given by
\begin{eqnarray}
r_{+} &=& \frac{\sqrt{-q}}{M_{5}}.~ \label{hor2}
\end{eqnarray}
This implies that in the absence of gravitational collapse of matter, there is an intriguing
possibility of string tidal effects in the early universe  {which} could lead to BH formation.
Then the tidal charge $q$  {will} be determined by both the brane source i.e. the mass
$M$, and any coulomb part of the bulk. Weyl tensor that survives when $M$ is set to
be zero. Now we shall discuss shortly the thermodynamic properties of tidal-charged BH.

\begin{figure}
\begin{center}
{\includegraphics[width=0.45\textwidth]{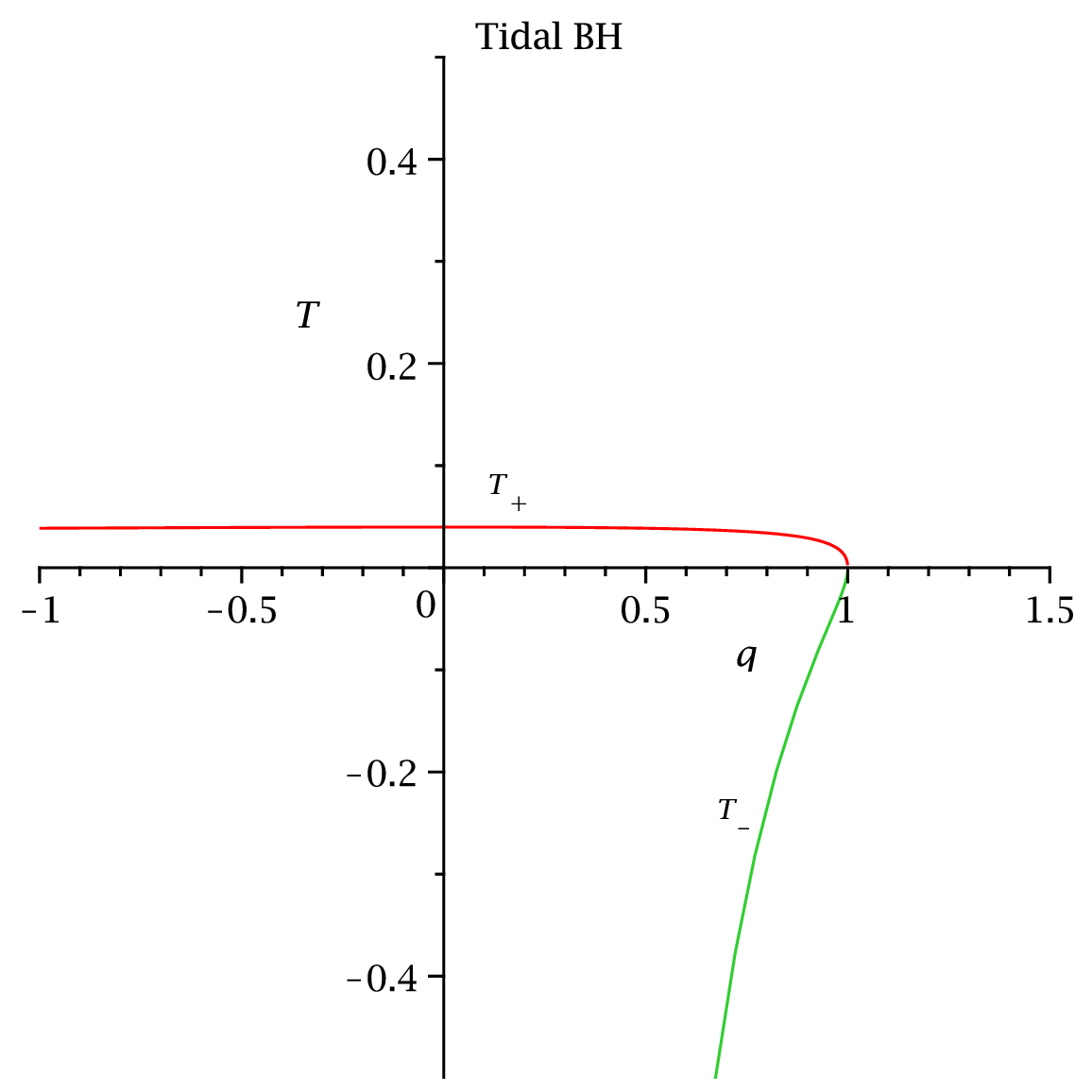}}
{\includegraphics[width=0.45\textwidth]{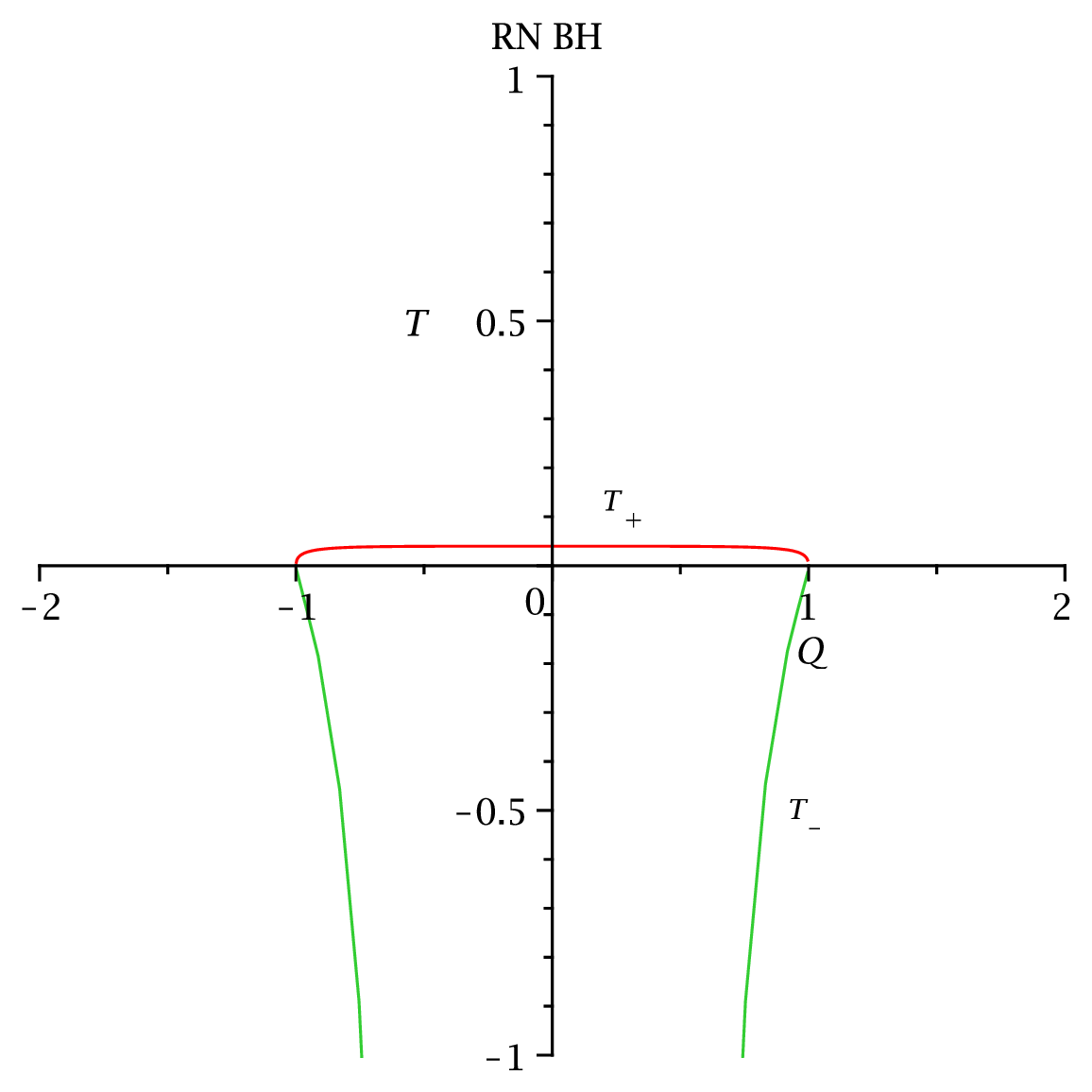}}
\end{center}
\caption{The figure depicts the qualitative behavior of Hawking temperature for RS tidal-charged BH and
RN BH while $M=M_{p}=M_{5}=1$.
\label{hawtmp}}
\end{figure}

\begin{figure}
\begin{center}
{\includegraphics[width=0.45\textwidth]{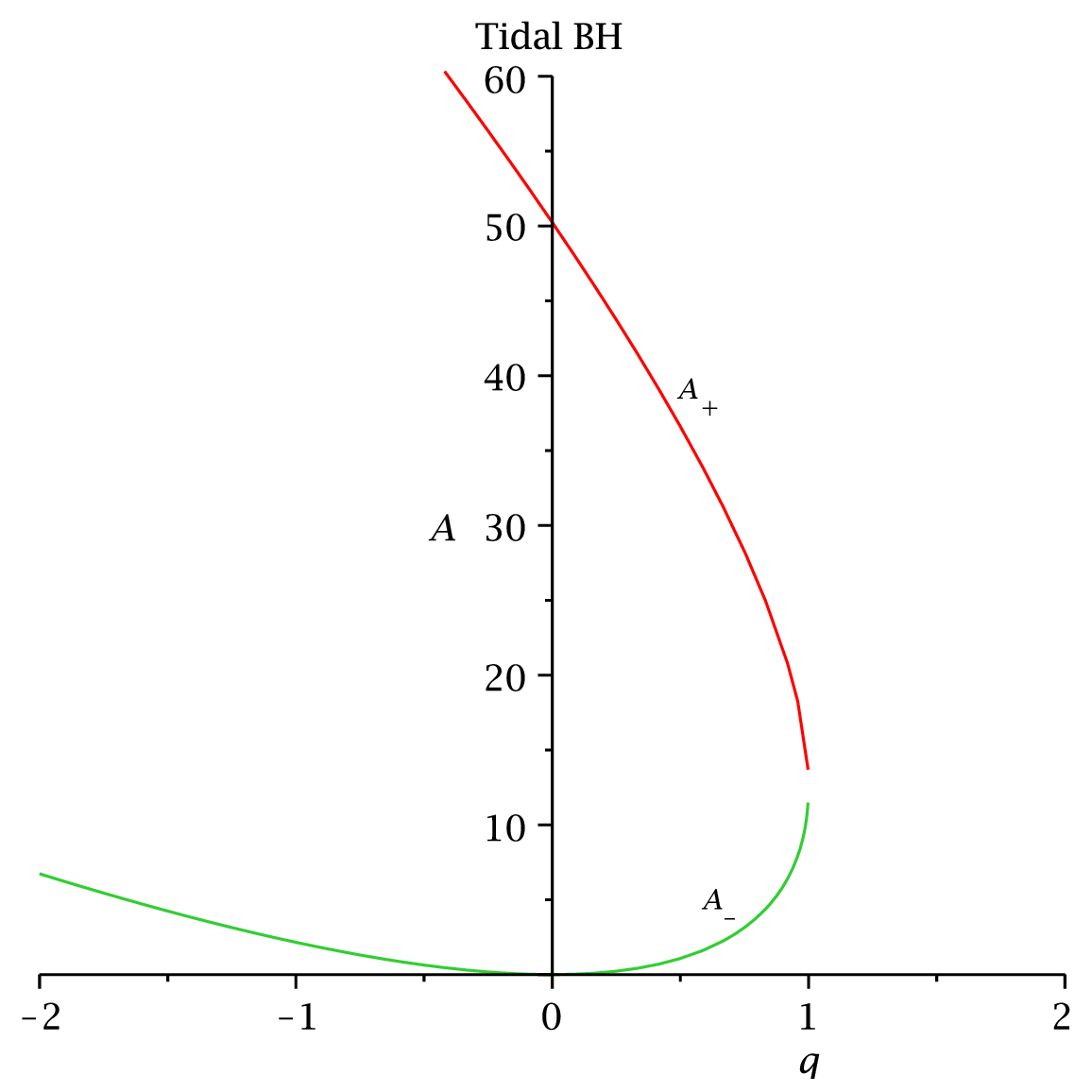}}
{\includegraphics[width=0.45\textwidth]{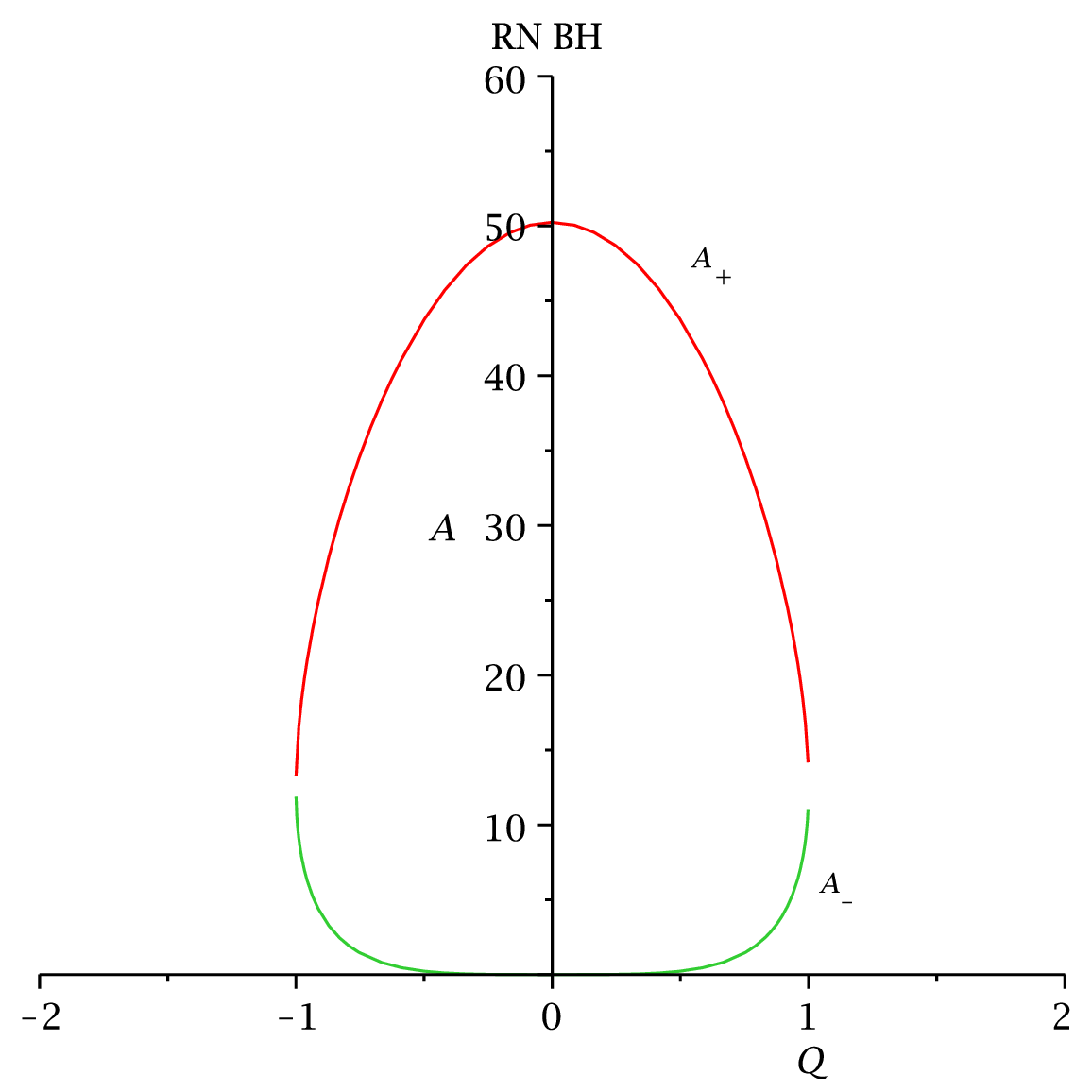}}
\end{center}
\caption{The figure shows the qualitative behavior of surface area for RS tidal-charged BH and
RN BH while $M=M_{p}=M_{5}=1$.
\label{sarea}}
\end{figure}

The  BH temperature or Hawking temperature of ${\cal H}^\pm$ is
\begin{eqnarray}
T_{\pm} &=&   \pm \frac{\sqrt{(\frac{M}{M_{p}^2})^2-\frac{q}{M_{5}^2}}}
{2\pi[2(\frac{M}{M_{p}^2})r_{\pm}-\frac{q}{M_{5}^2}]}~\label{bhtt}
\end{eqnarray}
The graphical plot of the Hawking temperature can be shown in the Fig. ~\ref{hawtmp}.
Then the area of both the horizons (${\cal H}^\pm$) are
\begin{eqnarray}
{\cal A}_{\pm} &=& 4\pi\left[2(\frac{M}{M_{p}^2})r_{\pm}-\frac{q}{M_{5}^2}\right]  ~\label{art}
\end{eqnarray}
The variation of the surface area can be shown in the Fig.~\ref{sarea}.
Finally, the entropy of ${\cal H}^\pm$ is
\begin{eqnarray}
{\cal S}_{\pm} &=&
\pi \left[2(\frac{M}{M_{p}^2})r_{\pm}-\frac{q}{M_{5}^2} \right] ~\label{etpt}
\end{eqnarray}

\section{Equatorial circular orbit in the DMPR BH}
In this section, we shall investigate the complete geodesic properties of the DMPR 
BH and to derive the effective potential for geodesic motion of the test particle. 
{Also} to derive the CM energy, first we need to know in detail the complete geodesic
structure of the DMPR BH in the equatorial plane i.e. $\theta=\frac{\pi}{2}$. We 
also compute ISCO or LSCO, MBCO and CPO of the said BH. They are very crucial in
BH accretion disk theory. To determine the geodesic motion of a test particle in 
the equatorial plane we impose the condition i.e. $u^{\theta}=\dot{\theta}=0$ and
$\theta=constant=\frac{\pi}{2}$ and follow the well known book of S. Chandrasekhar~\cite{sch}. 

The radial equation that governs the complete geodesic structure of DMPR BH is given by
\begin{eqnarray}
\dot{r}^{2} &=& E^{2}-V_{eff}=E^{2}-\left(\frac{L^{2}}{r^2}-\epsilon \right){\cal B}(r) ~\label{radial}
\end{eqnarray}
where, the standard effective potential for DMPR BH  {becomes}
\begin{eqnarray}
V_{eff} &=& \left(\frac{L^{2}}{r^2}-\epsilon \right){\cal B}(r) ~\label{vrn}
\end{eqnarray}
Where,  $\epsilon=-1$ for time-like geodesics, $\epsilon=0$ for light-like geodesics and
$\epsilon=+1$ for space-like geodesics.  {The parameters} $E$ and $L$ are the energy per unit mass  and 
angular momentum per unit mass of the test particle.

\subsection{Particle Case}
For massive particles,  the effective potential becomes
\begin{eqnarray}
V_{eff} &=& \left(1+\frac{L^{2}}{r^{2}}\right)\left(1-\frac{r_{+}}{r}\right)\left(1-\frac{r_{-}}{r}\right)
 \nonumber\\
&=&  \left(1+\frac{L^{2}}{r^{2}}\right)\left(1-\frac{2M}{M_{p}^2r}+\frac{q}{M_{5}^2r}\right)
~\label{vrt}
\end{eqnarray}

\begin{figure}
\begin{center}
{\includegraphics[width=0.45\textwidth]{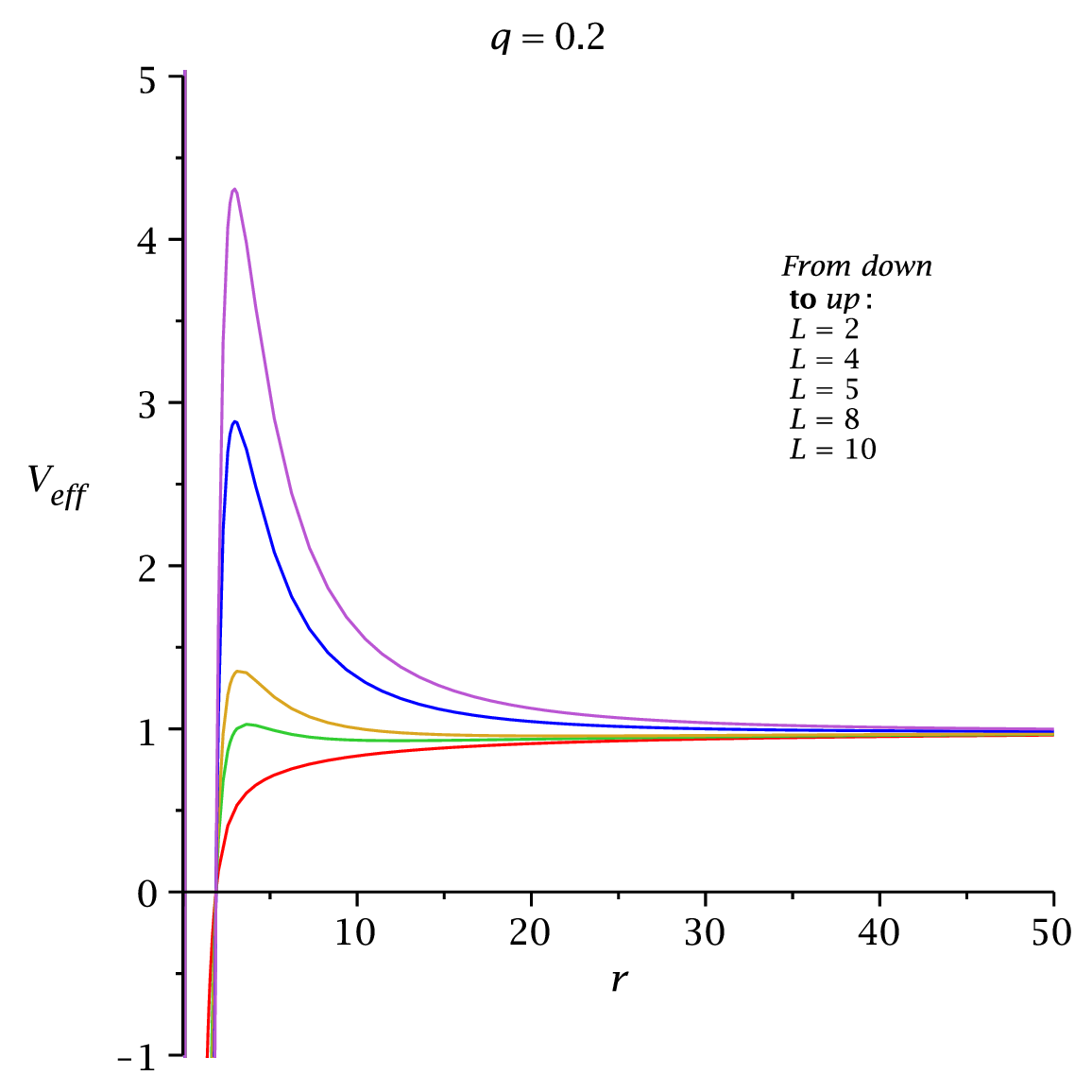}}
{\includegraphics[width=0.45\textwidth]{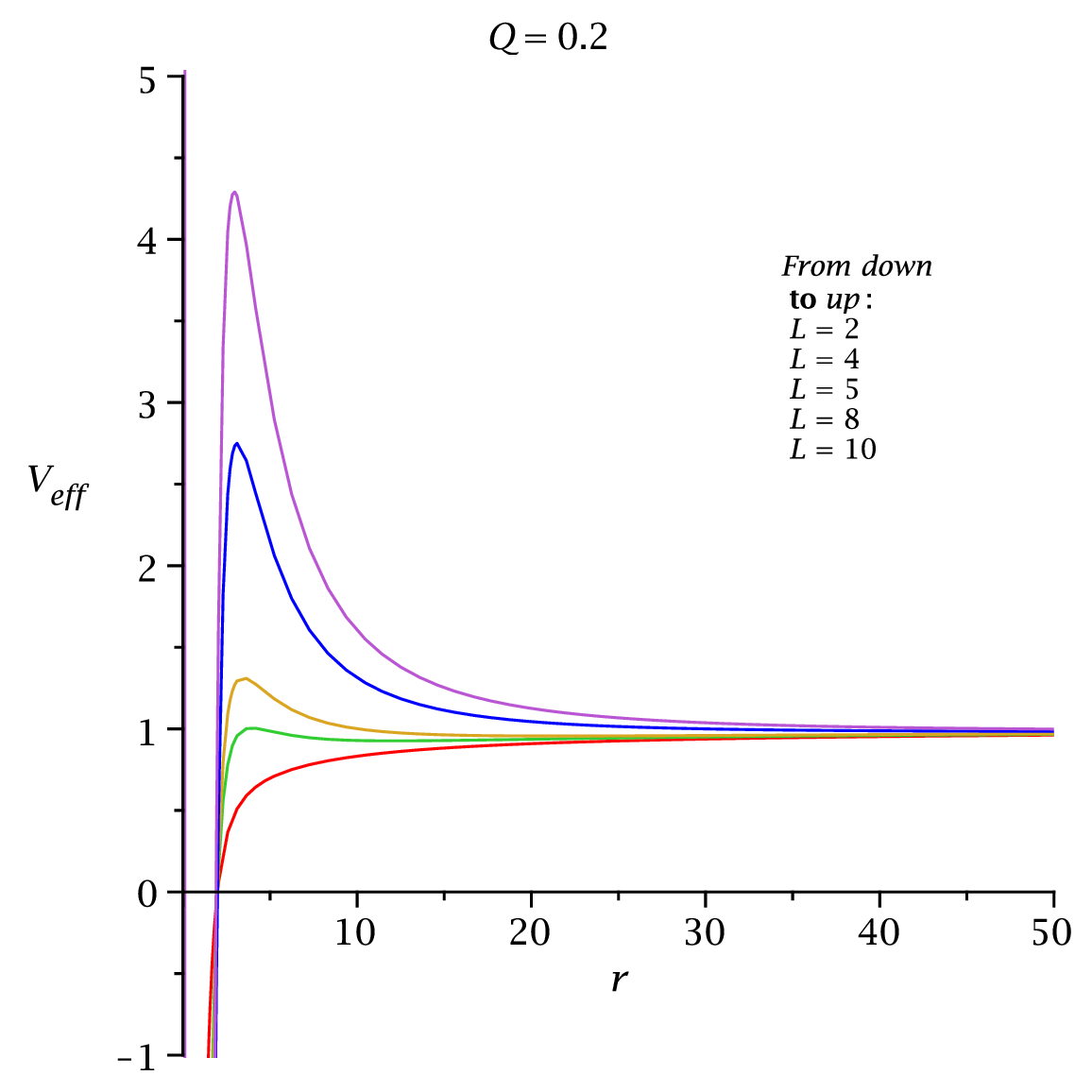}}
\end{center}
\caption{The figure shows the variation  of $V_{eff}$  with $r$ for DMPR BH and RN BH with $M=M_{p}=M_{5}=1$.
 \label{veff}}
\end{figure}

\begin{figure}
\begin{center}
{\includegraphics[width=0.45\textwidth]{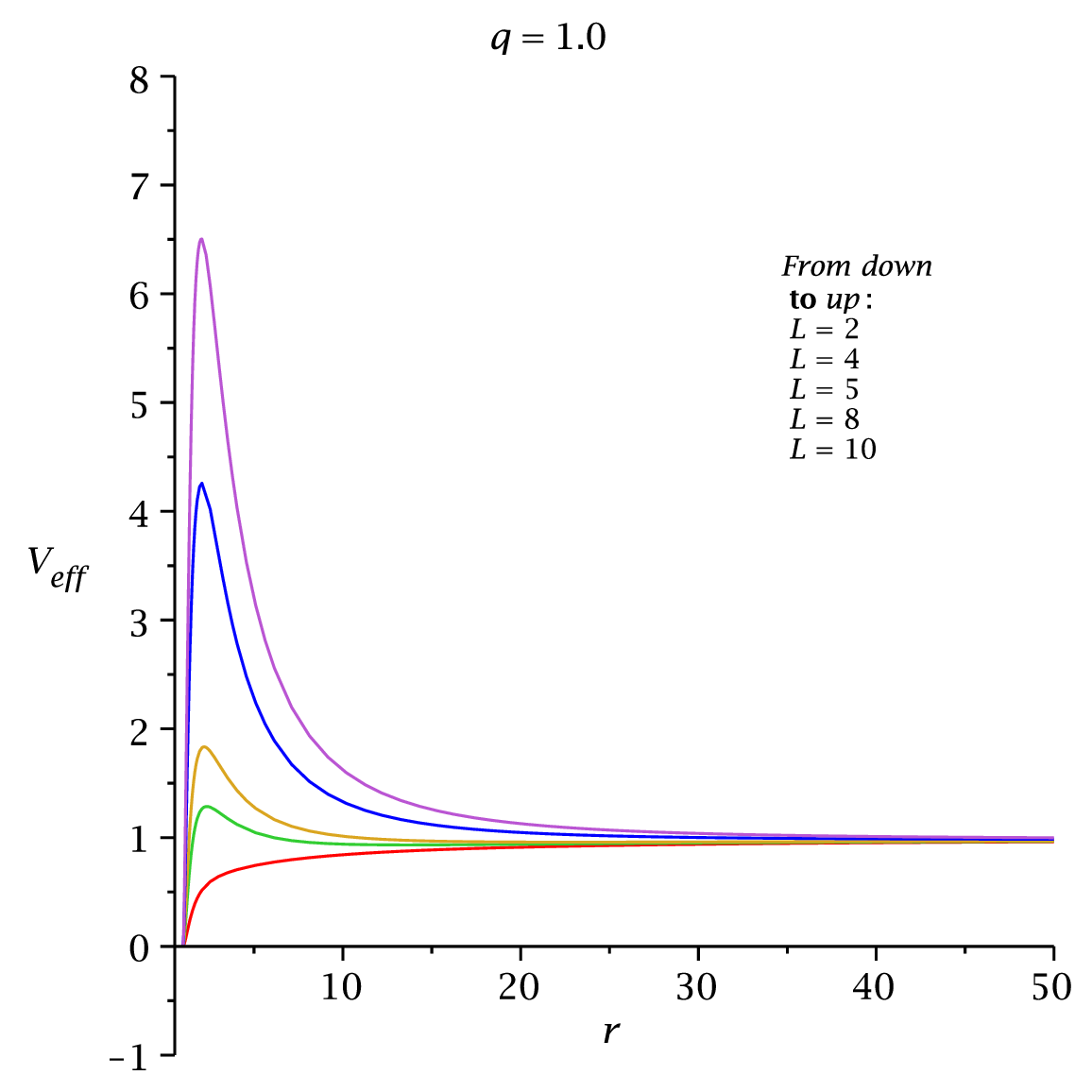}}
{\includegraphics[width=0.45\textwidth]{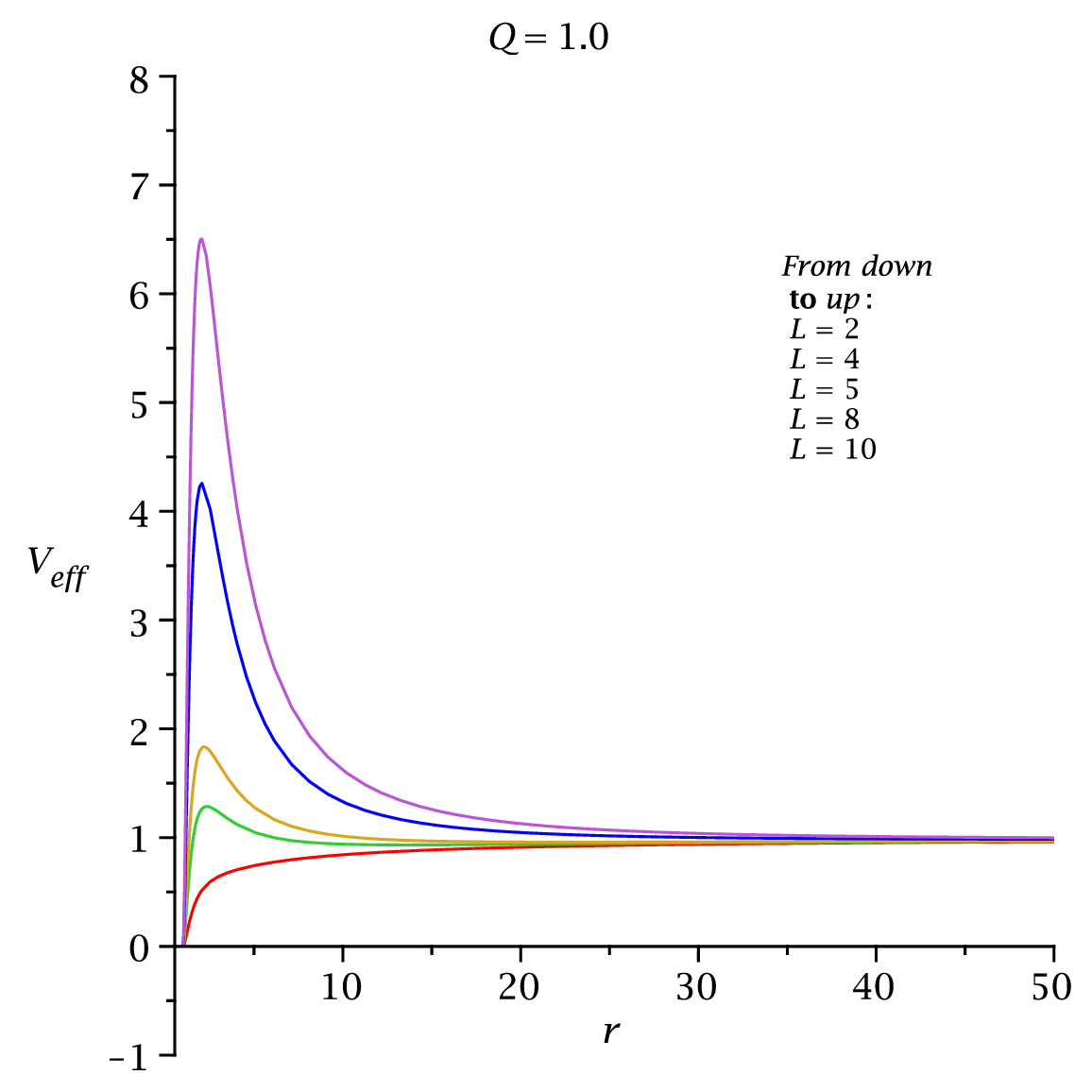}}
\end{center}
\caption{The figure shows the variation  of $V_{eff}$  with $r$ for  DMPR BH and RN BH
in the extremal limit i.e. $q=Q=1$ with $M=M_{p}=M_{5}=1$.
 \label{veff2}}
\end{figure}

\begin{figure}
\begin{center}
{\includegraphics[width=0.45\textwidth]{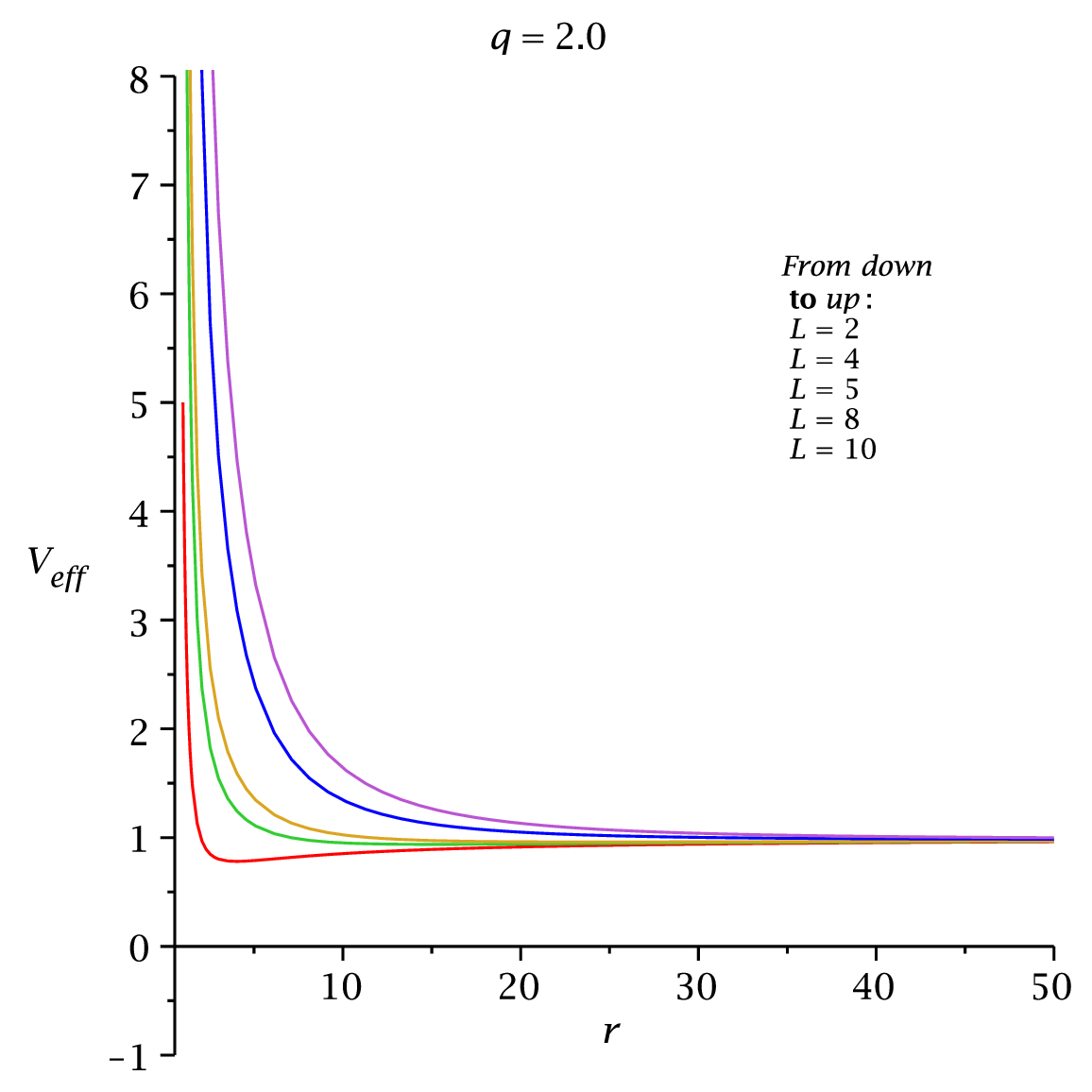}}
{\includegraphics[width=0.45\textwidth]{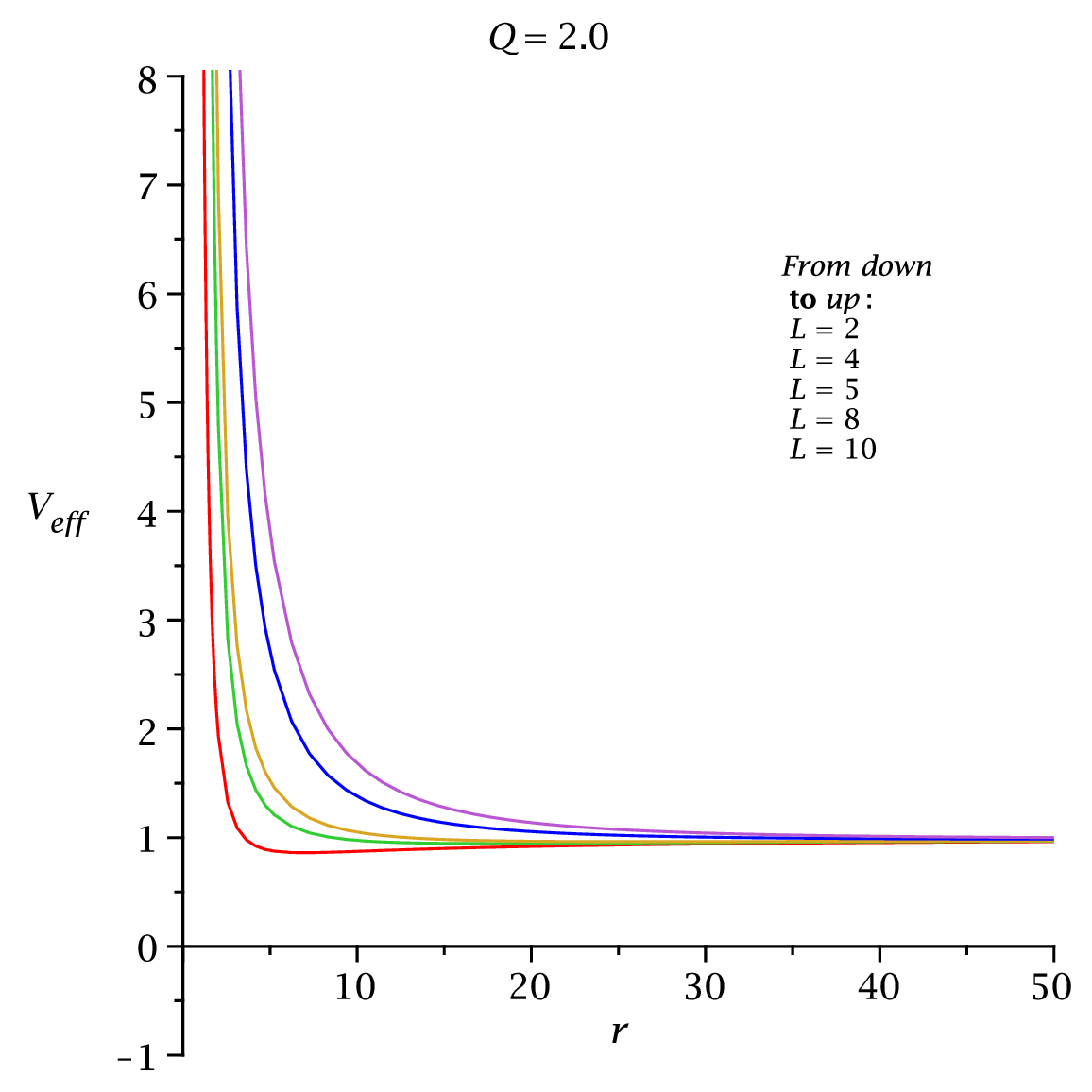}}
\end{center}
\caption{The figure shows the variation  of $V_{eff}$  with $r$ for  DMPR BH and RN BH with $M=M_{p}=M_{5}=1$.
 \label{veff3}}
\end{figure}

Figures.~\ref{veff}, \ref{veff2}, \ref{veff3} describe the qualitative differences
between RS tidal-charged BH and RN BH in the effective potential for various values of angular
momentum. Note that the structure of the effective potential is exactly same in the extremal
limit both for tidal-charged BH and RN BH, as can be seen from Fig.~\ref{veff2}.


In the extremal limit, the effective potential (See Fig.~\ref{veff2}) is found to be
\begin{eqnarray}
V_{eff} &=& \left(1+\frac{L^{2}}{r^{2}}\right)\left(1-\frac{r_{+}}{r}\right)^2~\label{vrtx}
\end{eqnarray}

For circular geodesic motion of the test particle of constant $r=r_{0}$, one obtains
\begin{eqnarray}
V_{eff} &=& E^{2} ~\label{vt}
\end{eqnarray}
and
\begin{eqnarray}
\frac{dV_{eff}}{dr} &=& 0 ~\label{dvt}
\end{eqnarray}

Therefore,  {one obtains} the conserved energy per unit mass and angular momentum per unit mass of the
test particle along the circular orbit
\begin{eqnarray}
E^{2}_{0} &=& \frac{2(r_{0}-r_{+})^2(r_{0}-r_{-})^2}{r_{0}^2[2r_{0}^{2}-3(r_{+}+r_{-})r_{0}+4r_{+}r_{-}]}
~\label{engt}
\end{eqnarray}
and
\begin{eqnarray}
L_{0}^{2} &=& \frac{r_{0}^{2}[r_{0}(r_{+}+r_{-})-2r_{+}r_{-}]}{2r_{0}^{2}-3(r_{+}+r_{-})r_{0}+4r_{+}r_{-}}
~ \label{angt}
\end{eqnarray}
For our record, we also note  {that} the energy and angular momentum for extremal tidal-charged BH
\begin{eqnarray}
E^{2}_{0} &=& \frac{(r_{0}-r_{+})^3}{r_{0}^2(r_{0}-2r_{+})} ~\label{engtx}
\end{eqnarray}
and
\begin{eqnarray}
L_{0}^{2} &=& \frac{r_{0}^{2}r_{+}}{r_{0}-2r_{+}} ~\label{angtx}
\end{eqnarray}

Circular motion of the test particle to be exists when both the energy and angular
momentum are real and finite. Thus one obtains
\begin{eqnarray}
2r_{0}^{2}-3(r_{+}+r_{-})r_{0}+4r_{+}r_{-} > 0 \,\, \mbox{and} \,\, r_{0}>2\frac{r_{+}r_{-}}{r_{+}+r_{-}}
~ \label{crt}
\end{eqnarray}
General relativity does not permit arbitrary circular radii, so the denominator of equations
(\ref{engt},\ref{angt}) real only if $2r_{0}^{2}-3(r_{+}+r_{-})r_{0}+4r_{+}r_{-}\geq 0$. The limiting case of 
equality gives circular orbit with infinite energy per unit mass, i.e. a  CPO. This photon 
orbit is the innermost boundary of the circular orbit for massive particles.

Comparing the above equation of particle orbits with (\ref{ph1t}) when
$r_{0}=r_{c}$, we can see that photon orbit is the limiting case
of time-like circular orbit.  It occurs at the radius
\begin{eqnarray}
r_{c} &=& r_{cpo}= \frac{1}{4}\left[3\left(r_{+}+r_{-}\right)\pm \sqrt{9\left(r_{+}+r_{-}\right)^{2}
 -32r_{+}r_{-}} \right] ~. \label{cpot}
\end{eqnarray}
In the extremal limit, one obtains
\begin{eqnarray}
 r_{cpo} &=& 2r_{+}=2r_{-}=2\frac{M}{M_{p}^2} ~. \label{erad}
\end{eqnarray}

\subsection{Photon Case}
For massless particles, the effective potential becomes
\begin{eqnarray}
U_{eff} &=& \frac{L^2}{r^2}{\cal B}(r)
=\frac{L^2}{r^{2}}\left(1-\frac{r_{+}}{r}\right)\left(1-\frac{r_{-}}{r}\right)\nonumber\\
&=&\frac{L^2}{r^{2}}\left(1-\frac{2M}{M_{p}^2r}+\frac{q}{M_{5}^2r}\right)~ \label{uef}
\end{eqnarray}

\begin{figure}
\begin{center}
{\includegraphics[width=0.45\textwidth]{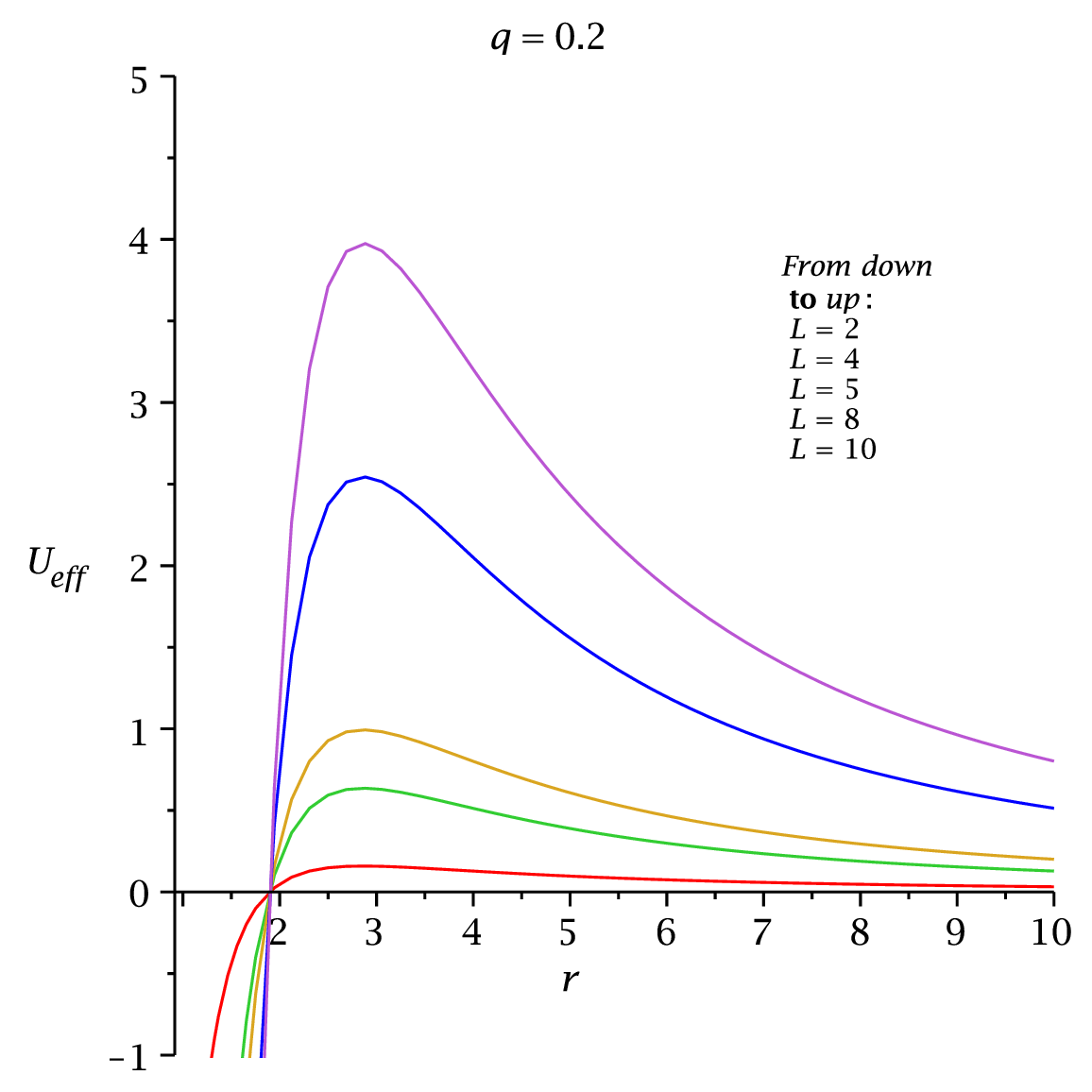}}
{\includegraphics[width=0.45\textwidth]{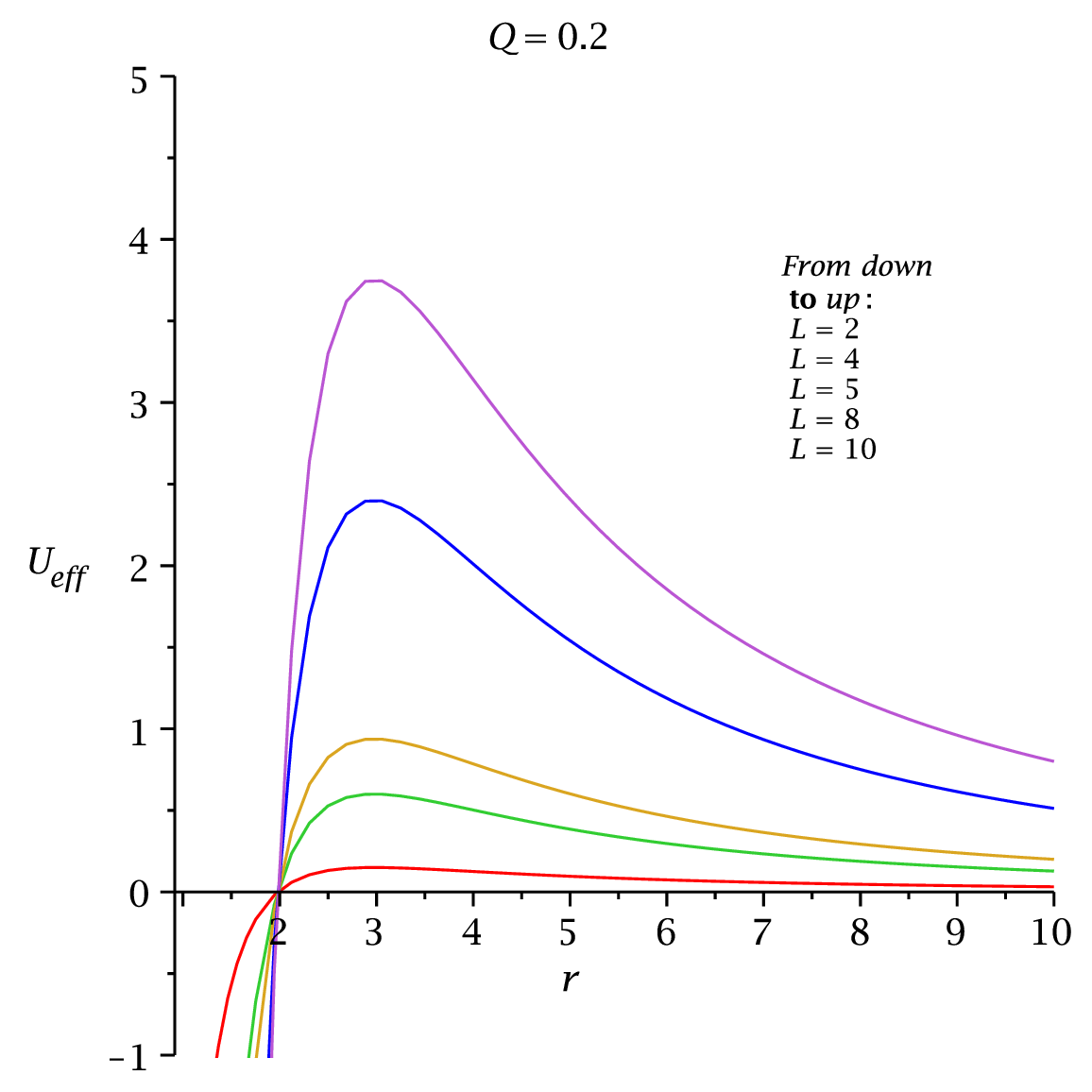}}
\end{center}
\caption{The figure shows the variation  of $U_{eff}$  with $r$ for  RS tidal-charged BH
 and RN BH with $M=M_{p}=M_{5}=1$.
 \label{ueff}}
\end{figure}

\begin{figure}
\begin{center}
{\includegraphics[width=0.45\textwidth]{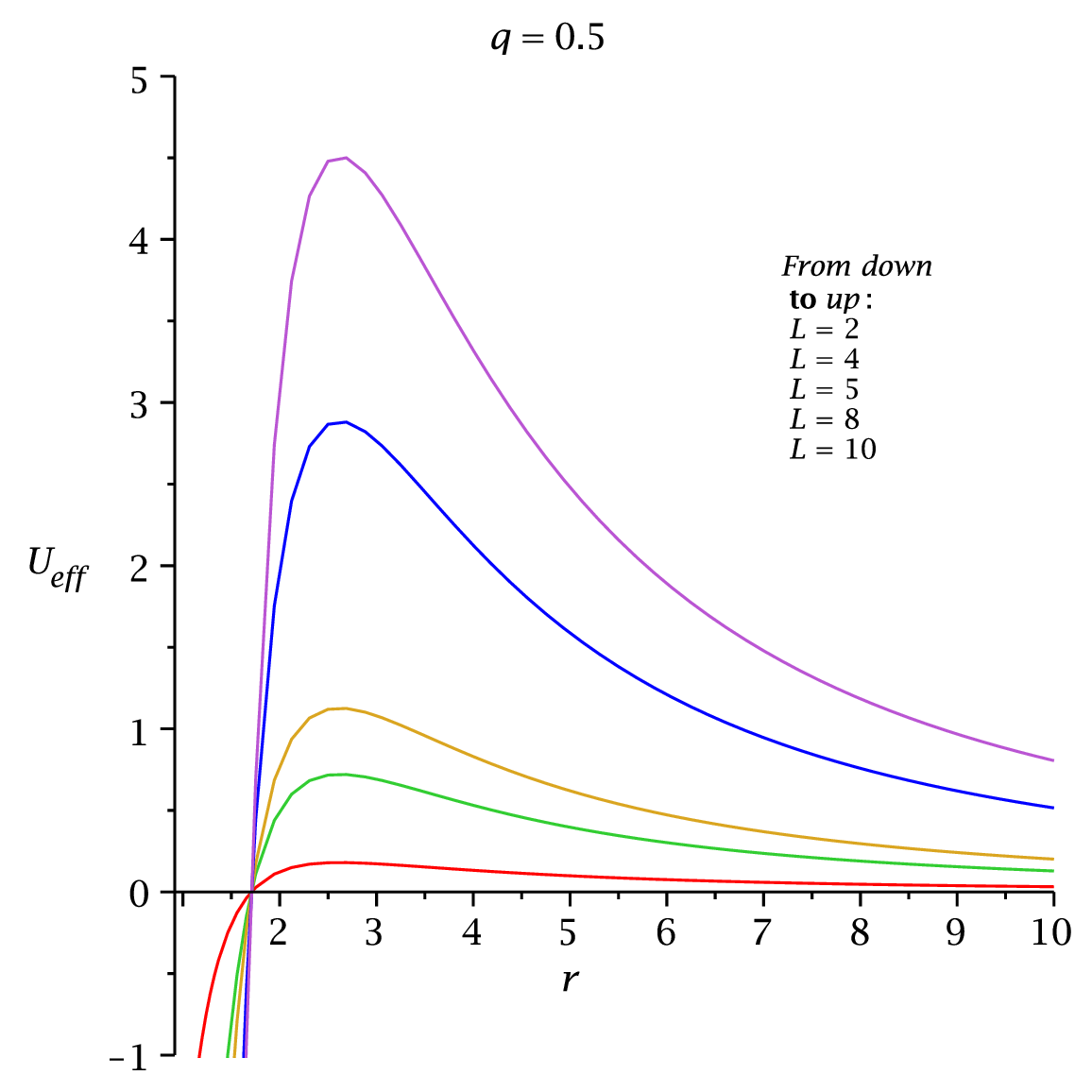}}
{\includegraphics[width=0.45\textwidth]{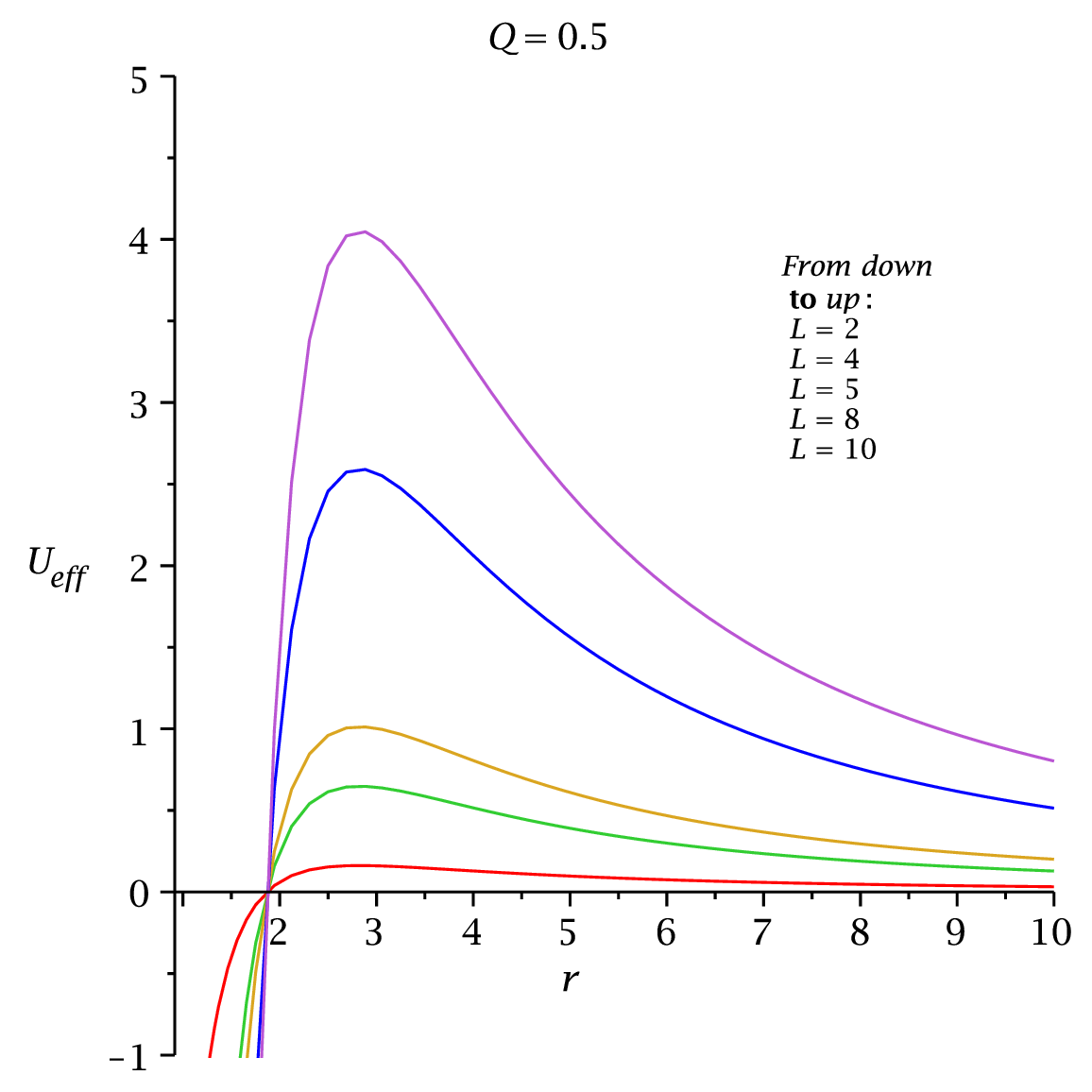}}
\end{center}
\caption{The figure shows the variation  of $U_{eff}$  with $r$ for  RS tidal-charged BH
and RN BH with $M=M_{p}=M_{5}=1$.
 \label{ueff1}}
\end{figure}

\begin{figure}
\begin{center}
{\includegraphics[width=0.45\textwidth]{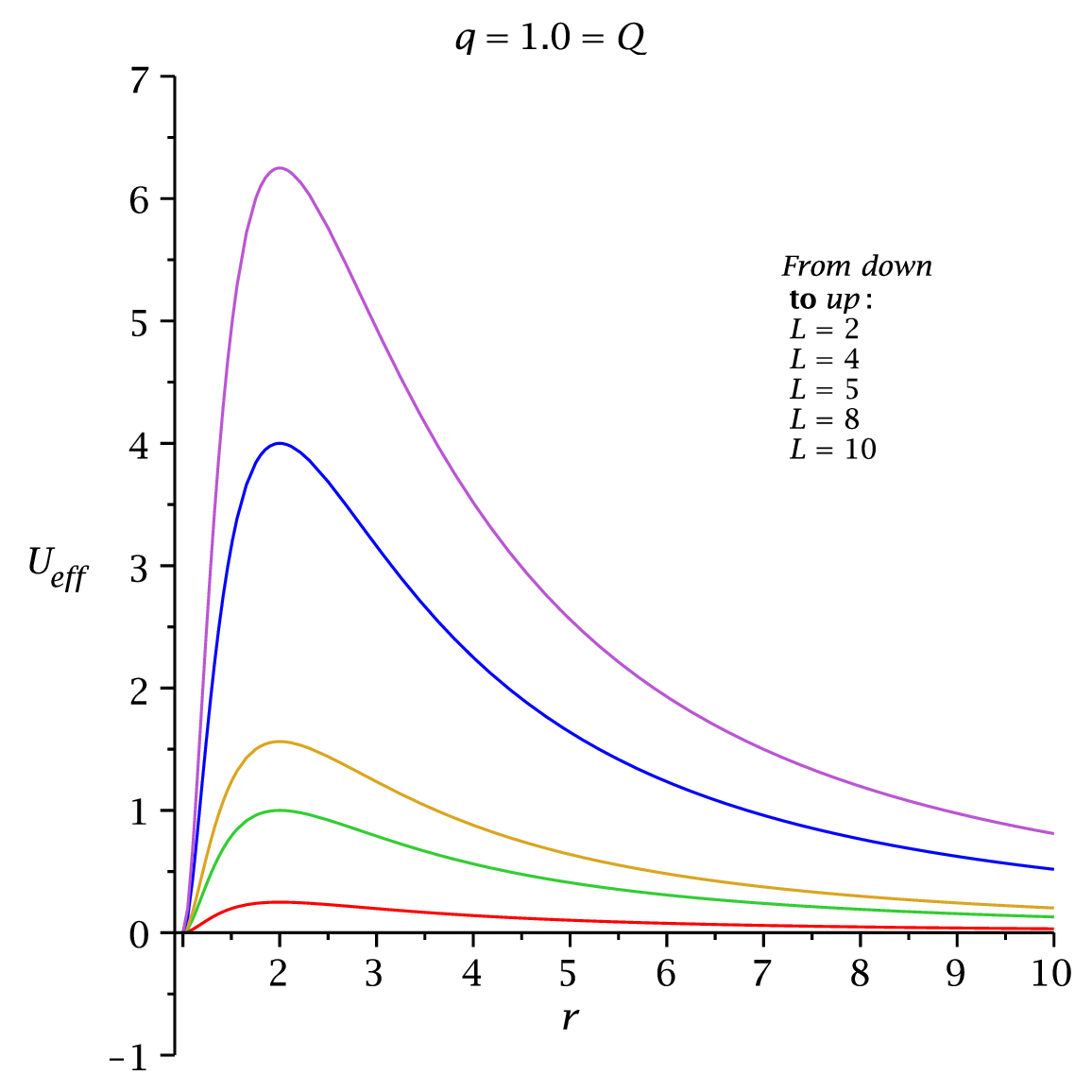}}
\end{center}
\caption{The figure shows the variation  of $U_{eff}$  with $r$ for  RS tidal-charged BH and RN BH
in the extremal limit with $M=M_{p}=M_{5}=1$.
 \label{ueff2}}
\end{figure}

\begin{figure}
\begin{center}
{\includegraphics[width=0.45\textwidth]{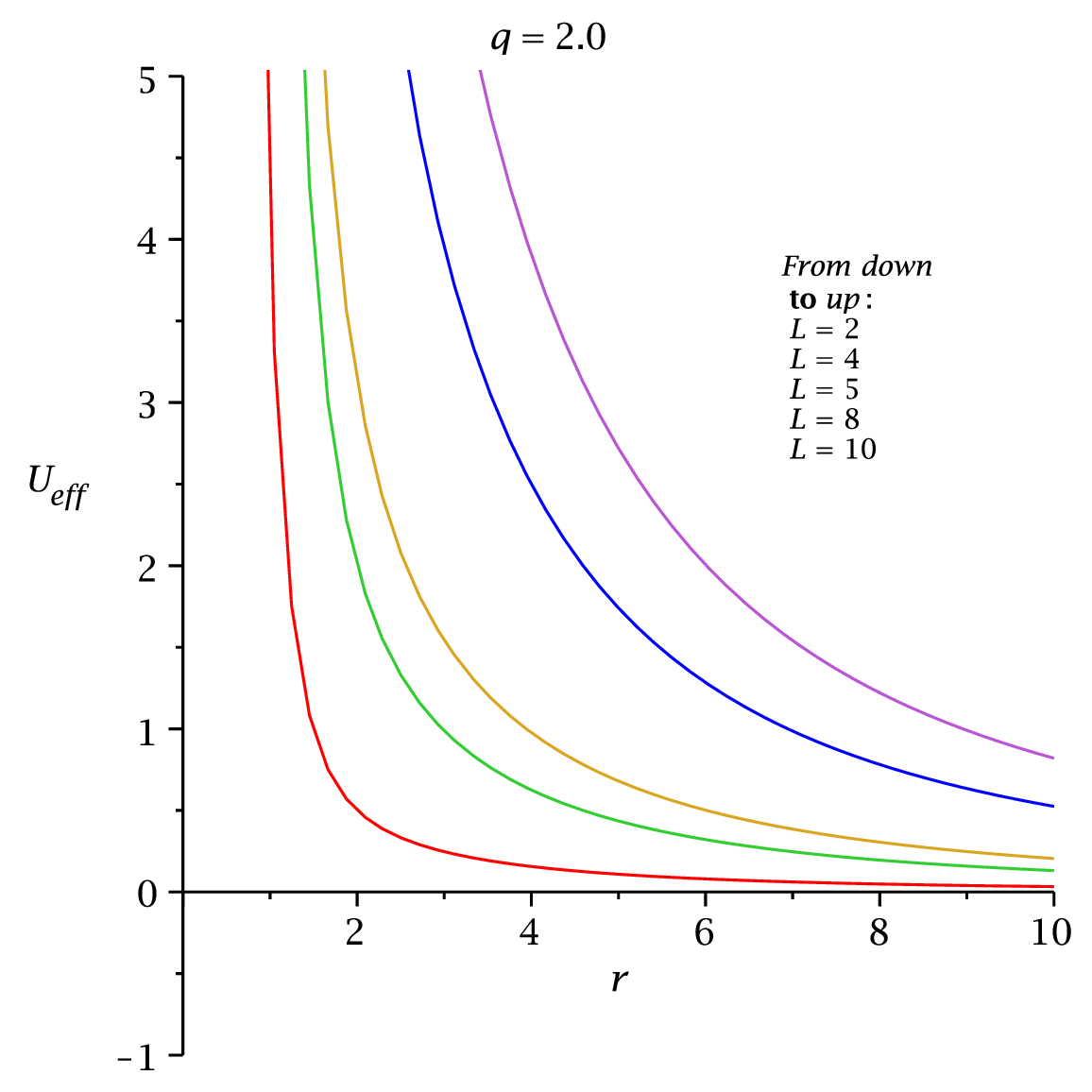}}
{\includegraphics[width=0.45\textwidth]{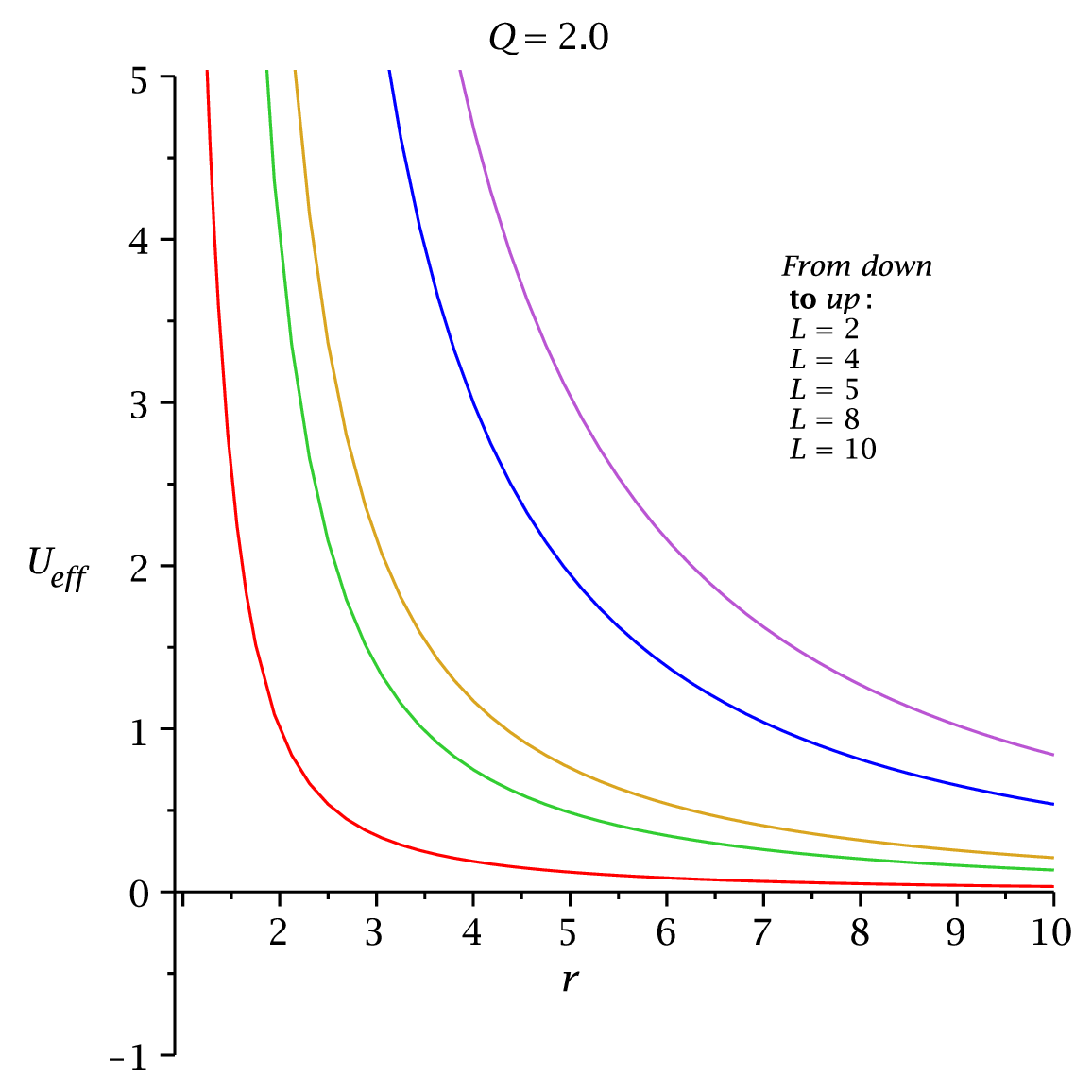}}
\end{center}
\caption{The figure shows the variation  of $U_{eff}$  with $r$ for  RS tidal-charged BH
and RN BH with $M=M_{p}=M_{5}=1$.
 \label{ueff3}}
\end{figure}

In Figures \ref{ueff},\ref{ueff1},\ref{ueff2},\ref{ueff3}, one can observed how the 
effective potential $U_{eff}$ of CPO changes for different values of angular momentum 
$L$ for two BHs.


In the extremal limit, one obtains
\begin{eqnarray}
U_{eff} &=& \frac{L^2}{r^2}{\cal B}(r)
=\frac{L^2}{r^{2}}\left(1-\frac{r_{+}}{r}\right)^{2} ~ \label{uefx}
\end{eqnarray}
For circular null geodesics at $r=r_{c}$, one obtains
\begin{eqnarray}
U_{eff} &=& E^2 ~ \label{uefr}
\end{eqnarray}
and
\begin{eqnarray}
 \frac{dU_{eff}}{dr} &=& 0 ~ \label{uefr1}
\end{eqnarray}
Therefore one can evaluate the ratio of energy and angular momentum of the
test particle at $r=r_{c}$ for CPO
\begin{eqnarray}
\frac{E_{c}}{L_{c}} &=& \frac{\sqrt{(r_{c}-r_{+})(r_{c}-r_{-})}}{r_{c}^2} ~\label{imft}
\end{eqnarray}
and
\begin{eqnarray}
2r_{c}^{2}-3(r_{+}+r_{-})r_{c}+4r_{+}r_{-}  &=& 0 ~\label{ph1t}
\end{eqnarray}
Let $r_{c}=r_{cpo}$ be the  solution of the Eq. (\ref{ph1t}) which gives
the radius of the CPO of the DMPR BH.

After introducing the impact parameter $D_{c}=\frac{L_{c}}{E_{c}}$, the above
equation  {can} be re-written as
\begin{eqnarray}
 \frac{1}{D_{c}} &=& \frac{E_{c}}{L_{c}}=\frac{\sqrt{(r_{c}-r_{+})(r_{c}-r_{-})}}{r_{c}^2}=\Omega_{c} ~\label{omegt}
\end{eqnarray}
where $\Omega_{c}$ is the angular frequency.

In the extremal limit $r_{+}=r_{-}$,  {one obtains}
\begin{eqnarray}
\Omega_{c} =\frac{1}{D_{c}} &=& \frac{E_{c}}{L_{c}}
=\frac{(r_{c}-r_{+})}{r_{c}^2} ~\label{omect}
\end{eqnarray}

Now  {one can derive} the classical capture cross-section   {which} is given by
\begin{eqnarray}
\sigma &=& \pi D_{c}^{2}=\pi \frac{r_{c}^{4}}{(r_{c}-r_{+})(r_{c}-r_{-})}~\label{sigmt}
\end{eqnarray}
In the extremal limit,  {the above expression reduces to}
\begin{eqnarray}
\sigma_{ex} &=& \pi D_{c}^{2}=\pi \frac{r_{c}^{4}}{(r_{c}-r_{+})^2}~\label{sigmx}
\end{eqnarray}

\subsection{MBCO}
Another important class of orbit is the MBCO  {can} be found
by setting $E_{0}^{2}=1$  in Eq.~(\ref{engt}), then  {one obtains} the MBCO
equation as
\begin{eqnarray}
(r_{+}+r_{-})r_{0}^{3}-2(r_{+}+r_{-})^{2}r_{0}^{2}+4r_{+}r_{-}(r_{+}+r_{-})r_{0}
-2(r_{+}r_{-})^{2} &=& 0
~ \label{mbcot}
\end{eqnarray}
or
\begin{eqnarray}
MM_{p}^2M_{5}^4 r_{0}^3-4M^2M_{5}^4r_{0}^2+4MM_{p}^2M_{5}^2qr_{0}-q^2M_{p}^4 &=& 0
~ \label{mbcot1}
\end{eqnarray}
Let $r_{0}=r_{mb}$ be the solution of the equation which gives the radius of
MBCO of brane-world BH.

In the extremal limit the above equation reduces to
\begin{eqnarray}
r_{0}^{3}-4r_{+}r_{0}^{2}+4r_{+}^{2}r_{0}-r_{+}^{3} &=& 0
~ \label{mbcot2}
\end{eqnarray}
or
\begin{eqnarray}
r_{0} &=& \left(\frac{3 + \sqrt{5}}{2}\right)r_{+} =\left(\frac{3 + \sqrt{5}}{2}\right) \frac{M}{M_{p}^{2}}
\end{eqnarray}

\subsection{ISCO}
\begin{figure}
\begin{center}
{\includegraphics[width=0.45\textwidth]{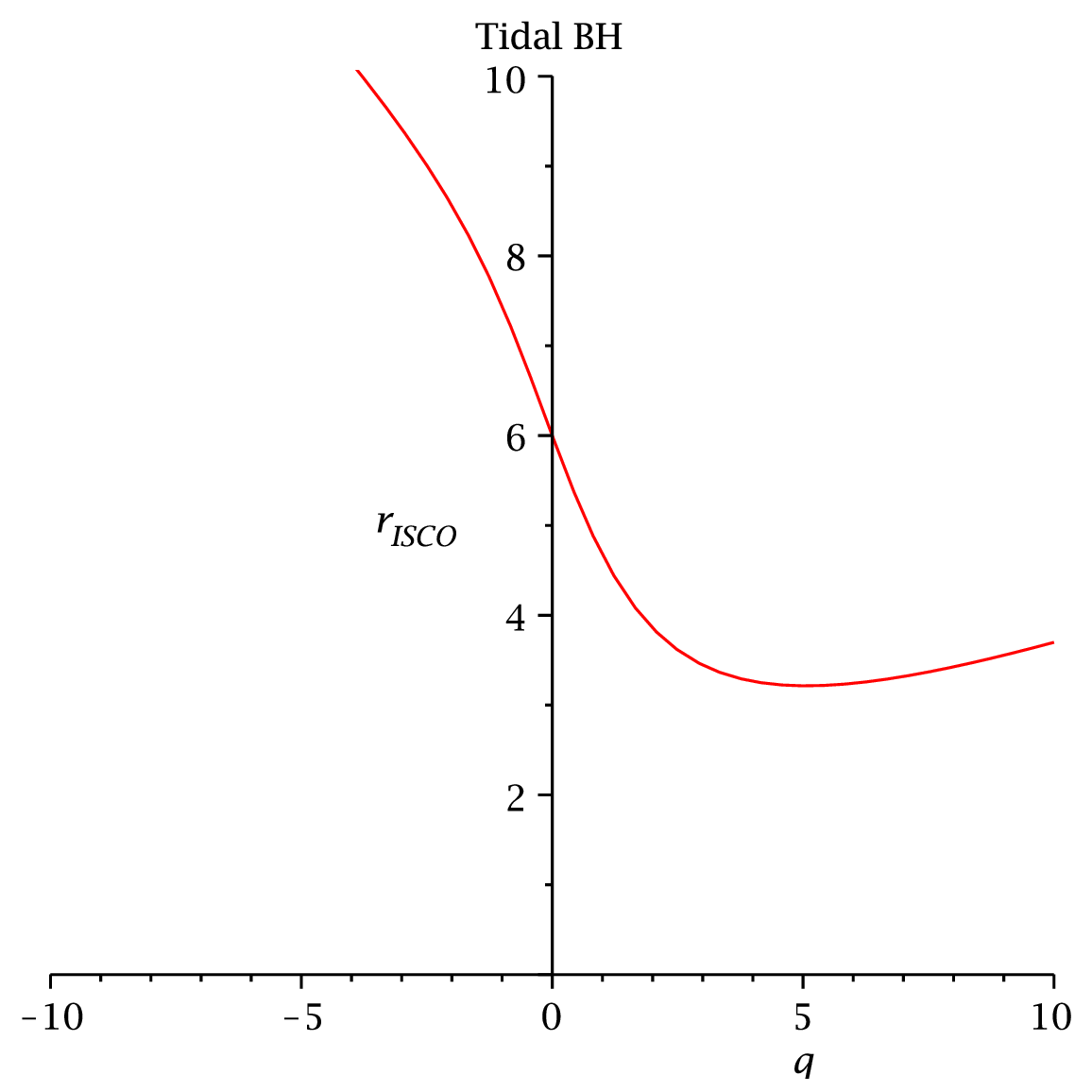}}
{\includegraphics[width=0.45\textwidth]{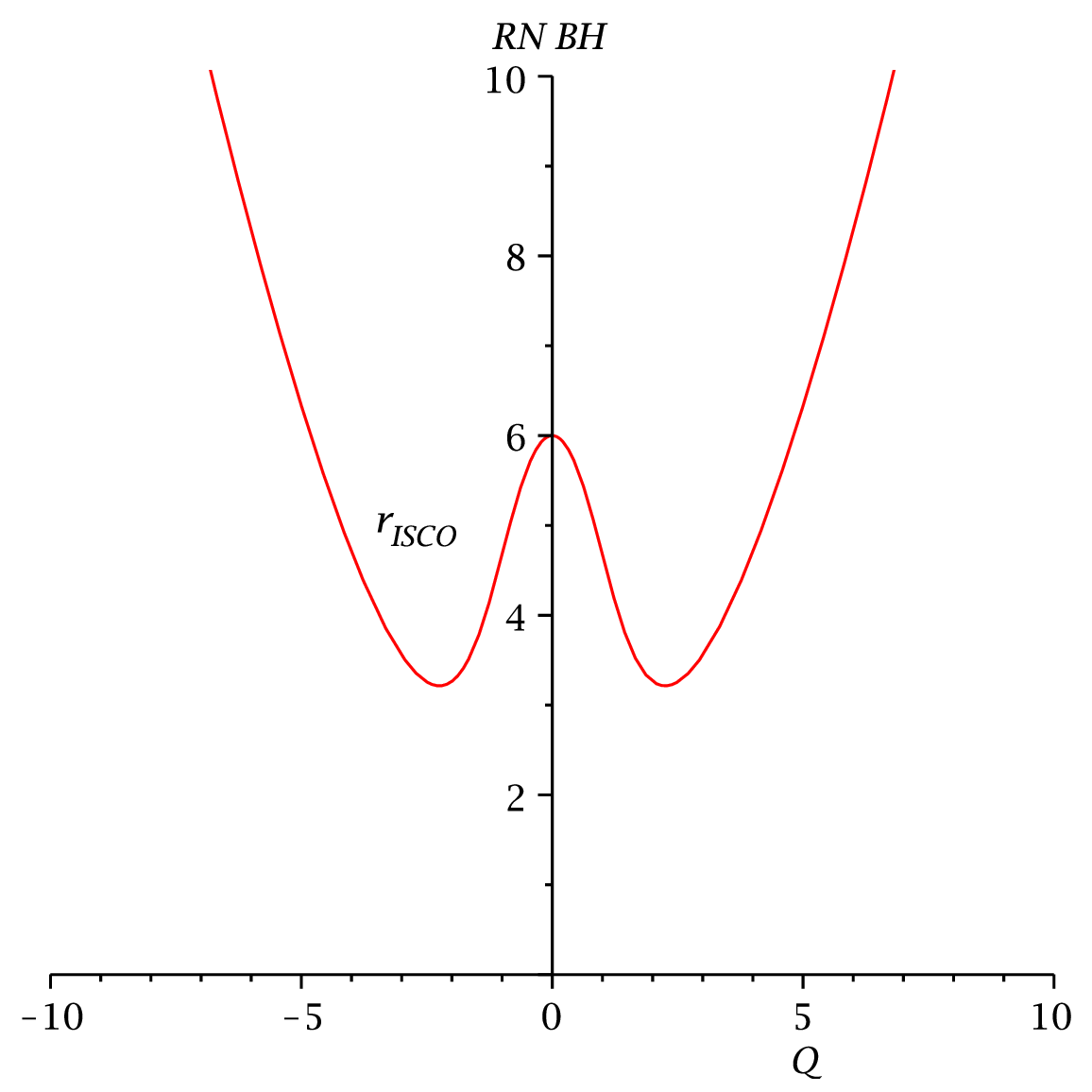}}
\end{center}
\caption{The figure shows the variation of $r_{ISCO}$ with charge parameter
for DMPR BH and RN BH with $M=M_{p}=M_{5}=1$. For DMPR BH the ISCO radius decreases with increasing the 
tidal parameter $q$ up to a certian value then it again slowly increases. For RN BH, the picture is quite 
different. Here the radius of ISCO falls quickly with increasing the charge  {parameter} $Q$ 
up to a certain value then it increases exponentially.
 \label{msco}}
\end{figure}

In astrophysics, there is an important class of orbit which is  {the} most relevant in accretion
disk theory  {which is so} called ISCO or the orbit of marginal stability  {can} be 
derived from the following criterion
\begin{eqnarray}
V_{eff} &=& E^2\,\,,  \mbox{and} \,\, \frac{dV_{eff}}{dr}=\frac{d^2V_{eff}}{dr^2} = 0 \label{stct}
\end{eqnarray}
Thus one obtains
\begin{eqnarray}
(r_{+}+r_{-})r_{0}^{3}-3(r_{+}+r_{-})^{2}r_{0}^{2}+9r_{+}r_{-}(r_{+}+r_{-})r_{0}
-8(r_{+}r_{-})^{2} &=& 0 ~\label{iscot}
\end{eqnarray}
or
\begin{eqnarray}
MM_{p}^2M_{5}^4 r_{0}^3-6M^2M_{5}^4r_{0}^2+9MM_{p}^2M_{5}^2qr_{0}-4q^2M_{p}^4 &=& 0
~ \label{iscot1}
\end{eqnarray}
Let $r_{0}=r_{ISCO}$ be the solution of the equation which gives the radius of the
ISCO
\begin{eqnarray}
\frac{r_{ISCO}}{M/M_{p}^2} &=& 2+X^{1/3}+\frac{(4-3\frac{q}{M_{5}^2})}{X^{1/3}}\\
X &=& \left[8-9(\frac{qM_{p}^4}{M_{5}^2}M^2)+2\frac{q^2M_{p}^8}{M^4M_{5}^4}+
 \sqrt{5\frac{q^2M_{p}^8}{M^4M_{5}^4}-9\frac{q^3M_{p}^{12}}{M^6M_{5}^6}+4\frac{q^4M_{p}^8}{M^8M_{5}^8}}\right]
 ~ \label{iscot2}
\end{eqnarray}
In Fig.~\ref{msco}, we show how the radius of ISCO changes with charge parameter both for DMPR BH
and RN BH.
For extremal DMPR BH, one could obtain the ISCO equation
\begin{eqnarray}
r_{0}^{3}-6r_{+}r_{0}^{2}+9r_{+}^{2}r_{0}-4r_{+}^{3} &=& 0
\end{eqnarray}
or
\begin{eqnarray}
r_{0} &=& 4r_{+} =4\frac{M}{M_{p}^{2}}
\end{eqnarray}

\section{Null Circular Geodesics and QNM for DMPR BH in the Eikonal limit}
When a BH is given to be an exterior perturbation it oscillates with a certain frequency. The frequency of 
oscillation is 
said to be a QNM  {frequency}. In this section, we briefly introduce the QNM in the eikonal 
limit for DMPR BH in comparison with RN BH following the work of Cardoso et al.~\cite{car}. 
In this work, the authors have derived the expressions for QNM frequency for higher dimensional BH spacetime.
The fact that the unstable circular photon 
geodesics is very useful to compute the characteristic modes of BH, which is popularly
known as QNMs~\cite{konoplya}. We have derived  in~\cite{pramana} the QNM frequency for RN BH and Schwarzschild BH 
in the eikonal approximation very recently. We have not discussed  the mathematical details here rather 
we write the explicit formula which was derived in~\cite{car}
\begin{eqnarray}
\omega_{QNM}=\ell \Omega_{c}- i \left(n+\frac{1}{2}\right){\lambda_{c}} ~.\label{tq3}
\end{eqnarray}
where $n$ denotes the overtone number, $\ell$ denots the angular momentum of the perturbation, $\Omega_{c}$ denotes 
the angular frequency measured by the asymptotic observers and  ${\lambda_{c}}$ denotes
the coordinate time Lyapunov exponent for CPO which is defined in terms of the effective 
potential for CPO
\begin{eqnarray}
\lambda_{c} & =&  \sqrt{-\frac{U_{eff}''}{2\dot{t}^{2}}} ~.\label{tpot1}
\end{eqnarray}
and 
\begin{eqnarray}
\Omega_{c}=\frac{\dot{\phi}}{\dot{t}}=\frac{1}{D_{c}} ~.\label{tom1}
\end{eqnarray}

For DMPR BH,  $\Omega_{c}$ and $\lambda_{c}$ are 
\begin{eqnarray}
\Omega_{c} &=& \sqrt{\frac{\left(\frac{Mr_{c}}{M_{p}^2}-\frac{q}{M_{5}^2}\right)}{r_{c}^4}}
~.\label{tom2}
\end{eqnarray}
and 
\begin{eqnarray}
\lambda_{c} & =& \sqrt{\frac{\left(\frac{Mr_{c}}{M_{p}^2}-\frac{q}{M_{5}^2}\right)
\left(\frac{3Mr_{c}}{M_{p}^2}-4\frac{q}{M_{5}^2}\right)}{r_{c}^6}}
~.\label{tpot2}
\end{eqnarray}
Substituting these values in Eq. (\ref{tom1}), one finds the QNM frequency for DMPR BH in the eikonal approximation
\begin{eqnarray}
\omega_{QNM} &=& 
\ell \sqrt{\frac{\frac{Mr_{c}}{M_{p}^2}-\frac{q}{M_{5}^2}}{r_{c}^4}} - 
i\left(n+\frac{1}{2}\right)\sqrt{\frac{\left(\frac{Mr_{c}}{M_{p}^2}-\frac{q}{M_{5}^2}\right)
\left(\frac{3Mr_{c}}{M_{p}^2}-4\frac{q}{M_{5}^2}\right)}{r_{c}^6}}
~.\label{tnq3}
\end{eqnarray}
In the limit $q=0$, one obtains the QNM frequency for Schwarzschild BH.
The implications of the above Eq.(\ref{tnq3})  are that in the eikonal approximation, the real and imaginary
parts of the QNMs of DMPR BH are given by the frequency and instability time scale of the
unstable CPO. This is one of the key investigation of this work.

\section{Conclusion}
This work deals with the study of the circular geodesics of spherically symmetric tidal-charged  BH.
We discussed in detail the equatorial causal geodesics of the DMPR BH in comparison with the 
spherically symmetric RN BH.  We discussed both the null circular geodesics and time-like circular geodesics. 
From the effective potential diagram, we derived the geodesic structure 
between two spacetimes. We also determined the location of ISCO , MBCO  
and CPO  for both the space-times. Moreover, we have computed the \emph{QNM frequency}
in the \emph{eikonal approximation} for both the spacetimes via Lyapunov exponent. 

In the Appendix section, we examined  {that} a spherically symmetric tidal-charged BH may act as  
particle accelerators with ultra-high CM energy in the limiting case of \emph{maximal BH tidal charge ($q$)} 
 {and} when two neutral particles are colliding near the horizon. 
For the value of tidal charge $q<0$, the BH can not act as  particle accelerators with arbitrarily high CM energy and 
the CM energy is finite in this situation. For the value of tidal charge $q>0$, the BH can act as  particle 
accelerators with arbitrarily high CM energy which is \emph{similar} to the spherically symmetric extreme RN BH. 
Finally, for $M=0$ and $q<0$, we get the CM energy is finite and the BH can not act as  particle accelerators 
with arbitrarily high CM energy.

\section{Appendix A.~CM Energy of Particle Collision near the horizon of the DMPR BH}

Now we shall calculate the CM energy of two particle collisions close to the horizon
of the brane-world BH. To proceeds, first we consider two particles are coming from
infinity with $\frac{E_{1}}{m_{\chi}}=\frac{E_{2}}{m_{\chi}}=1$ approaching the tidal
charged BH with different angular momenta $L_{1}$ and $L_{2}$ and colliding at
some radius $r$. Later, we consider the collision point $r$ to approach the
horizon $r=r_{+}$. Also we have assumed that the particles to be at rest at
infinity.

The CM energy is computed by using the following formula which was
first derived by BSW~\cite{bsw} reads 
\begin{eqnarray}
\left(\frac{E_{cm}}{\sqrt{2}m_{\chi}}\right)^{2} &=&
1-g_{\mu\nu}u^{\mu}_{(1)}u^{\nu}_{(2)} ~\label{cm}
\end{eqnarray}

We shall also assume throughout the work, the geodesic motion of the colliding particles
confined in the equatorial plane. Since the spacetime has a time-like isometry
followed by the time-like Killing vector field $\xi$ whose projection along the
four velocity ${\bf u}$ of geodesics $\xi.{\bf u}=-E$, is conserved along such
geodesics (where $\xi\equiv \partial_{t}$). Analogously, there is also the
rotational symmetry for which the `angular momentum' $L=\zeta.{\bf u}$ is also
conserved (where $\zeta\equiv \partial_{\phi})$.

\begin{figure}
\begin{center}
{\includegraphics[width=0.45\textwidth]{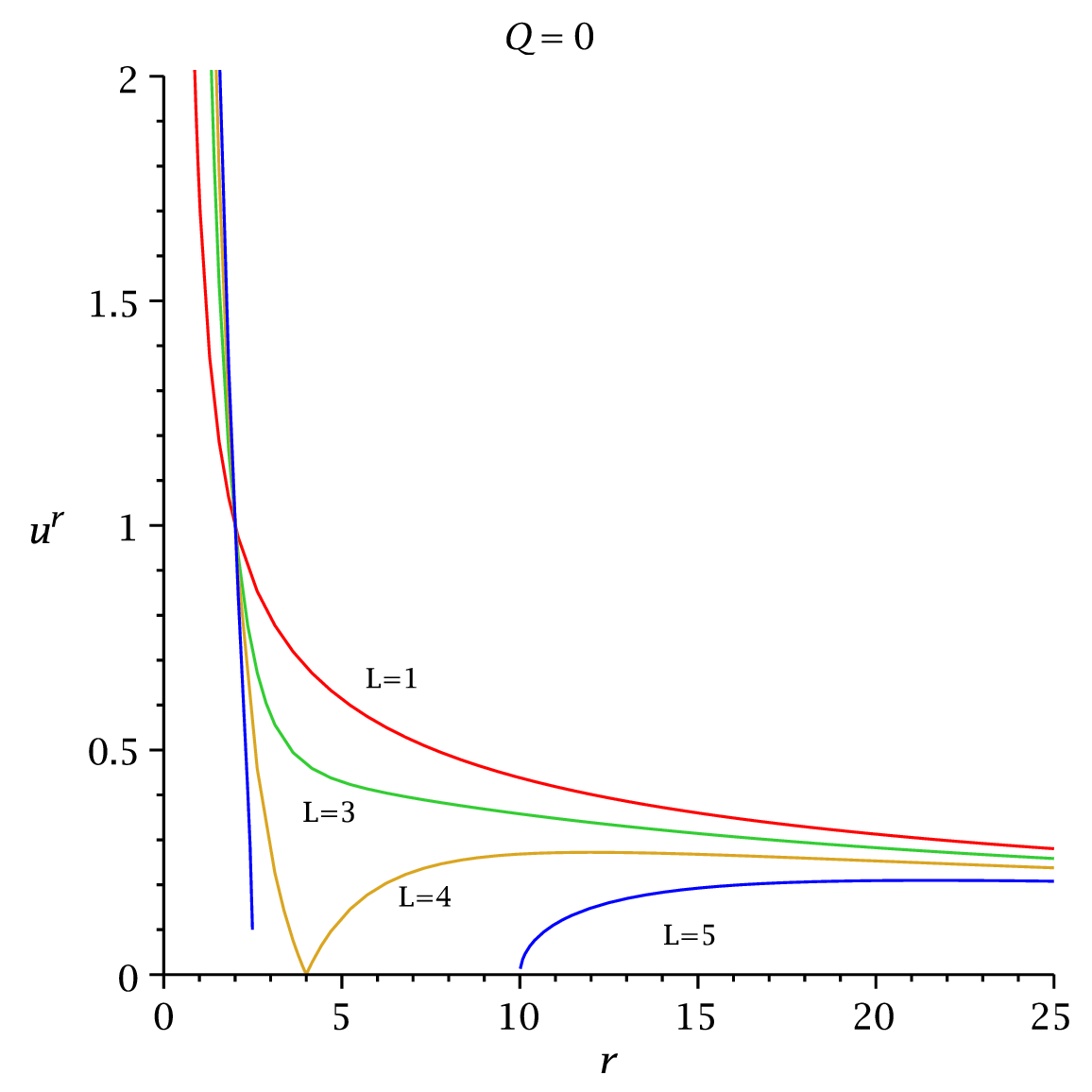}}
\end{center}
\caption{The figure shows the variation of $u^{r}=\dot{r}$  with $r$ for  DMPR BH in the limit 
$q=0$ with $M=M_{p}=M_{5}=1$.
\label{tdcase}}
\end{figure}

\begin{figure}
\begin{center}
{\includegraphics[width=0.45\textwidth]{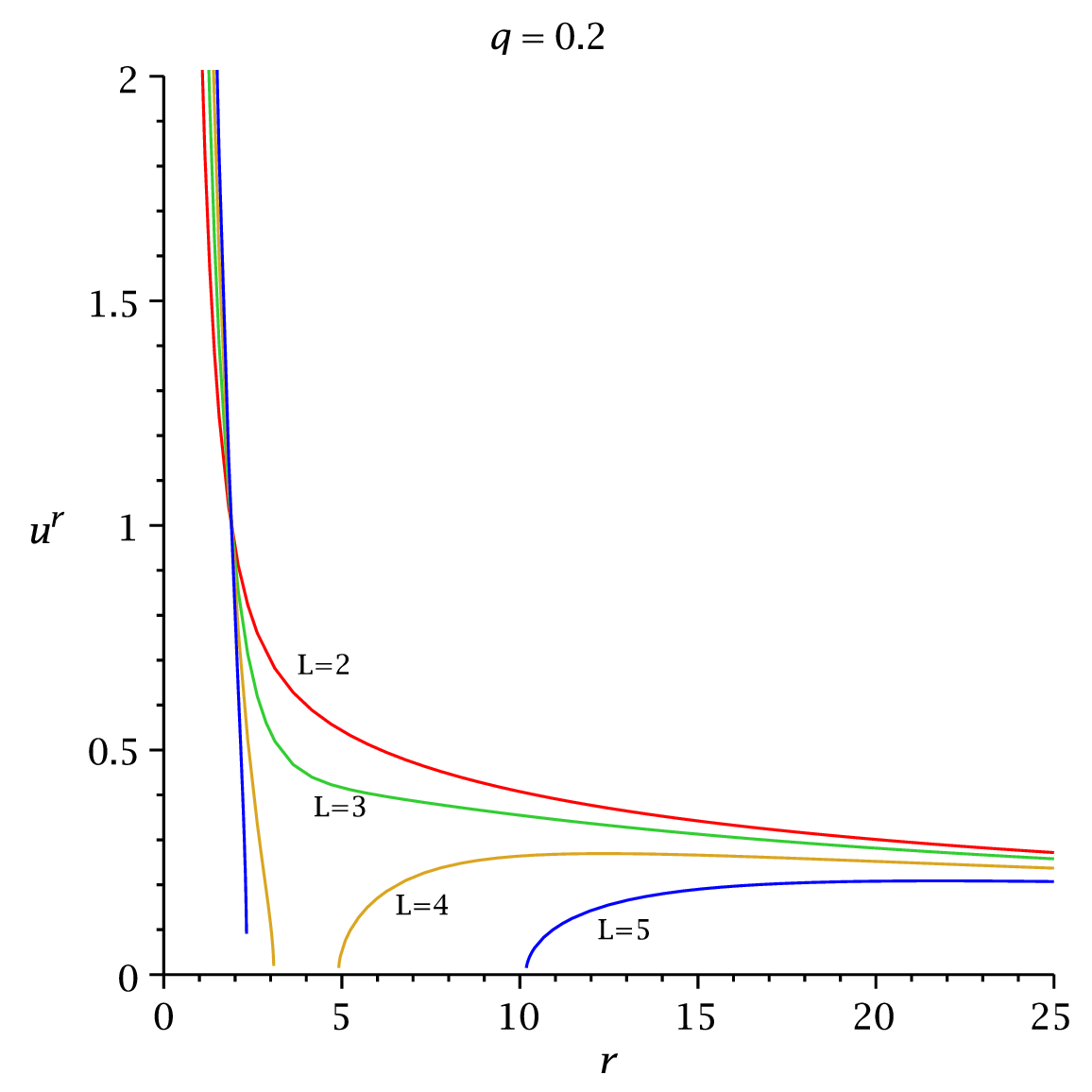}}
{\includegraphics[width=0.45\textwidth]{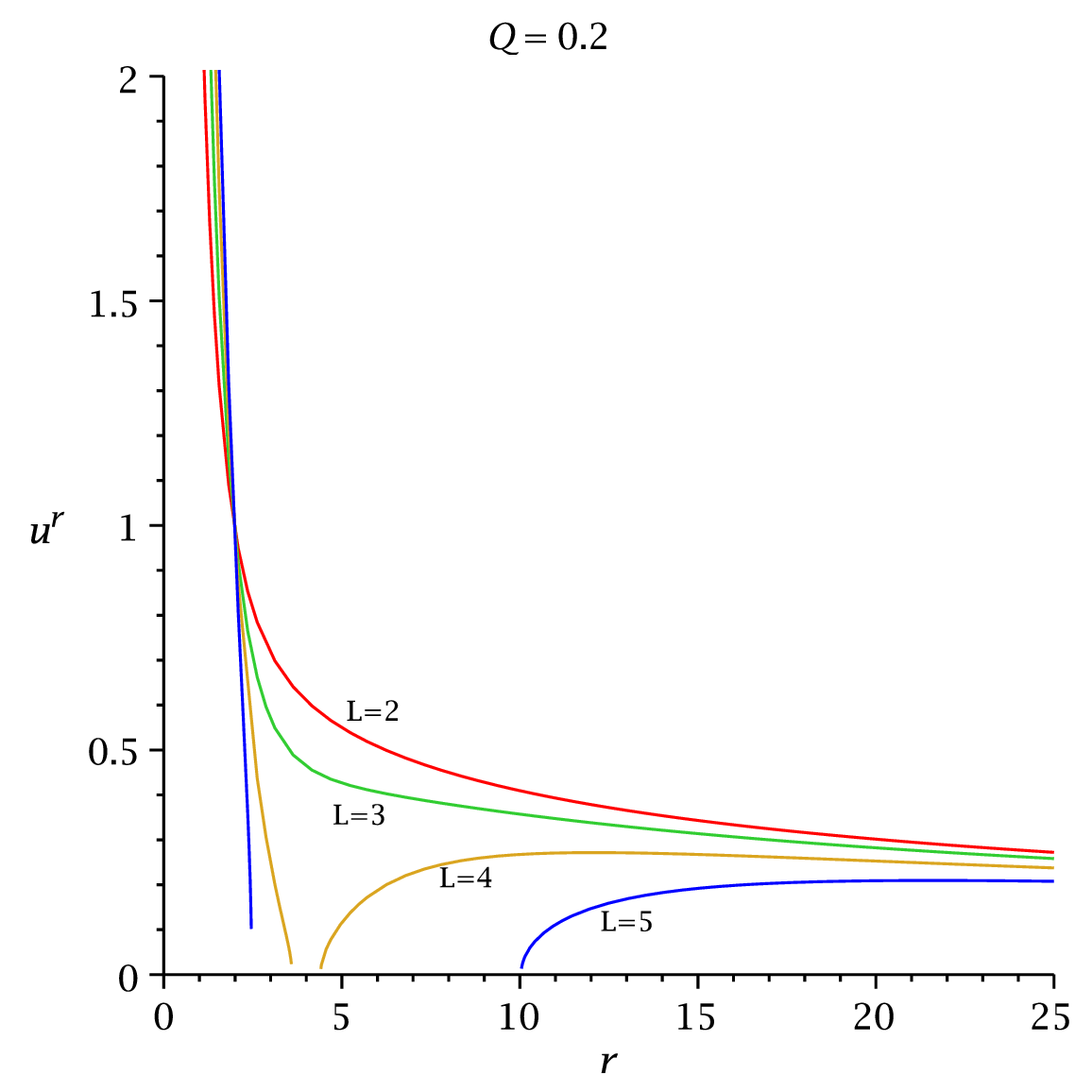}}
\end{center}
\caption{The figure shows the variation  of $u^{r}=\dot{r}$  with $r$ for  DMPR BH and RN BH with $M=M_{p}=M_{5}=1$.
 \label{tidc1}}
\end{figure}

\begin{figure}
\begin{center}
{\includegraphics[width=0.45\textwidth]{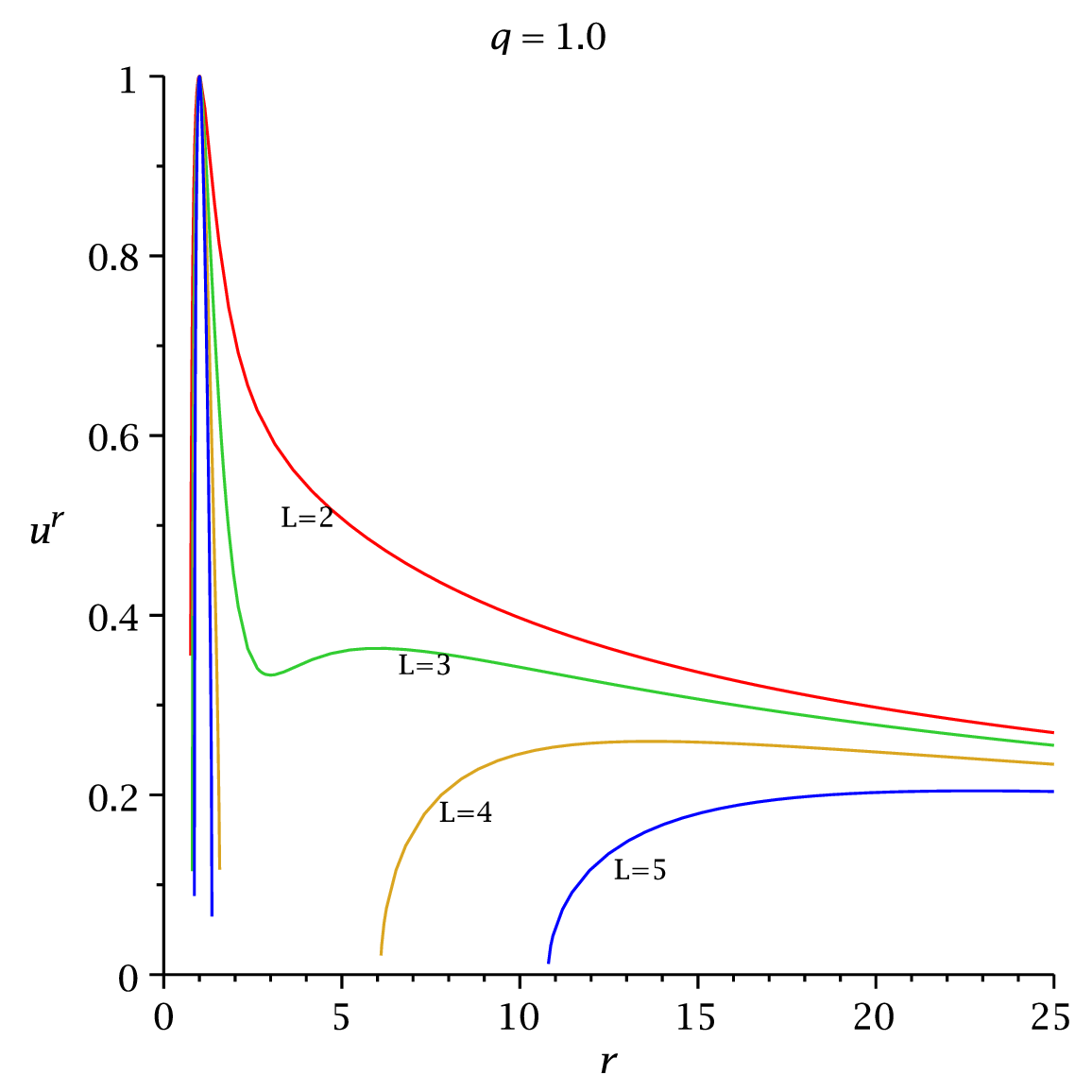}}
{\includegraphics[width=0.45\textwidth]{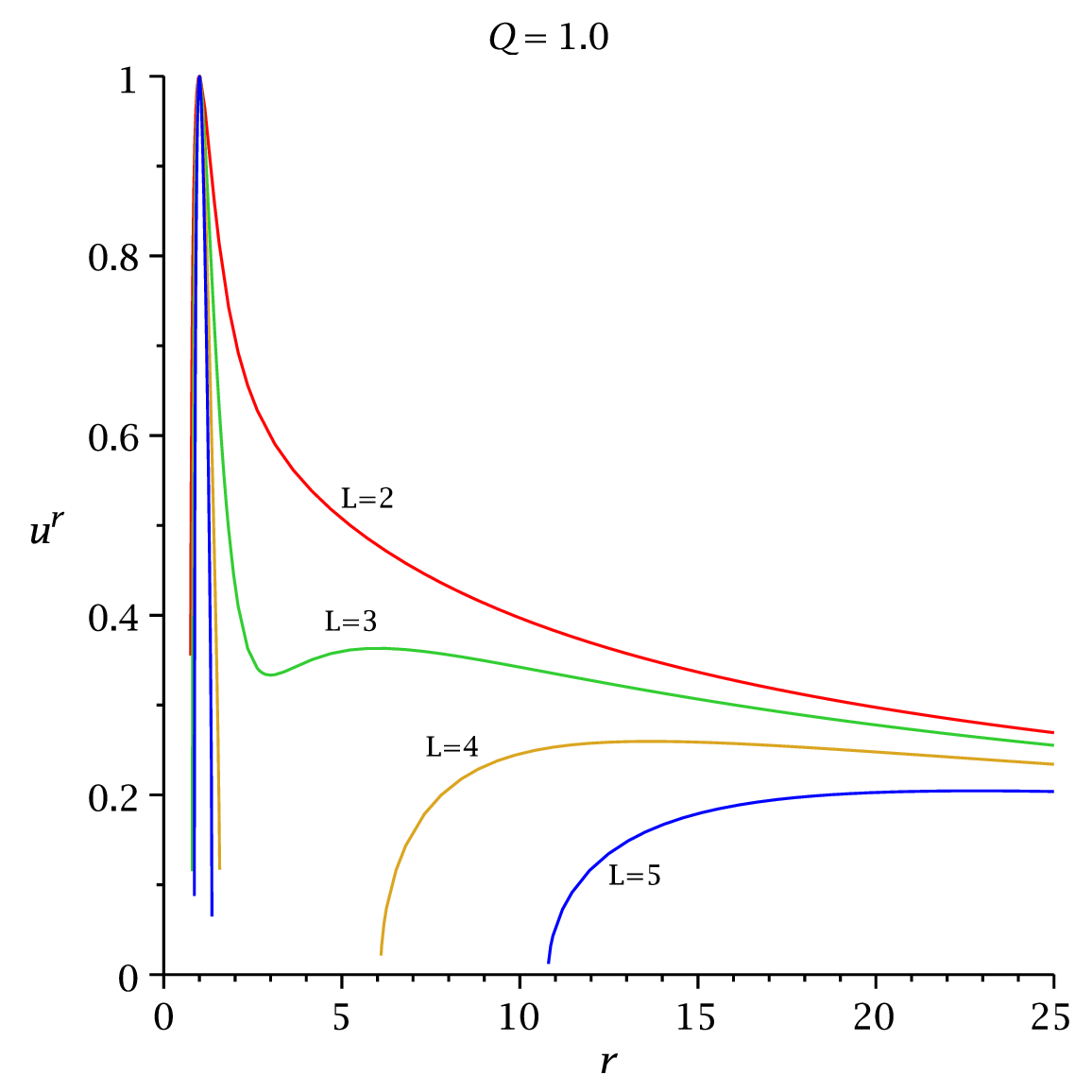}}
\end{center}
\caption{The figure shows the variation  of $u^{r}=\dot{r}$  with $r$ for  DMPR BH and RN BH with $M=M_{p}=M_{5}=1$.
 \label{tidc3}}
\end{figure}

\begin{figure}
\begin{center}
{\includegraphics[width=0.45\textwidth]{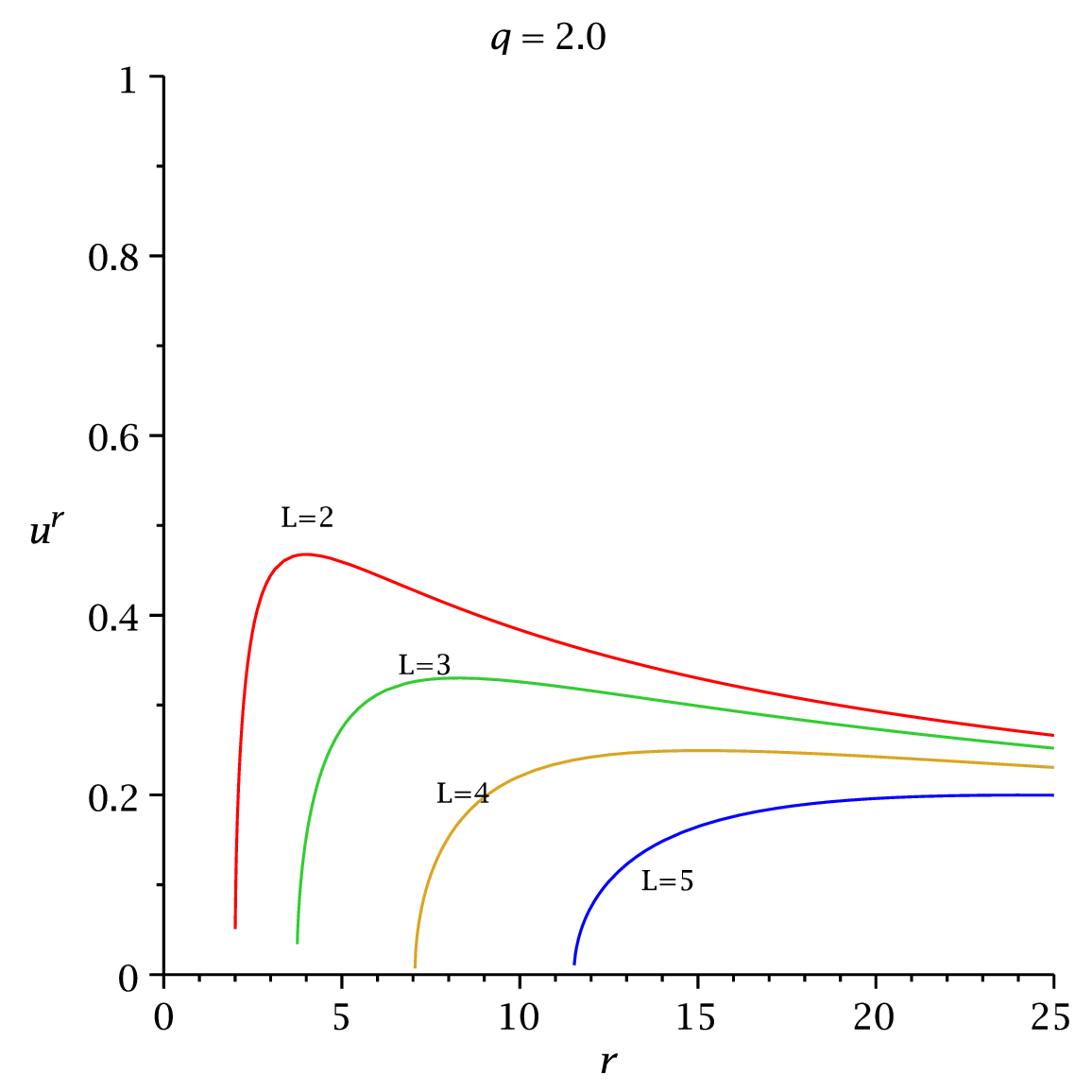}}
{\includegraphics[width=0.45\textwidth]{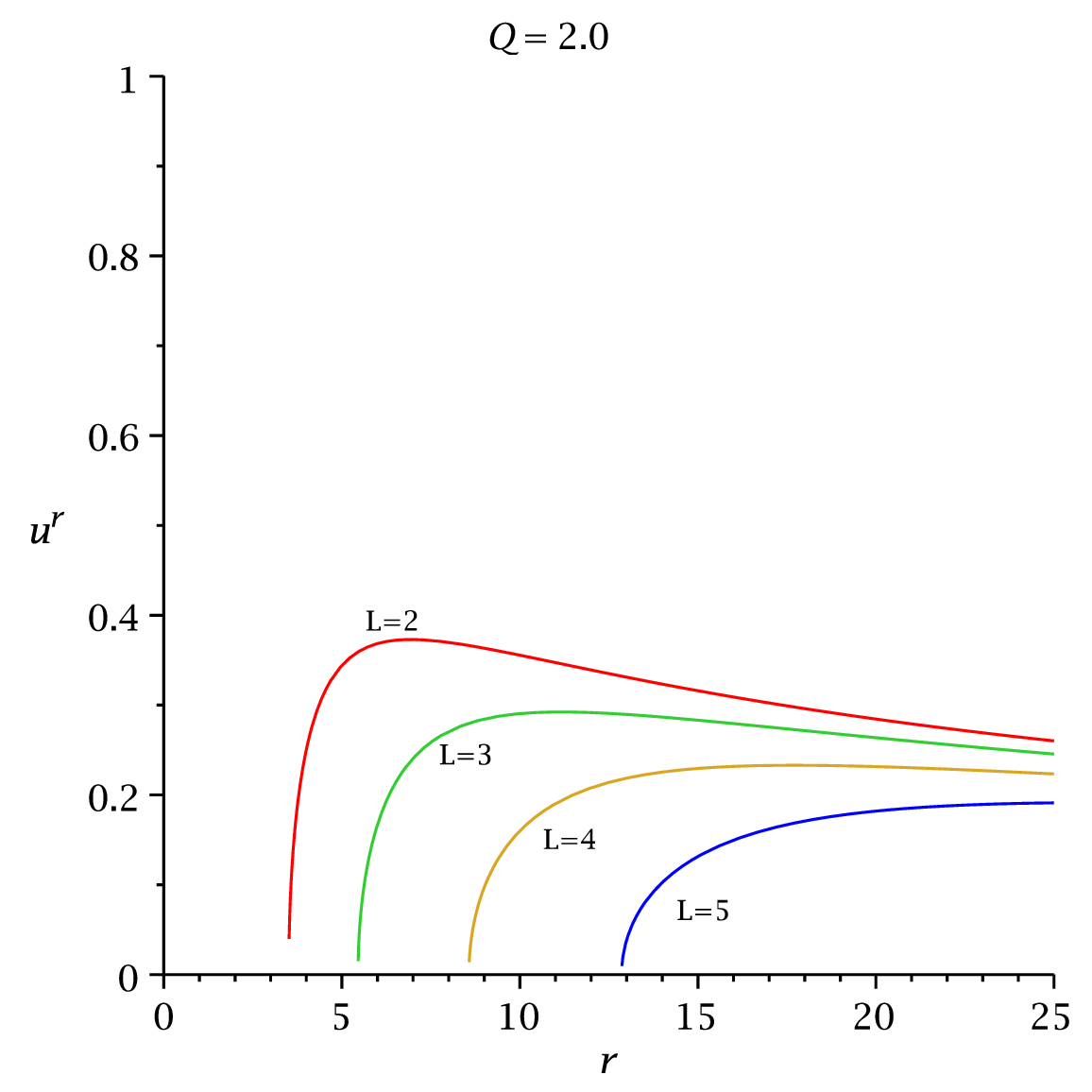}}
\end{center}
\caption{The figure shows the variation  of $u^{r}=\dot{r}$  with $r$ for  DMPR BH and RN BH with 
$M=M_{p}=M_{5}=1$.
 \label{tidc4}}
\end{figure}

Thus for massive particles, the components of four velocity are
\begin{eqnarray}
 u^{t} &=& \dot{t}= \frac{{E}}{{\cal B}(r)}  \\
 u^{r} &=& \dot{r}=\pm \sqrt{E^{2}-{\cal B}(r) \left(1+\frac{L^{2}}{r^2}\right)}\label{eff}\\
 u^{\theta} &=& \dot{\theta} = 0 \\
 u^{\phi} &=& \dot{\phi} \frac{L}{r^2} ~\label{utur}
\end{eqnarray}
and
\begin{eqnarray}
u^{\mu}_{(1)} &=& \left( \frac{E_{1}}{{\cal B}(r)},~ -X_{1},~ 0,~\frac{L_{1}}{r^2}\right) ~\label{u1}\\
u^{\mu}_{(2)} &=& \left( \frac{E_{2}}{{\cal B}(r)},~ -X_{2},~ 0,~\frac{L_{2}}{r^2}\right) ~\label{u2}
\end{eqnarray}
In Figs. \ref{tdcase}, \ref{tidc1}, \ref{tidc3}, \ref{tidc4},   we have plotted
the radial component of the four velocity for various values of angular momentum for DMPR BH in contrast 
with RN BH.

Thus we get the CM energy for extreme RN BH is diverging due to one of the two particles
have diverging angular momentum at the horizon. Therefore
\begin{eqnarray}
E_{cm}\mid_{r\rightarrow M}
&=& \sqrt{2}m_{\chi}\sqrt{\frac{4M^2+(L_{1}-L_{2})^{2}}{2M^2}} \rightarrow \infty ~\label{cm12}
\end{eqnarray}

Substituting this in (\ref{cm}),  {one obtains} the CM energy
\begin{eqnarray}
\left(\frac{E_{cm}}{\sqrt{2}m_{\chi}}\right)^{2} &=&  1 +\frac{E_{1}
E_{2}}{{\cal B}(r)}-
\frac{X_{1}X_{2}}{{\cal B}(r)}-\frac{L_{1}L_{2}}{r^2} ~\label{cm1}
\end{eqnarray}
where,
$$
X_{1} = \sqrt{E_{1}^{2}-{\cal B}(r)\left(1+\frac{L_{1}^{2}}{r^2}\right)}, \,\,
\\
X_{2} = \sqrt{E_{2}^{2}-{\cal B}(r)\left(1+\frac{L_{2}^{2}}{r^2}\right)}
$$

For simplicity, we have taken $E_{1}=E_{2}=1$ and substituting the value of ${\cal B}(r)$

{\bf Case I:}
when $q\geq 0$, we get the CM energy near the event horizon ($r_{+}$) of the tidal-charged
BH:
\begin{eqnarray}
E_{cm}\mid_{r\rightarrow r_{+}} &=& \sqrt{2}m_{\chi}\sqrt{2+(L_{1}-L_{2})^{2}
\frac{M_{p}^{2}M_{5}^{2}}{2(2MM_{5}^2r_{+}-qM_{p}^2)}}
 ~\label{cmt1}
\end{eqnarray}
and near the Cauchy horizon ($r_{-}$) is given by
\begin{eqnarray}
E_{cm}\mid_{r\rightarrow r_{-}} &=& \sqrt{2}m_{\chi}\sqrt{2+(L_{1}-L_{2})^{2}
\frac{M_{p}^{2} M_{5}^{2}}{2(2M M_{5}^2r_{-}-q M_{p}^2)}}
 ~\label{cmt2}
\end{eqnarray}
Which  {implies} that the CM energy is finite and depend upon the values of the
angular momentum parameter.

It was discussed in~\cite{bsw} that the maximum CM energy strictly  {depends} upon the values of
critical angular momentum such that the particles can reach the outer horizon with maximum 
tangential velocity. Therefore the critical values of angular momentum and the critical radius 
 {can} be derived from the radial effective potential. Now we impose the following 
condition for determining the critical values of angular momentum i.e.
\begin{eqnarray}
 \dot{r}^2r^4 &=& (E^2-1)r^4+2\frac{M}{M_{p}^2}r^3-(L^2+\frac{q}{M_{5}^2})r^2+2\frac{M}{M_{p}^2}L^2r
 -\frac{q}{M_{5}^2}L^2 = 0  ~\label{radrs}
\end{eqnarray}
Now setting $E^2=1$ for marginal case, the equation turns out to be
\begin{eqnarray}
2\frac{M}{M_{p}^2}r^3-(L^2+\frac{q}{M_{5}^2})r^2+2\frac{M}{M_{p}^2}L^2r-\frac{q}{M_{5}^2}L^2 &=& 0 ~\label{crs}
\end{eqnarray}
For non-extremal tidal-charged BH, one  {can} obtain the critical values of angular momentum by
solving the following equation
$$
 (\frac{M^2}{M_{p}^4}-\frac{q}{M_{5}^2})L^6-(3\frac{q^2}{M_{5}^4}-20\frac{M^2}{M_{p}^4}
 \frac{q}{M_{5}^2}+16\frac{M^4}{M_{p}^8})L^4
$$ 
\begin{eqnarray} 
 -(8\frac{M^2}{M_{p}^4}+3\frac{q}{M_{5}^2})
 \frac{q^2}{M_{5}^4}L^2-\frac{q^4}{M_{5}^8} &=& 0  ~\label{crs1}
\end{eqnarray}
But it is very difficult to find the roots of the above Eq. in the non-extremal limit. Now 
we discuss different simple cases.

In the limit $q=0$,  {one obtains} the critical values of angular momentum for Schwarzschild BH
\begin{eqnarray}
 L &=& \pm 4\frac{M}{M_{p}^2}  ~\label{crs2}
\end{eqnarray}
In the extremal limit, the above Eq. reduces to the following form
\begin{eqnarray}
 L^4-11 \frac{M^2}{M_{p}^4}L^2- \frac{M^4}{M_{p}^8} &=& 0  ~\label{crs3}
\end{eqnarray}
It has four solutions
\begin{eqnarray}
L_{1, 2} &=& -\sqrt{\frac{11\pm 5\sqrt{5}}{2}}\frac{M}{M_{p}^2} \\
L_{3, 4} &=& +\sqrt{\frac{11\pm 5\sqrt{5}}{2}}\frac{M}{M_{p}^2}   ~\label{crs4}
\end{eqnarray}
Thus the ranges of the angular momentum $L$ for in-falling geodesics are
\begin{eqnarray}
-\sqrt{\frac{11-5\sqrt{5}}{2}}\frac{M}{M_{p}^2} < L_{2}, L_{3} < \sqrt{\frac{11+5\sqrt{5}}{2}} \frac{M}{M_{p}^2}
~\label{crs5}
\end{eqnarray}
Thus if the particle is in this range, it never turns back and falling into an extremal BH.

Alternatively, the Eq. (\ref{crs})  {can} be rewritten as
\begin{eqnarray}
L(r) &=& \pm \sqrt{\frac{r^2(2\frac{M}{M_{p}^2}r-\frac{q}{M_{5}^2})}{r^2-2\frac{M}{M_{p}^2}r+\frac{q}{M_{5}^2}}}
~\label{crs6}
\end{eqnarray}
In the extremal limit, it reduces  {to the following form}
\begin{eqnarray}
L(r) &=& \pm \sqrt{\frac{\frac{M}{M_{p}^2}r^2(2r-\frac{M}{M_{p}^2})}{(r-\frac{M}{M_{p}^2})^2}} ~\label{crs7}
\end{eqnarray}
Again we have from Eq. (\ref{cmt1}) for extremal DMPR BH, the values of CM energy near the horizon
\begin{eqnarray}
E_{cm}\mid_{r\rightarrow \frac{M}{M_{p}^2}} &=& \sqrt{2}m_{\chi}\sqrt{\frac{4\frac{M^2}{M_{p}^4}+(L_{1}-L_{2})^{2}}
{2\frac{M^2}{M_{p}^4}}} ~\label{cm8}
\end{eqnarray}
Since, we have chosen the collision point is at $r=\frac{M}{M_{p}^2}$ and substituting the above 
 {values} of angular momenta we get the CM energy at this collision point
\begin{eqnarray}
E_{cm}\mid_{r\rightarrow \frac{M}{M_{p}^2}}  \rightarrow \infty  ~\label{cm9}
\end{eqnarray}
Thus for the extremal DMPR BH, we obtain the diverging CM energy. This implies 
that extremal DMPR BH  {can} act as natural Planck scale particle 
 {accelerators}.

{\bf Case II:}
When $q<0$, which gives the single horizon $r_{+}$ described by Eq. (\ref{hor1}) lying exterior
to the Schwarzschild horizon and thus the CM energy in this case is
\begin{eqnarray}
E_{cm}\mid_{r\rightarrow r_{+}} &=& \sqrt{2}m_{\chi}\sqrt{2+(L_{1}-L_{2})^{2}
\frac{M_{p}^{2}M_{5}^{2}}{2(2MM_{5}^2r_{+}-qM_{p}^2)}}
 ~\label{cmt3}
\end{eqnarray}
and which is also finite and in this case the BH can not be act as particle accelerators.

{\bf Case III:}
For $M=0$ and $q<0$, we get the CM energy:
\begin{eqnarray}
E_{cm}\mid_{r\rightarrow r_{+}} &=& \sqrt{2}m_{\chi}\sqrt{2-(L_{1}-L_{2})^{2}
\frac{M_{5}^{2}}{2q}} ~\label{cmt4}
\end{eqnarray}
This is also a finite quantity and in this case also the BH can not be act as particle 
accelerators.

It should be noted that in the limit $q\rightarrow 0$, the above expression reduces to
\begin{eqnarray}
E_{cm} &=& \sqrt{2}m_{\chi}\sqrt{\frac{16\frac{M^2}{M_{p}^4}+(L_{1}-L_{2})^{2}}{8\frac{M^2}{M_{p}^4}}}~\label{cmsch}
\end{eqnarray}
which is the CM energy of the Schwarzschild BH. In fact this is indeed a finite quantity which was
first observed in \cite{bsw}.

\begin{figure}[h]
  \begin{center}
\subfigure[]{
\includegraphics[width=2.1in,angle=0]{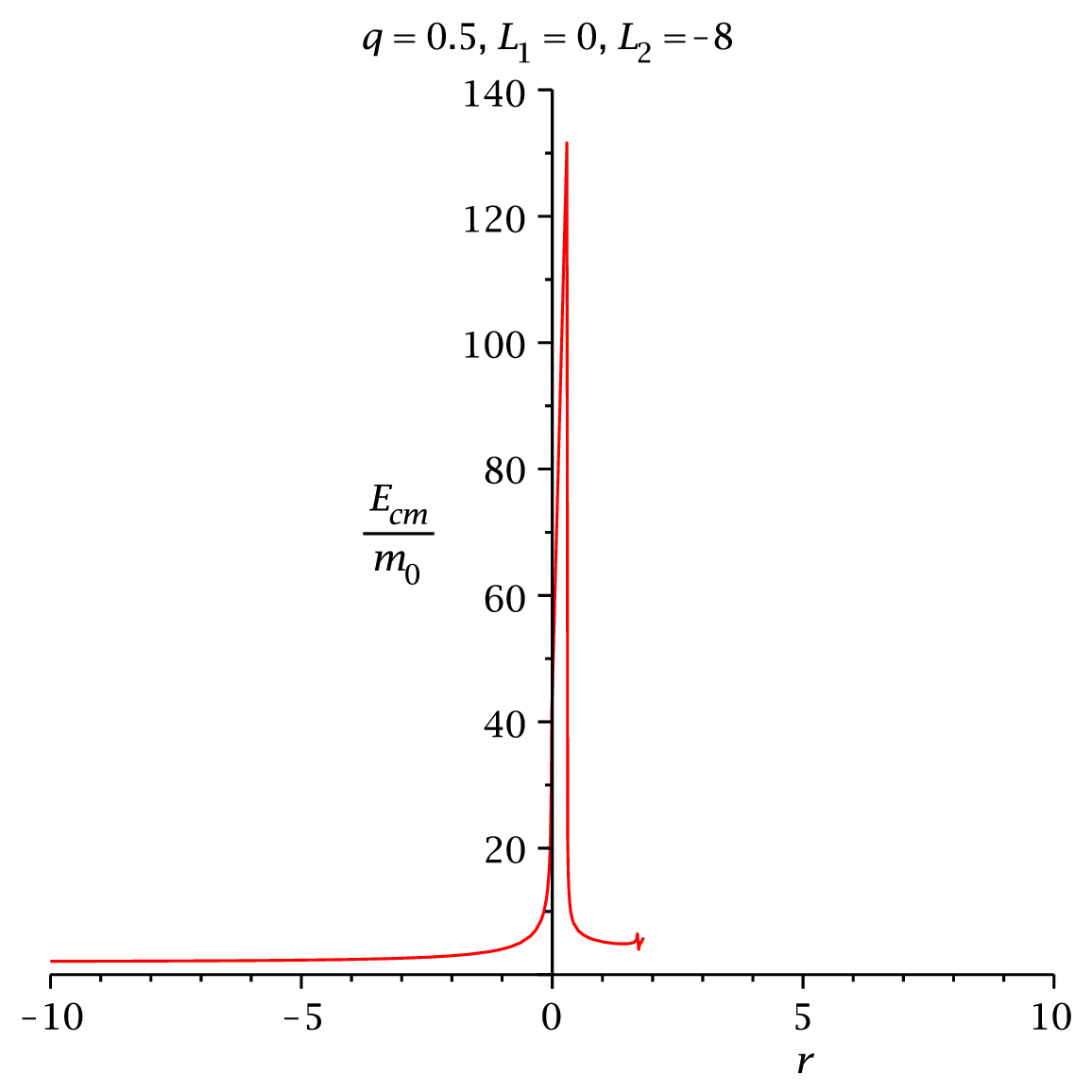}} 
\subfigure[]{
 \includegraphics[width=2.1in,angle=0]{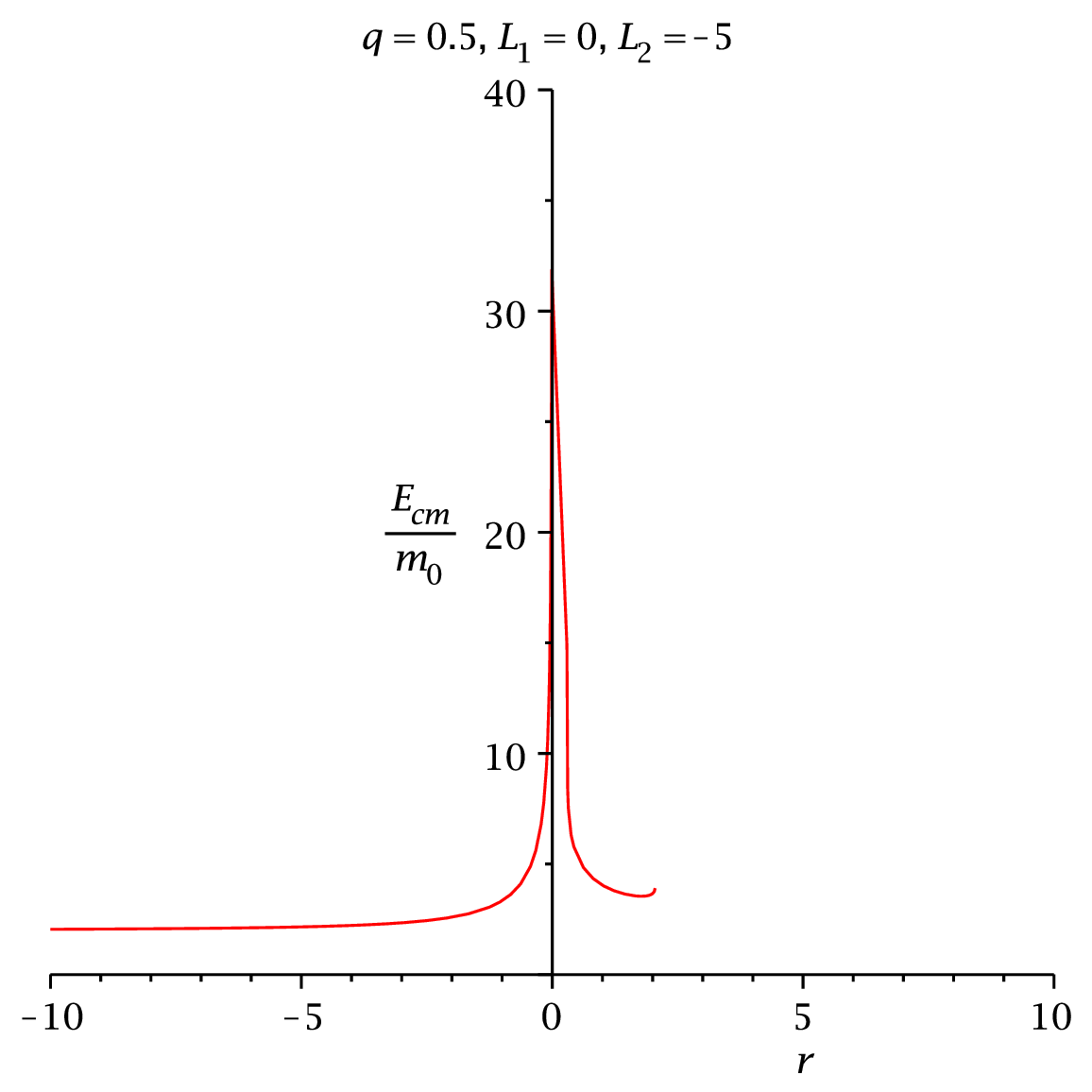}}
\subfigure[]{
 \includegraphics[width=2.1in,angle=0]{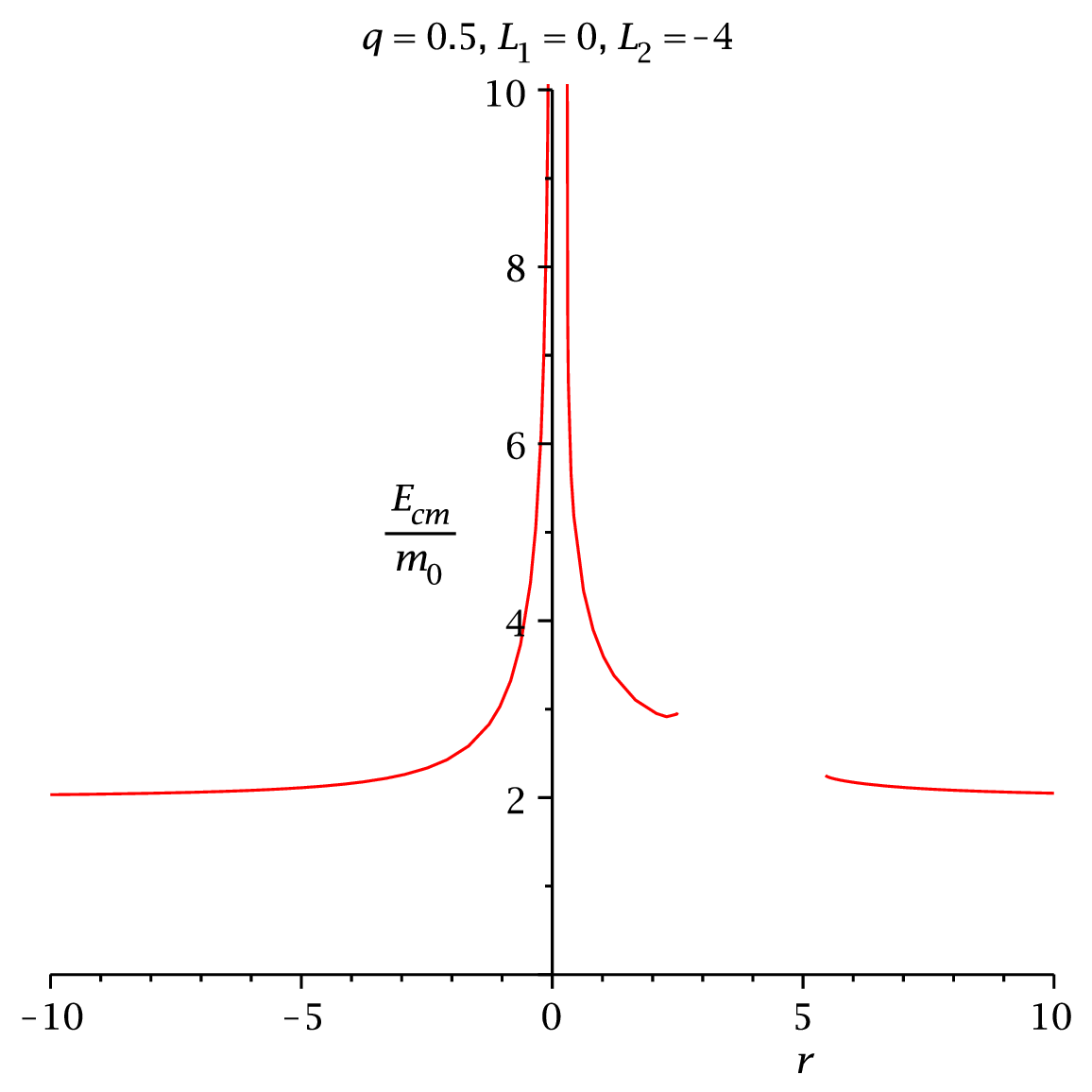}}
 \subfigure[]{
 \includegraphics[width=2.1in,angle=0]{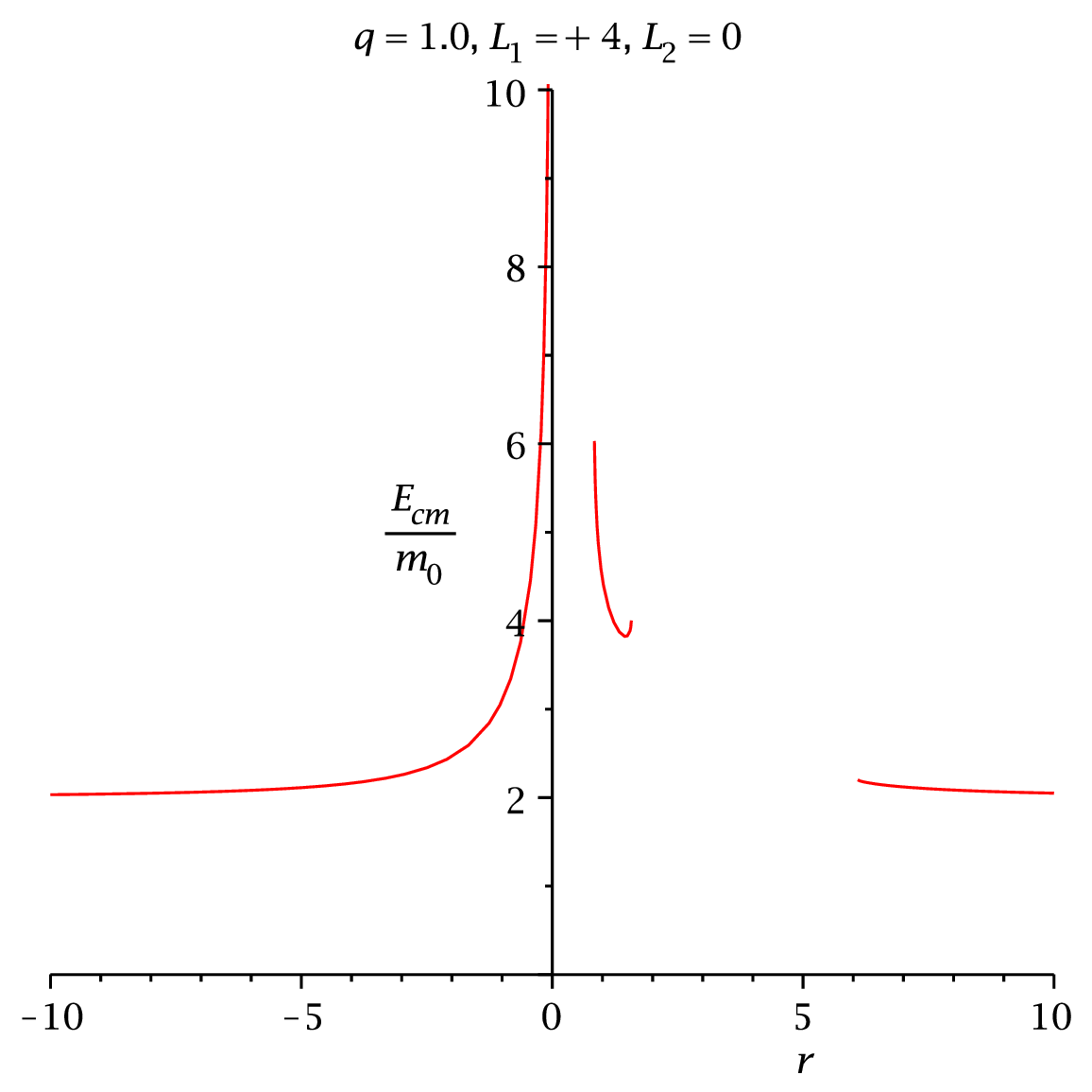}}
\caption{\label{tcm}\textit{The figure shows the variation  of $E_{cm}$  with $r$ for DMPR BH
with $M=M_{p}=M_{5}=1$. }}
\end{center}
\end{figure}

\begin{figure}[h]
  \begin{center}
\subfigure[]{
\includegraphics[width=2.1in,angle=0]{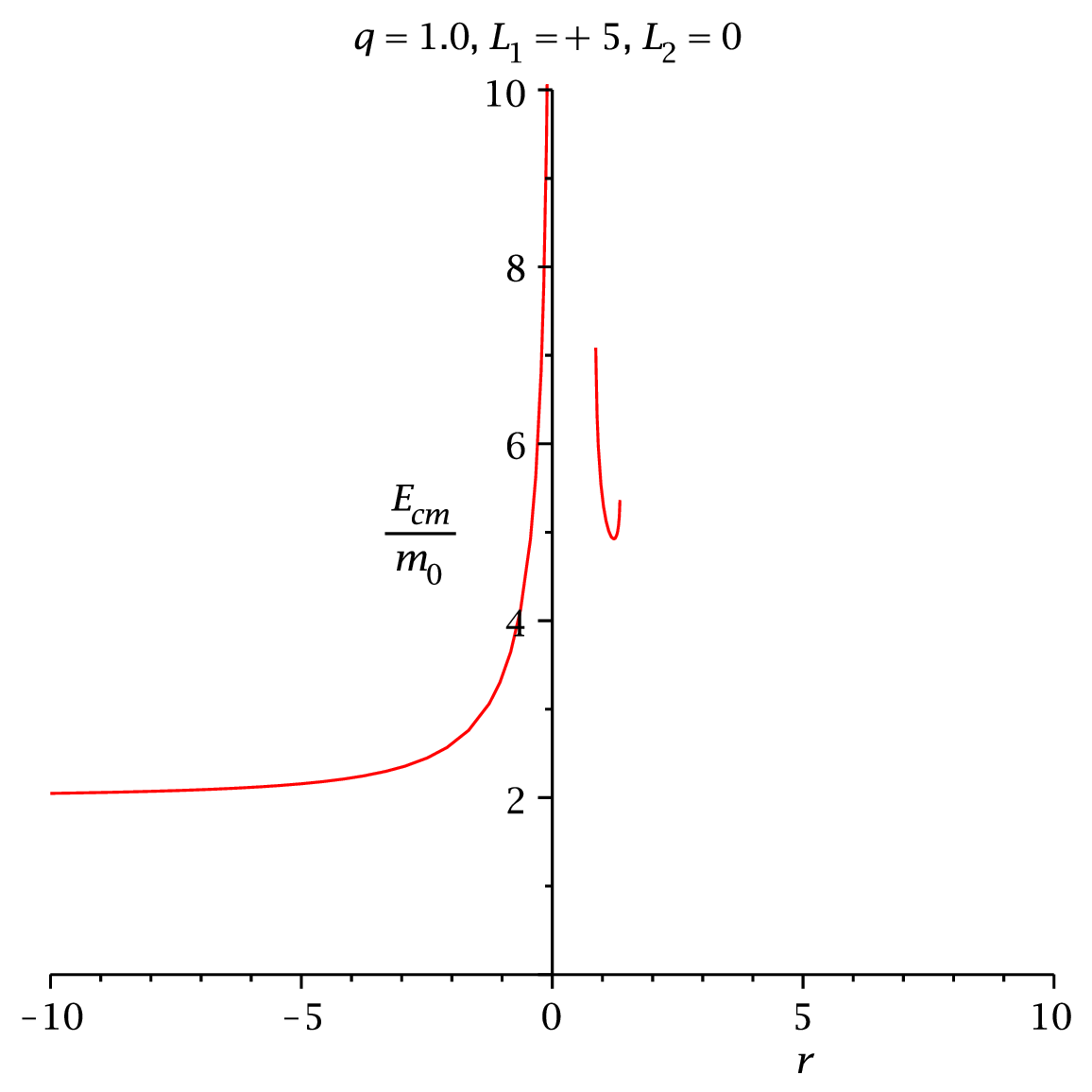}} 
\subfigure[]{
 \includegraphics[width=2.1in,angle=0]{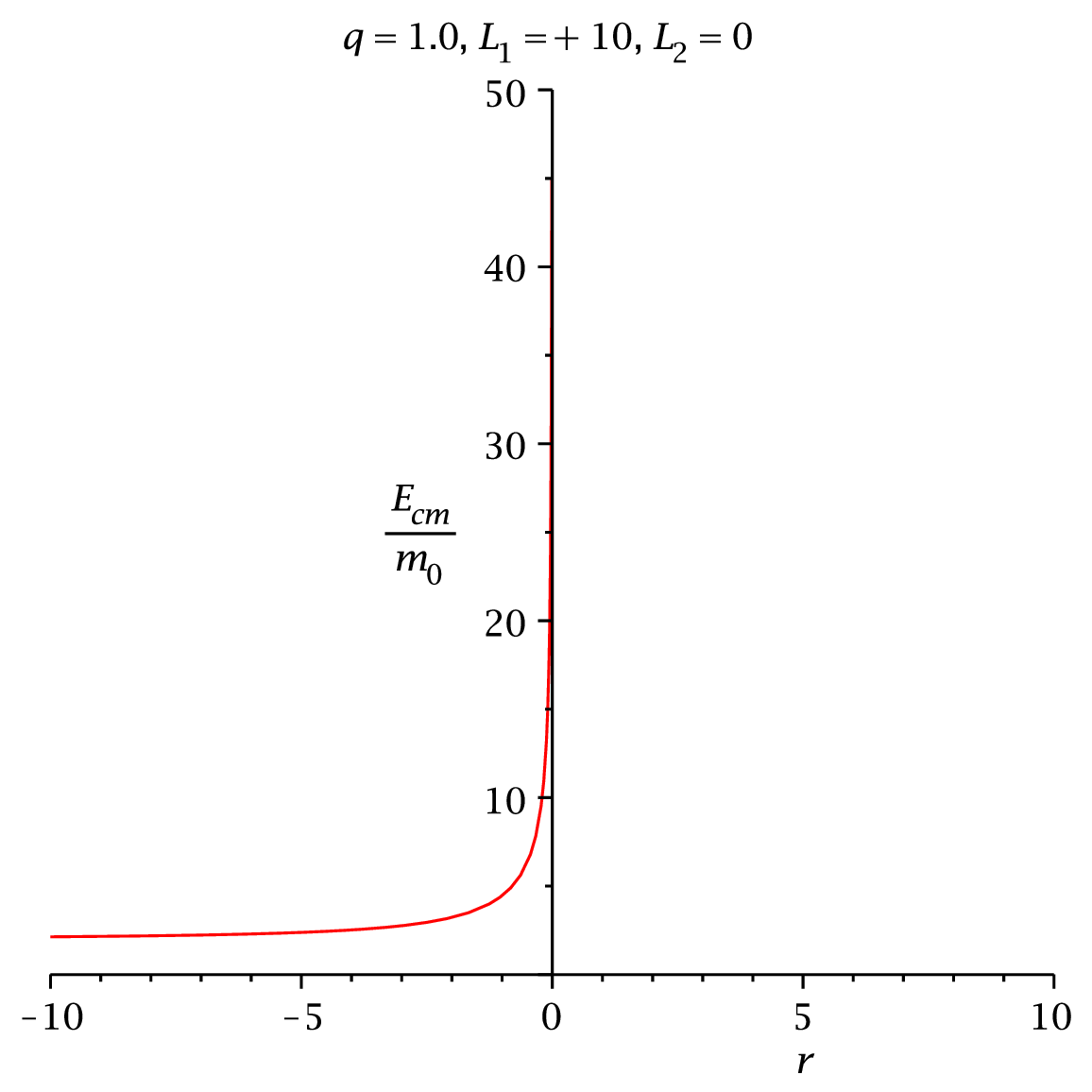}}
\subfigure[]{
 \includegraphics[width=2.1in,angle=0]{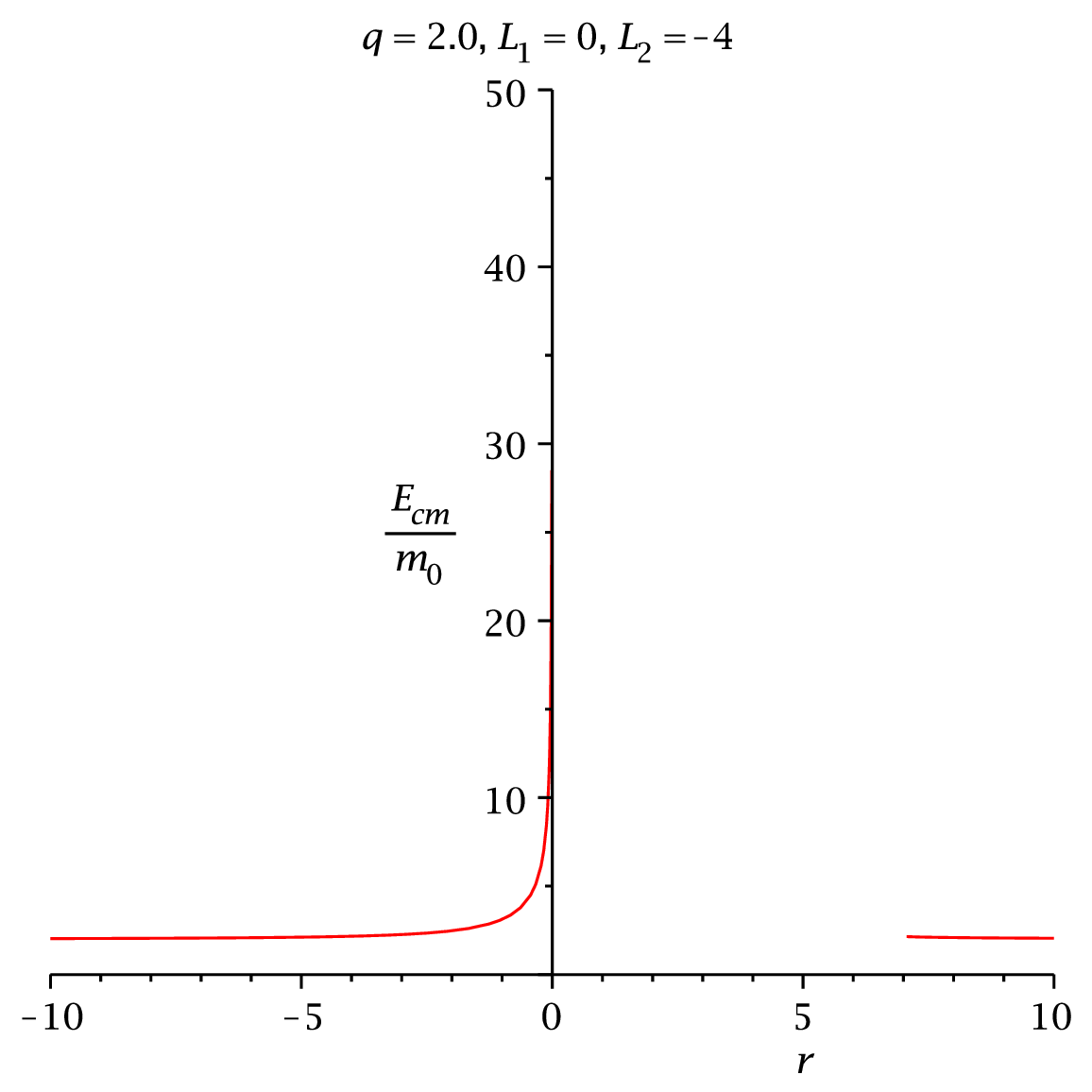}}
 \subfigure[]{
 \includegraphics[width=2.1in,angle=0]{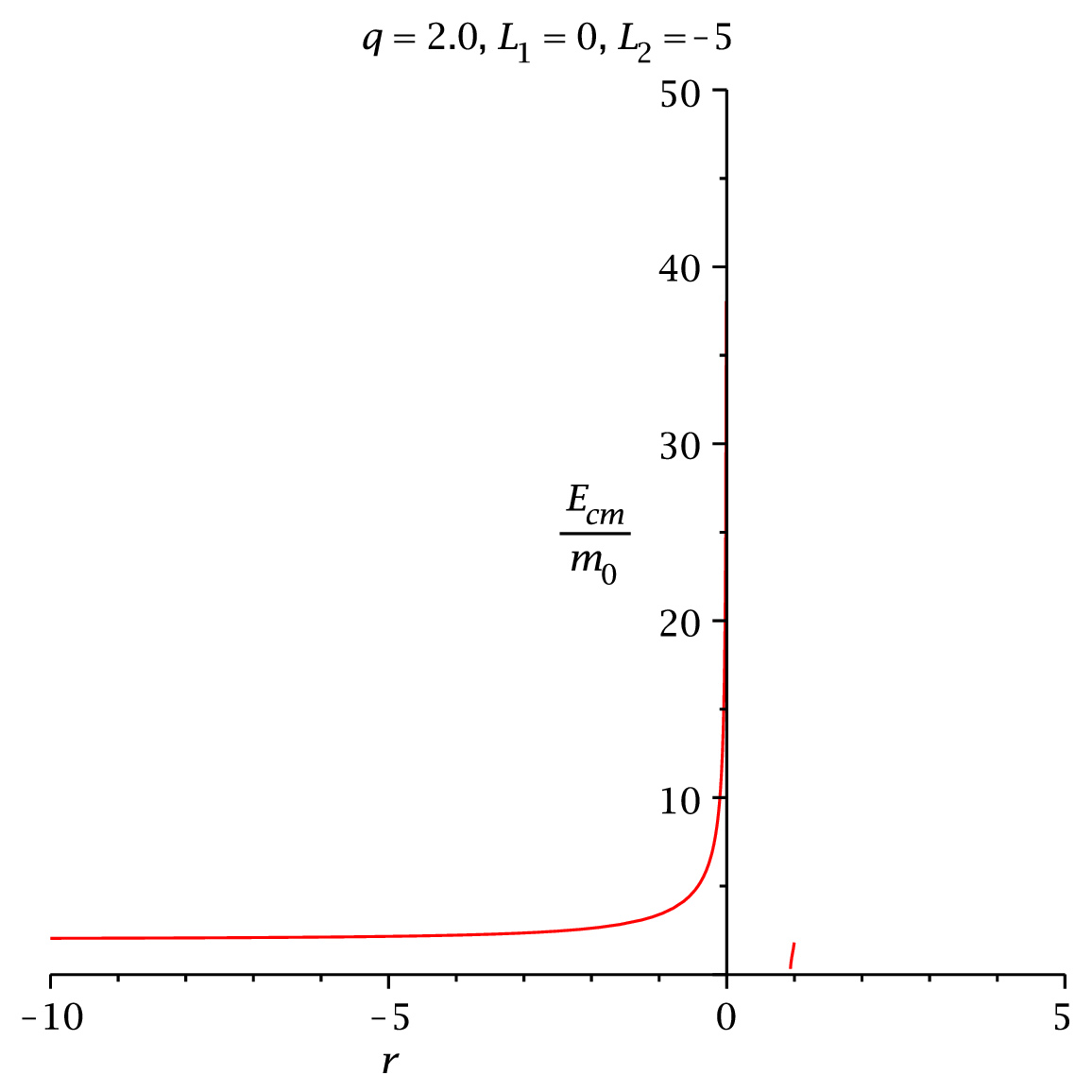}}
\caption{\label{tcm1}\textit{The figure shows the variation  of $E_{cm}$  with $r$ for DMPR BH
with $M=M_{p}=M_{5}=1$. }}
\end{center}
\end{figure}

\begin{figure}[h]
  \begin{center}
\subfigure[]{
\includegraphics[width=2.1in,angle=0]{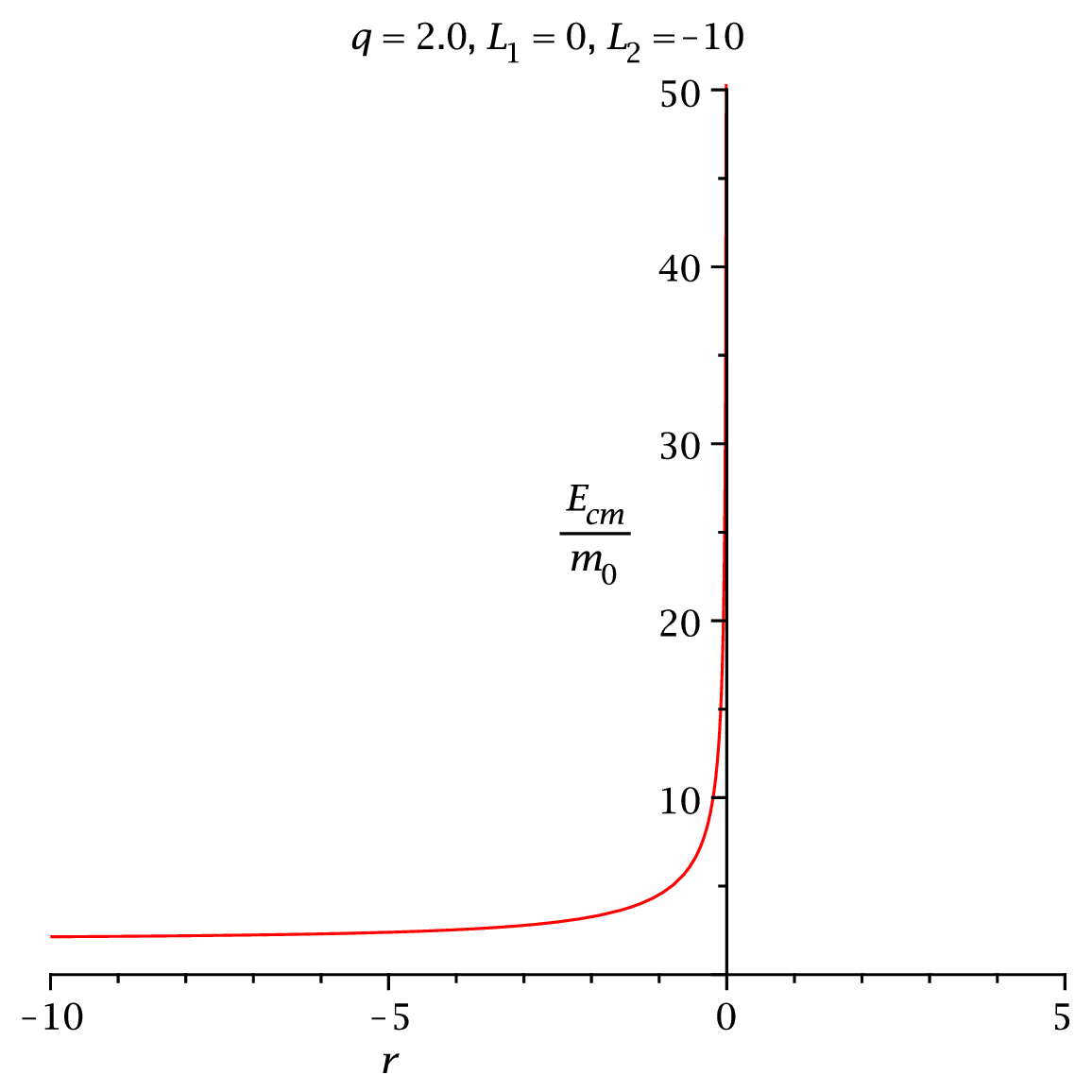}} 
\subfigure[]{
 \includegraphics[width=2.1in,angle=0]{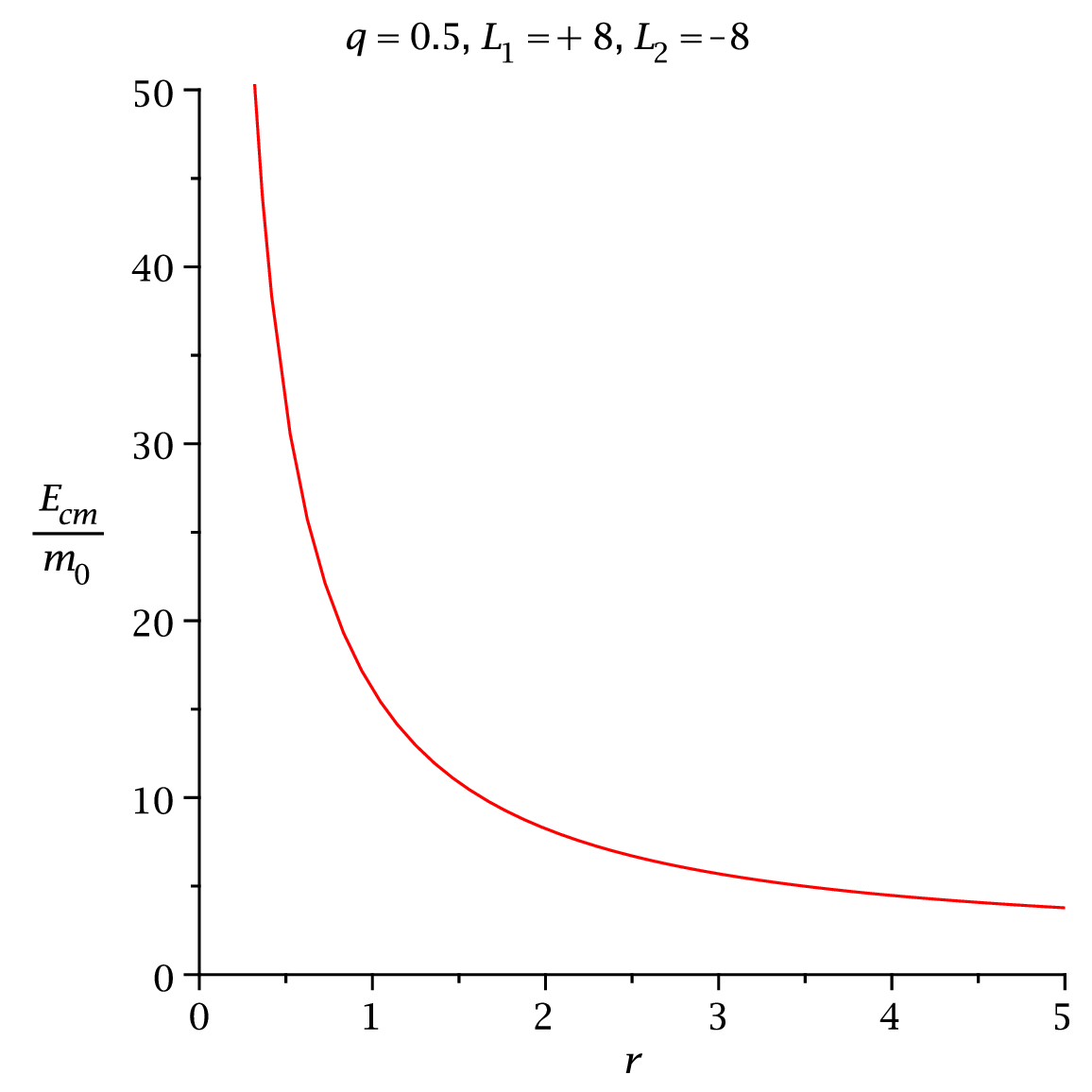}}
\subfigure[]{
 \includegraphics[width=2.1in,angle=0]{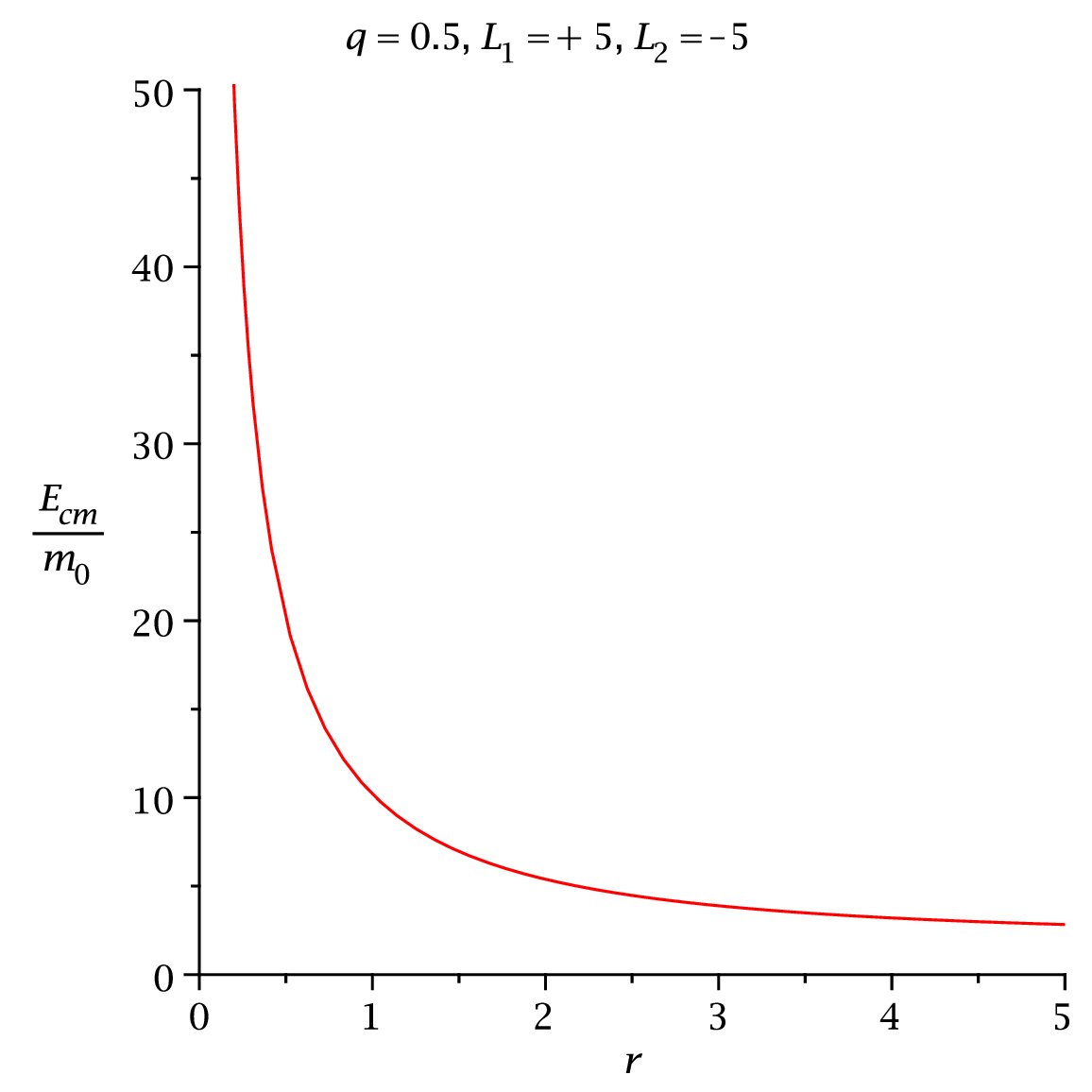}}
 \subfigure[]{
 \includegraphics[width=2.1in,angle=0]{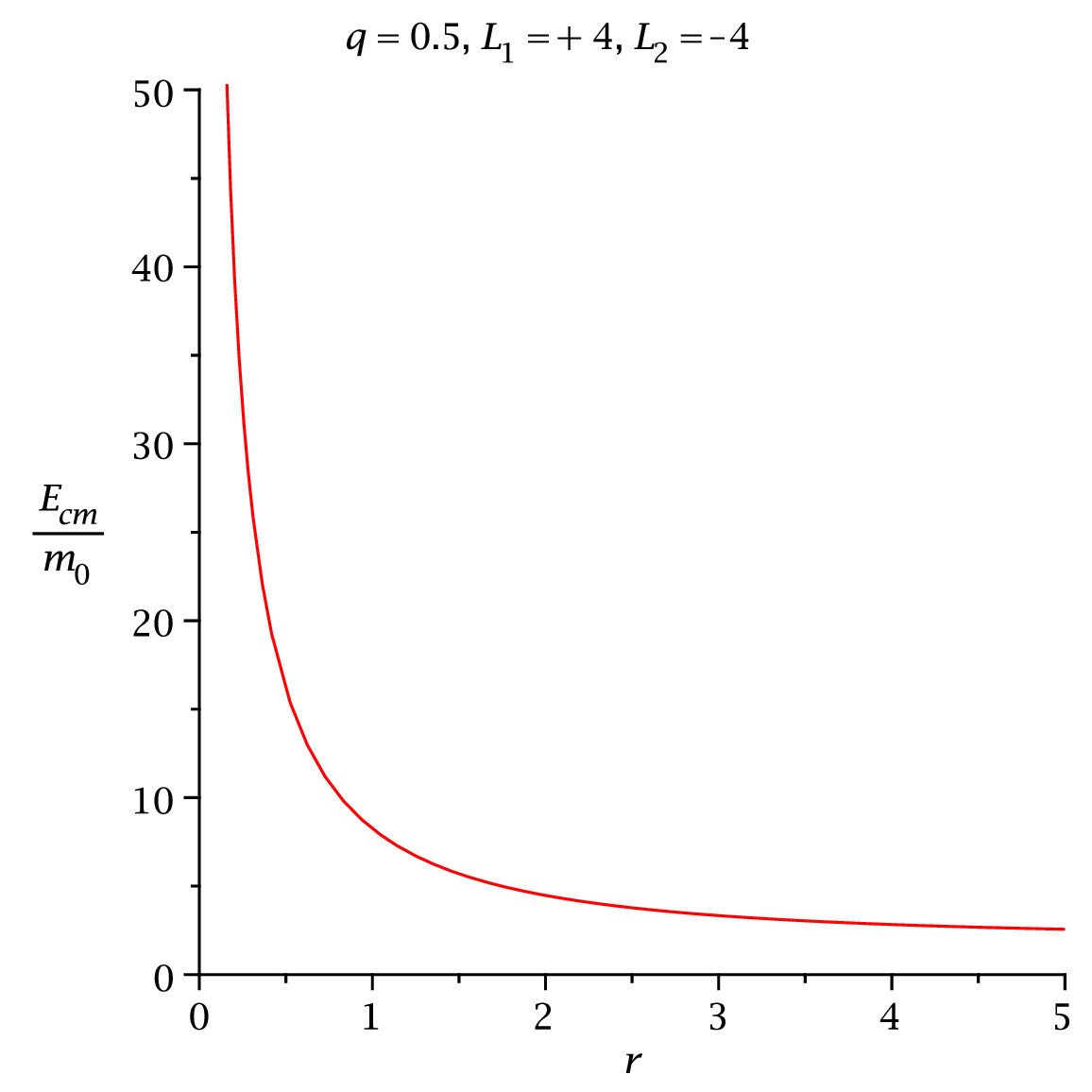}}
\caption{\label{tcm2}\textit{The figure shows the variation  of $E_{cm}$  with $r$ for DMPR BH
with $M=M_{p}=M_{5}=1$. }}
\end{center}
\end{figure}

\begin{figure}[h]
  \begin{center}
\subfigure[]{
\includegraphics[width=2.1in,angle=0]{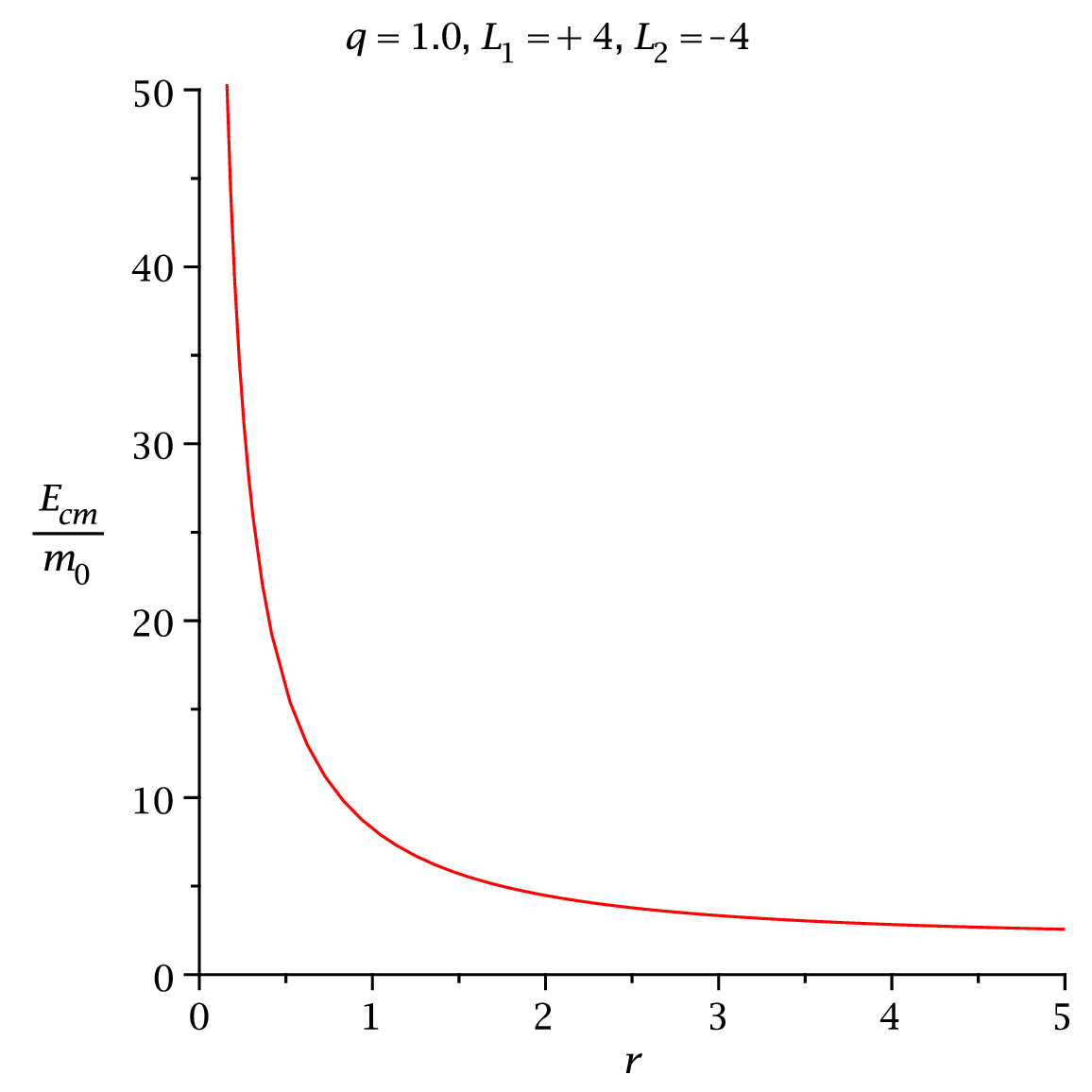}} 
\subfigure[]{
 \includegraphics[width=2.1in,angle=0]{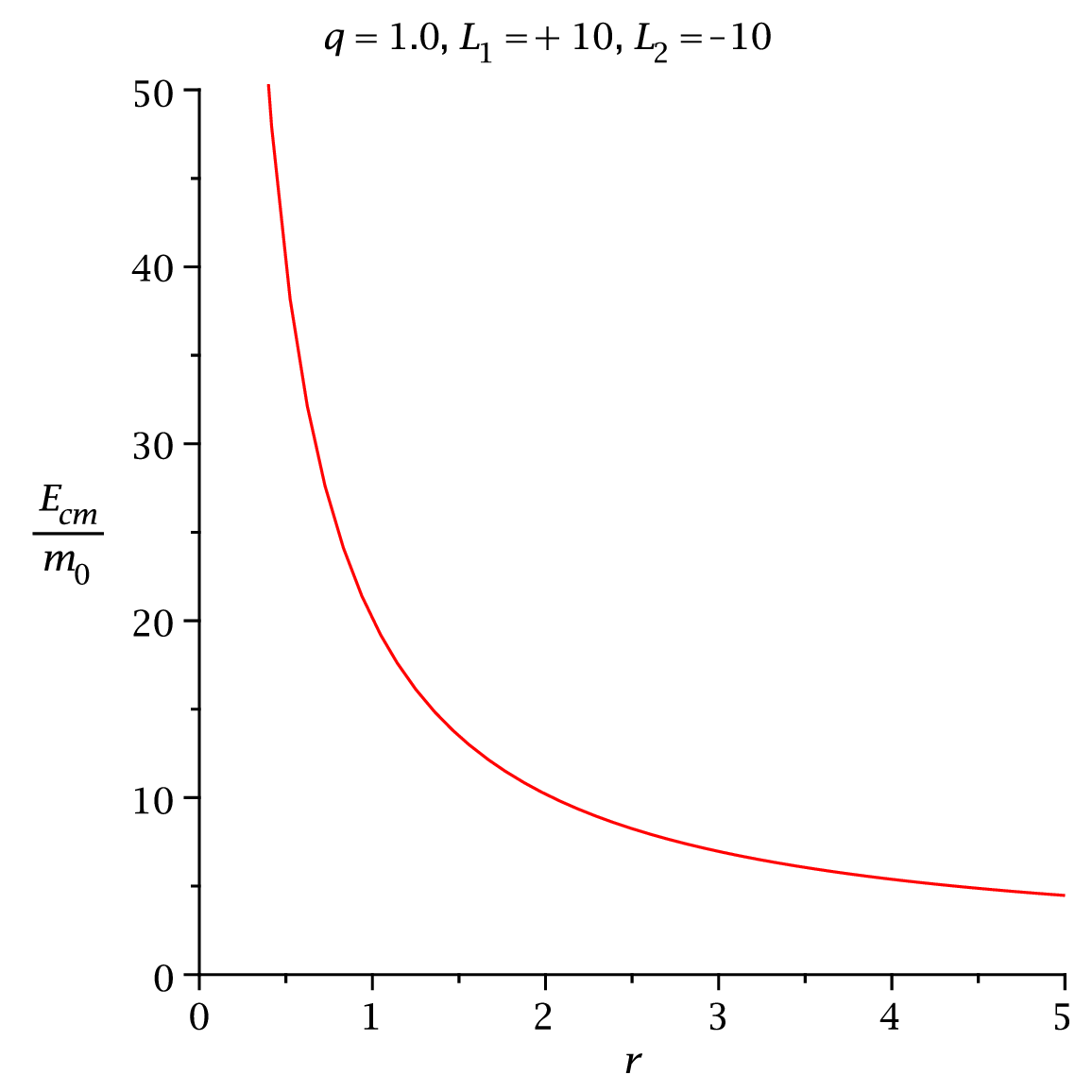}}
\subfigure[]{
 \includegraphics[width=2.1in,angle=0]{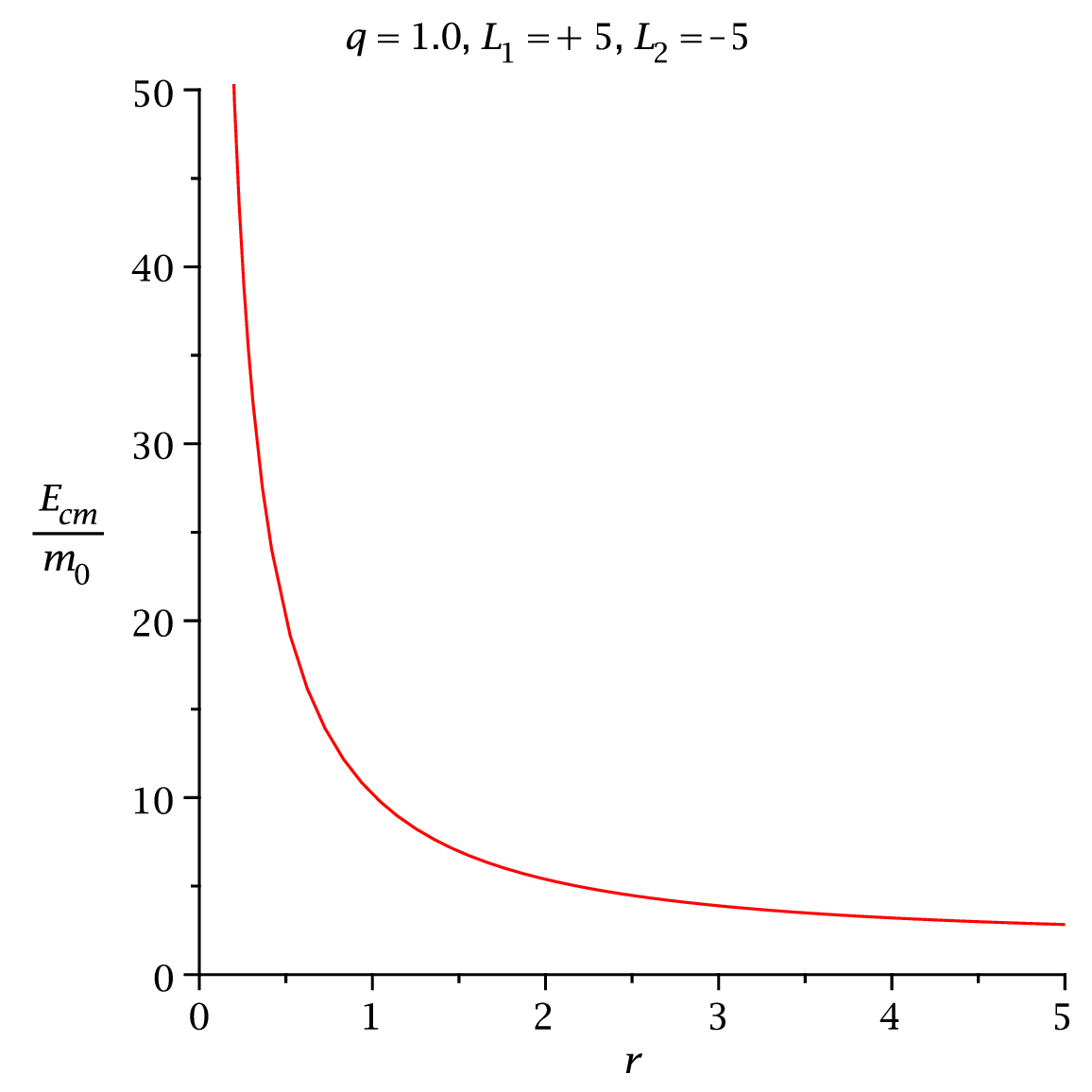}}
 \subfigure[]{
 \includegraphics[width=2.1in,angle=0]{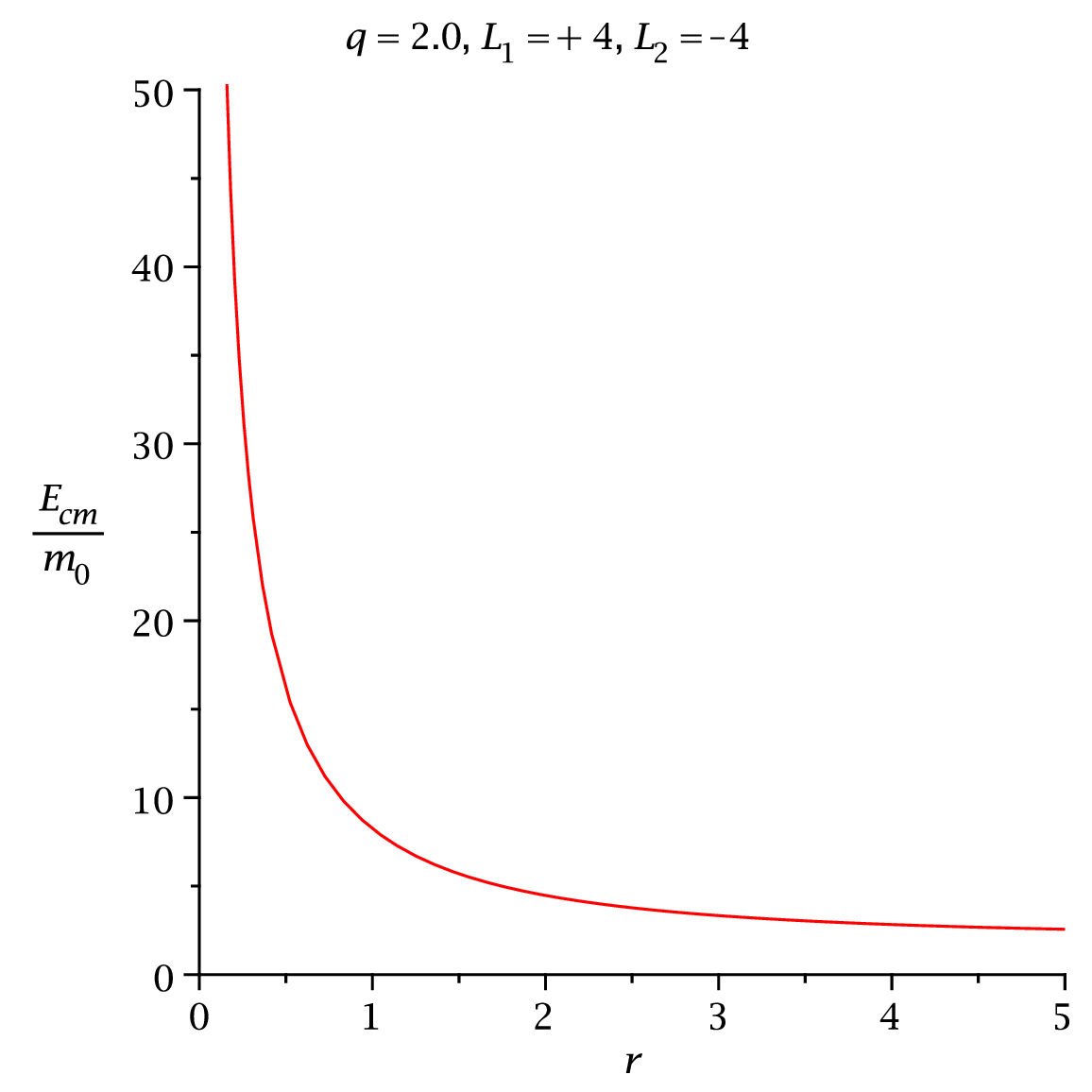}}
\caption{\label{tcm3}\textit{The figure shows the variation  of $E_{cm}$  with $r$ for DMPR BH
with $M=M_{p}=M_{5}=1$. }}
\end{center}
\end{figure}

\begin{figure}
\begin{center}
{\includegraphics[width=0.45\textwidth]{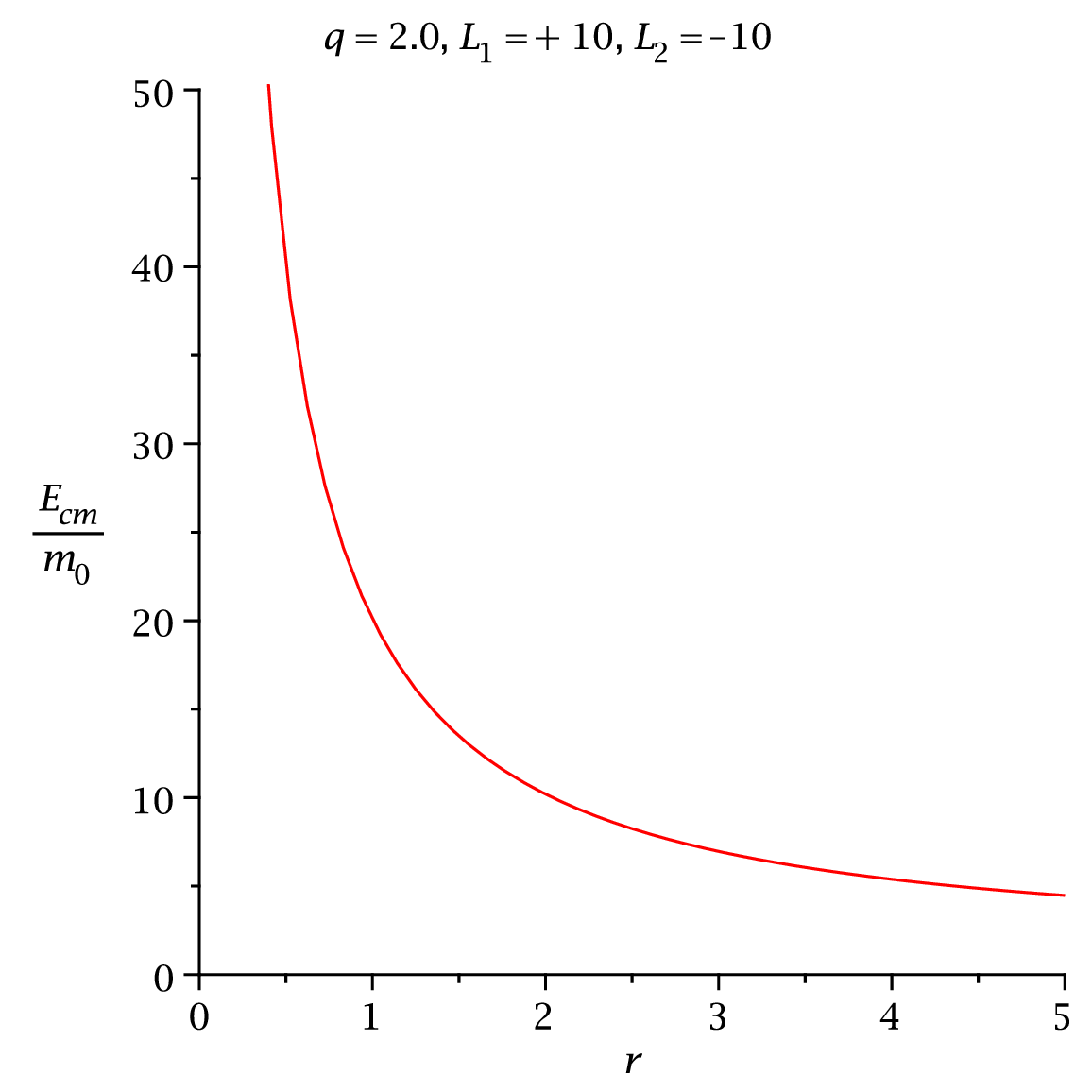}}
{\includegraphics[width=0.45\textwidth]{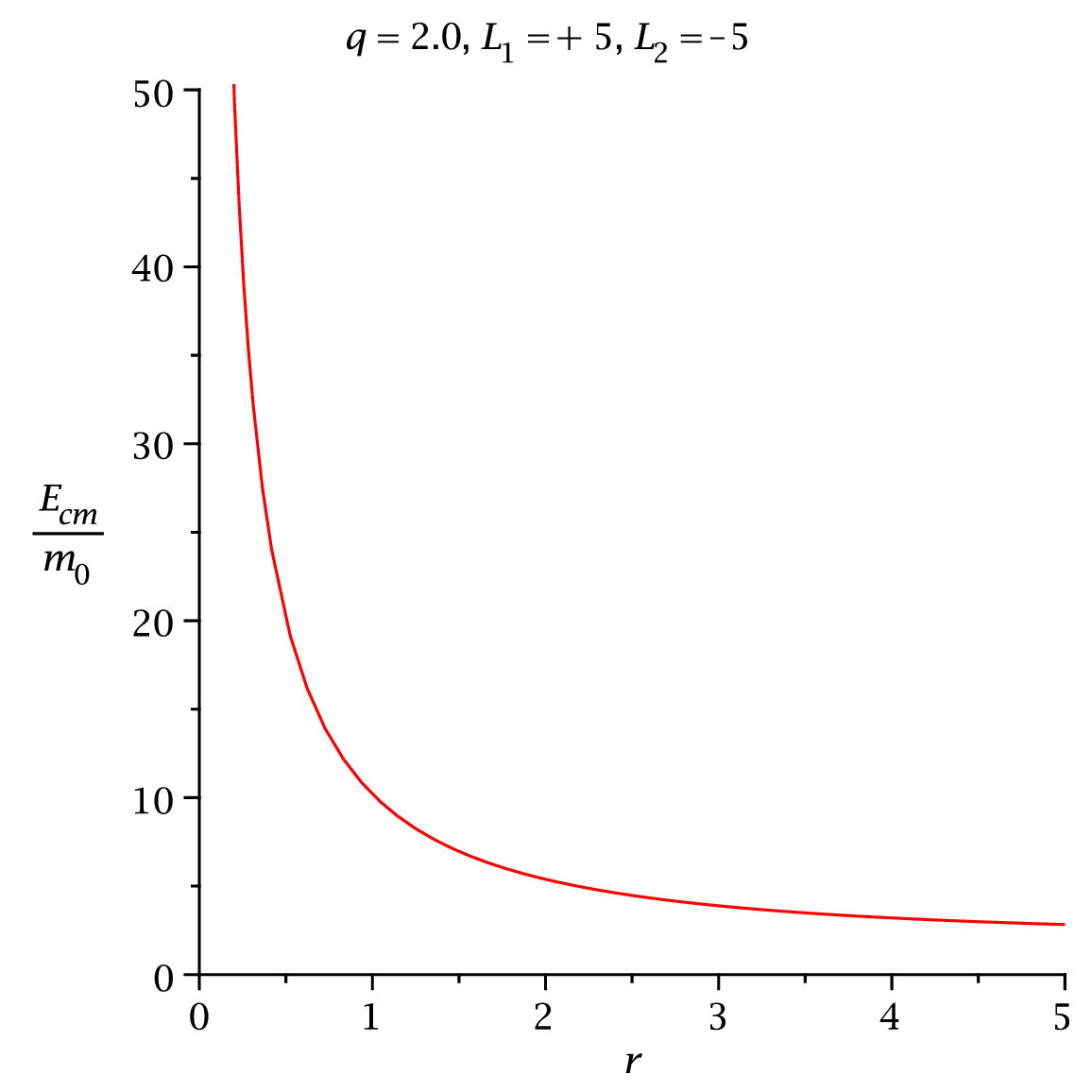}}
\end{center}
\caption{The figure shows the variation  of $E_{cm}$  with $r$ for DMPR BH
with $M=M_{p}=M_{5}=1$.
 \label{tcm4}}
\end{figure}

In Figs. (\ref{tcm},\ref{tcm1},\ref{tcm2},\ref{tcm3},\ref{tcm4}), we have shown that how the CM energy $E_{cm}$ changes 
in case of DMPR BH with the radial coordinate $r$ for various combination of angular momentum parameters $L_{1}$ and $L_{2}$.

\end{document}